\documentclass[prb,twocolumn,amssymb]{revtex4}

\usepackage{graphicx}

\begin{document}
\title{Topological Field Theory of Time-Reversal Invariant Insulators}
\author{Xiao-Liang Qi, Taylor Hughes and Shou-Cheng Zhang
\\
\normalsize{Department of Physics, Stanford University,
Stanford, CA 94305}\\
}
\begin{abstract}
We show that the fundamental time reversal invariant (TRI)
insulator exists in  $4+1$ dimensions, where the effective field
theory is described by the $4+1$ dimensional Chern-Simons theory
and the topological properties of the electronic structure is
classified by the second Chern number. These topological
properties are the natural generalizations of the time reversal
breaking (TRB) quantum Hall insulator in $2+1$ dimensions. The TRI
quantum spin Hall insulator in $2+1$ dimensions and the
topological insulator in $3+1$ dimension can be obtained as
descendants from the fundamental TRI insulator in $4+1$ dimensions
through a dimensional reduction procedure. The effective
topological field theory, and the $Z_2$ topological classification
for the TRI insulators in $2+1$ and $3+1$ dimensions are naturally
obtained from this procedure. All physically measurable
topological response functions of the TRI insulators are
completely described by the effective topological field theory.
Our effective topological field theory predicts a number of novel
and measurable phenomena, the most striking of which is the
topological magneto-electric effect, where an electric field
generates a magnetic field in the same direction, with an
universal constant of proportionality quantized in odd multiples
of the fine structure constant $\alpha=e^2/\hbar c$. Finally, we
present a general classification of all topological insulators in
various dimensions, and describe them in terms of a unified
topological Chern-Simons field theory in phase space.
\end{abstract}

\maketitle
\tableofcontents
\section{Introduction}

Most states or phases of condensed matter can be described by
local order parameters and the associated broken symmetries.
However, the quantum Hall (QH)
state\cite{klitzing1980,tsui1982,laughlin1981,laughlin1983} gives
the first example of  topological states of matter which have
topological quantum numbers different from ordinary states of
matter, and are described in the low energy limit by topological
field theories. Soon after the discovery of the integer QH effect,
the quantization of Hall conductance in units of $e^2/h$ was shown
to be a general property of two-dimensional time reversal breaking
(TRB) band insulators\cite{thouless1982}. The integral of the
curvature of the Berry's phase gauge field defined over the
magnetic Brillouin zone (BZ) was shown to be a topological
invariant called the first Chern number, which is physically
measured as the quanta of the Hall conductance. In the presence of
many-body interactions and disorder, the Berry curvature and the
first Chern number can be defined over the space of twisted
boundary conditions\cite{niu1985}. In the long wave length limit,
both the integer and the fractional QH effect can be described by
the topological Chern-Simons field theory\cite{zhang1989} in $2+1$
dimensions. This effective topological field theory captures all
physically measurable topological effects, including the
quantization of the Hall conductance, the fractional charge, and
the statistics of quasi-particles\cite{zhang1992}.

Insulators in $1+1$ dimensions can also have unique topological
effects. Solitons in charge density wave insulators can have
fractional charge or spin-charge separation\cite{su1979}.  The
electric polarization of these insulators can be expressed in
terms of the integral of the Berry's phase gauge field in momentum
space\cite{kingsmith1993,ortiz1994}. During an adiabatic pumping
cycle, the change of electric polarization, or the net charge
pumped across the 1D insulator, is given by the integral of the
Berry curvature over the hybrid space of momentum and the
adiabatic pumping parameter. This integral is quantized to be a
topological integer\cite{thouless1983}. Both the charge of the
soliton and the adiabatic pumping current can be obtained from the
Goldstone-Wilczek formula\cite{goldstone1981}.

In this paper we shall show that the topological effects in the
$1+1$ dimensional insulator can be obtained from the QH effect of
the $2+1$ dimensional TRB insulator by a procedure called
dimensional reduction. In this procedure one of the momenta is
replaced by an adiabatic parameter, or field, and the
Goldstone-Wilczek formula, and thus, all topological effects of
the $1+1$ dimensional insulators, can be derived from the $2+1$
dimensional QH effect. The procedure of dimensional reduction can
be generalized to the higher dimensional TRI insulators and
beyond, which is the key result of this paper.

In recent years, the QH effect of the $2+1$ dimensional TRB
insulators has been generalized to TRI insulators in various
dimensions. The first example of a topologically non-trivial TRI
state in condensed matter context was the 4D generalization of the
QH effect (4DQH) proposed in Ref. \onlinecite{zhang2001}.  The
effective theory of this model is given by the Chern-Simons
topological field theory in $4+1$ dimensions\cite{bernevig2002}.
The quantum spin Hall (QSH) effect has been proposed in $2+1$
dimensional TRI quantum models\cite{kane2005A,bernevig2006a}. The
QSH insulator state has a gap for all bulk excitations, but has
topologically protected gapless edge states, where opposite spin
states counter-propagate\cite{kane2005A,wu2006,xu2006}. Recently
the QSH state has been theoretically predicted\cite{bernevig2006d}
and experimentally observed in HgTe quantum
wells\cite{koenig2007}. TRI topological insulators have also been
classified in $3+1$ dimensions\cite{fu2007b,moore2007,roy2006a}.
These 3D states all carry spin Hall current in the insulating
state\cite{murakami2004A}.

The topological properties of the $4+1$ dimensional TRI insulator
can be described by the second Chern number defined over  four
dimensional momentum space. On the other hand, TRI insulators in
$2+1$ and $3+1$ dimensions are described by a $Z_2$ topological
invariant defined over momentum
space\cite{kane2005A,kane2005B,fu2006,fu2007a,fu2007b,moore2007,roy2006a,roy2006b,roy2006c}.
In the presence of interactions and disorder, the momentum space
$Z_2$ invariant is not well defined, however, one can define a
more general $Z_2$ topological invariant in terms of spin-charge
separation associated with a $\pi$ flux\cite{qi2008,ran2008}. One
open question in this field concerns the relationship between the
classification of the $4+1$ dimensional TRI insulator by the
second Chern number and the classification of the $3+1$ and $2+1$
dimensional TRI insulators by the $Z_2$ number.

The effective theory of the $4+1$ dimensional TRI insulator is
given by the topological Chern-Simons field
theory\cite{bernevig2002,niemi1983}. While the $2+1$ dimensional
Chern-Simons theory describes a linear topological response to an
external $U(1)$ gauge field\cite{zhang1989,zhang1992}, the $4+1$
dimensional Chern-Simons theory describes a nonlinear topological
response to an external $U(1)$ gauge field. The key outstanding
theoretical problem in this field is the search for the
topological field theory describing the TRI insulators in $2+1$
and $3+1$ dimensions, from which all measurable topological
effects can be derived.

In this paper, we solve this outstanding problem by constructing
topological field theories for the $2+1$ and $3+1$ dimensional TRI
insulators using the procedure of dimensional reduction. We show
that the $4+1$ dimensional topological insulator is the
fundamental state from which all lower dimensional TRI insulators
can be derived. This procedure is analogous to the dimensional
reduction from the $2+1$ dimensional TRB topological insulator to
the $1+1$ dimensional insulators. There is a deep reason why the
fundamental TRB topological insulator exists in $2+1$ dimensions,
while the fundamental TRI topological insulator exists in $4+1$
dimensions. The reason goes back to the Wigner-von Neumann
classification\cite{neumann1929} of level crossings in TRB unitary
quantum systems and the TRI symplectic quantum systems.
Generically three parameters need to be tuned to obtain a level
crossing in a TRB unitary system, while five parameters need to be
tuned to obtain a level crossing in a TRI symplectic system. These
level crossing singularities give rise to the non-trivial
topological curvatures on the 2D and 4D parameter surfaces which
enclose the singularities. Fundamental topological insulators are
obtained in space dimensions where all these parameters are
momentum variables. Once the fundamental TRI topological insulator
is identified in $4+1$ dimensions, the lower dimensional versions
of TRI topological insulators can be easily obtained by
dimensional reduction. In this procedure, one or two momentum
variables of the $4+1$ dimensional topological insulator are
replaced by adiabatic parameters or fields, and the $4+1$
dimensional Chern-Simons topological field theory is reduced to
topological field theories involving both the external $U(1)$
gauge field and the adiabatic fields. For the $3+1$ TRI
insulators, the topological field theory is given by that of the
``axion Lagrangian", or the $3+1$ dimensional $\theta$ vacuum
term, familiar in the context of quantum chromodynamics (QCD),
where the adiabatic field plays the role of the axion field or the
$\theta$ angle. From these topological field theories, all
physically measurable topological effects of the $3+1$ and the
$2+1$ dimensional TRI insulators can be derived. We predict a
number of novel topological effects in this paper, the most
striking of which is the topological magneto-electric (TME)
effect, where an electric field induces a magnetic field in the
same direction, with a universal constant of proportionality
quantized in odd multiples of the fine structure constant
$\alpha=e^2/\hbar c$. We also present an experimental proposal to
measure this novel effect in terms of Faraday rotation. Our
dimensional reduction procedure also naturally produces the $Z_2$
classification of the $3+1$ and the $2+1$ dimensional TRI
topological insulators in terms of the integer second Chern class
of the $4+1$ dimensional TRI topological insulators.

The remaining parts of the paper are organized as follows. In Sec.
II we review the physical consequences of the first Chern number,
namely the $(2+1)$-d QH effect and $(1+1)$-d fractional charge and
topological pumping effects. We begin with the $(2+1)$-d time
reversal breaking insulators and study the topological transport
properties. We then present a dimensional reduction procedure that
allows us to consider related topological phenomena in $(1+1)$-d
and $(0+1)$-d. Subsequently, we define a $Z_2$ classification of
these lower dimensional descendants which relies on the presence
of a discrete particle-hole symmetry. This will serve as a review
and a warm-up exercise for the more complicated phenomena we
consider in the later sections. In Secs. III, IV, and V we discuss
consequences of a non-trivial second Chern number beginning with a
parent $(4+1)$-d topological insulator in Sec. III. In Secs. IV
and V we continue studying the consequences of the second Chern
number but in the physically realistic $(3+1)$-d and $(2+1)$-d
models which are the descendants of the initial $(4+1)$-d system.
We present effective actions describing all of the physical
systems and their responses to applied electromagnetic fields.
This provides the first effective field theory for the TRI
topological insulators in $(3+1)$-d and $(2+1)$-d. For these two
descendants of the $(4+1)$-d theory, we show that the $Z_2$
classification of the decedents are obtained from the 2nd Chern
number classification of the parent TRI insulator. Finally, in
Sec. VI we unify all of the results into families of topological
effective actions defined in a phase space formalism. From this we
construct a family tree of all topological insulators, some of
which are only defined in higher dimensions, and with topological
$Z_2$ classifications which repeat every $8$ dimensions.

This paper contains many new results on topological insulators,
but it can also be read by advanced students as a pedagogical and
self-contained introduction of topology applied to condensed
matter physics. Physical models are presented in the familiar
tight-binding forms, and all topological results can be derived by
exact and explicit calculations, using techniques such as response
theory already familiar in condensed matter physics. During the
course of reading this paper, we suggest the readers to consult
Appendix A which covers all of our conventions.

\section{TRB topological insulators in $2+1$ dimensions and its dimensional
reduction}\label{sec:firstChern}

In this section, we review the physics of the TRB topological
insulators in $2+1$ dimensions. We shall use the example of a
translationally invariant tight-binding model\cite{qi2006} which
realizes the QH effect without Landau levels. We discuss the
procedure of dimensional reduction, from which all topological
effects of the $1+1$ dimensional insulators can be obtained. This
section serves as a simple pedagogical example for the more
complex case of the TRI insulators presented in Sec. III and IV.

\subsection{The first Chern number and topological response function
in $(2+1)$-d }\label{sec:response2d}

In general, the tight-binding Hamiltonian of a $(2+1)$-d band
insulator can be expressed as
\begin{eqnarray}
H=\sum_{m,n;\alpha,\beta}c_{m\alpha}^\dagger
h_{mn}^{\alpha\beta}c_{n\beta}\label{H2d}
\end{eqnarray}
with $m,n$ the lattice sites and $\alpha,\beta=1,2,..N$ the band
indices for a $N$-band system. With translation symmetry
$h_{mn}^{\alpha\beta}=h^{\alpha\beta}\left(\vec{r}_m-\vec{r}_n\right)$,
the Hamiltonian can be diagonalized in a Bloch wavefunction basis:
\begin{eqnarray}
H=\sum_{\bf k}c_{{\bf k}\alpha}^\dagger h^{\alpha\beta}\left({\bf
k}\right)c_{{\bf k}\beta}\label{H2dk}
\end{eqnarray}
The minimal coupling to an external electro-magnetic field is
given by $h_{mn}^{\alpha\beta}\rightarrow
h_{mn}^{\alpha\beta}e^{iA_{mn}}$ where $A_{mn}$ is a gauge
potential defined on a lattice link with sites $m,n$ at the end.
To linear order, the Hamiltonian coupled to the electro-magnetic
field is obtained as
\begin{eqnarray}
H\simeq\sum_{\bf k}c_{{\bf k}}^\dagger h\left({\bf
k}\right)c_{{\bf k}}+\sum_{\bf k,q}A^i(-{\bf q})c_{\bf
k+q/2}^\dagger \frac{\partial h({\bf k})}{\partial k_i} c_{\bf
k-q/2}\nonumber
\end{eqnarray}
with the band indices omitted. The DC response of the system to
external field $A^i({\bf q})$ can be obtained by the standard Kubo
formula:
\begin{eqnarray}
\sigma_{ij}&=&\lim_{\omega\rightarrow
0}\frac i\omega Q_{ij}(\omega+i\delta)\ , \nonumber\\
Q_{ij}(i\nu_m)&=&\frac{1}{\Omega\beta}\sum_{{\bf k},n}{\rm
tr}\left(J_{i}({\bf k})G({\bf
k},i(\omega_n+\nu_m))\right. \nonumber\\
& &\left.\cdot J_{j}({\bf k})G({\bf k},i\omega_n)\right)\ ,
\label{kubo}
\end{eqnarray}
with the DC current $J_{i}({\bf k})=\partial h({\bf k})/\partial
k_i,~i,j=x,y$, Green's function $G({\bf
k},i\omega_n)=\left[i\omega_n-h({\bf k})\right]^{-1}$, and
$\Omega$ the area of the system. When the system is a band
insulator with $M$ fully-occupied bands, the longitudinal
conductance vanishes, {\it i.e.} $\sigma_{xx}=0$, as expected,
while $\sigma_{xy}$ has the form shown in Ref.
\onlinecite{thouless1982}:
\begin{eqnarray}
\sigma_{xy}&=&\frac{e^2}{h}\frac1{2\pi}\int dk_x \int dk_y
f_{xy}\left({\bf k}\right)\label{sigmaxy}\\
\text{with~}f_{xy}\left({\bf k}\right)&=&\frac{\partial a_y({\bf
k})}{\partial
k_x}-\frac{\partial a_x({\bf k})}{\partial k_y}\nonumber\\
a_i({\bf k})&=&-i\sum_{\alpha\in ~{\rm occ}}\left\langle
\alpha{\bf k}\right|\frac{\partial}{\partial k_i}\left|\alpha{\bf
k}\right\rangle,~i=x,y.\nonumber
\end{eqnarray}
 Physically, $a_i({\bf k})$
is the $U(1)$ component of the Berry's phase gauge field
(adiabatic connection) in momentum space. The quantization of the
first Chern number
\begin{eqnarray}
C_1=\frac1{2\pi}\int dk_x\int dk_y f_{xy}({\bf k})\in \mathbb{Z}
\end{eqnarray}
is satisfied for any continuous states $\left|\alpha{\bf
k}\right\rangle$ defined on the BZ.

Due to charge conservation, the QH response
$j_i=\sigma_H\epsilon^{ij}E_j$ also induces another response
equation:
\begin{eqnarray}
j_i&=&\sigma_H\epsilon^{ij}E_j\label{current}\\
& \Rightarrow &\frac{\partial \rho}{\partial t}=-\nabla\cdot
{\textbf{j}}=-\sigma_H\nabla\times{\bf
E}=\sigma_H\frac{\partial B}{\partial t}\nonumber\\
& \Rightarrow & \rho(B)-\rho_0=\sigma_H B\label{density}
\end{eqnarray}
where $\rho_0=\rho(B=0)$ is the charge density in the ground
state. Equations (\ref{current}) and (\ref{density}) can be
combined together in a covariant way:
\begin{eqnarray}
j^\mu=\frac{C_1}{2\pi}\epsilon^{\mu\nu\tau}\partial_\nu
A_\tau\label{response2d}
\end{eqnarray}
where $\mu,\nu,\tau=0,1,2$ are temporal and spatial indices. Here
and below we will take the units $e=\hbar=1$ so that $e^2/h=1/2\pi$.

The response equations (\ref{response2d}) can be described by the
topological Chern-Simons field theory of the external field
$A_\mu$:
\begin{eqnarray}
S_{\rm eff}&=&\frac{C_1}{4\pi}\int d^2x\int dt
A_\mu\epsilon^{\mu\nu\tau}\partial_\nu A_\tau,\label{Seff2d}
\end{eqnarray}
in the sense that $\delta S_{\rm eff}/\delta A_\mu=j^\mu$ recovers
the response equations (\ref{response2d}). Such an effective
action is topologically invariant, in agreement with the
topological nature of the first Chern number. All topological
responses of the QH state are contained in the Chern-Simons
theory\cite{zhang1992}.

\subsection{Example: two band models}\label{twobandmodel}

\begin{figure}[t!]
\begin{center}
\includegraphics[width=3in] {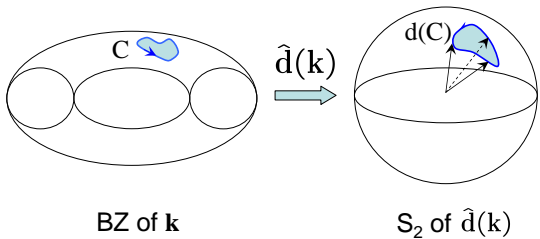}
\end{center}
\caption{Illustration of the Berry's phase curvature in a two-band
model. The Berry's phase $\oint_C {\bf A\cdot dr}$ around a path $C$
in the BZ is half of the solid angle subtended by the image path
$d(C)$ on the sphere $S_2$.} \label{berryphase}
\end{figure}

To make the physical picture clearer, the simplest case of a two
band model can be studied as an example\cite{qi2006}. The
Hamiltonian of a two-band model can be generally written as
\begin{eqnarray}
h({\bf k})=\sum_{a=1}^3d_a({\bf k})\sigma^a+\epsilon({\bf
k})\mathbb{I}\label{twobandH}
\end{eqnarray}
where $\mathbb{I}$ is the $2\times 2$ identity matrix and
$\sigma^a$ are the three Pauli matrices. Here we assume that the
$\sigma^a$ represent a spin or pseudo-spin degree of freedom. If
it is a real spin then the $\sigma^a$ are thus odd under time
reversal. If If the $d_a({\bf k})$ are odd in ${\bf k}$ then the
Hamiltonian is time-reversal invariant. However, if any of the
$d_a$ contain a constant term then the model has explicit
time-reversal symmetry breaking. If the $\sigma^a$ are a
pseudo-spin then one has to be more careful. Since, in this case,
${\cal{T}}^2=1$ then only $\sigma^y$ is odd under time-reversal
(because it is imaginary) while $\sigma^x,\sigma^z$ are even. The
identity matrix is even under time-reversal and $\epsilon({\bf
k})$ must be even in ${\bf k}$ to preserve time-reversal. The
energy spectrum is easily obtained: $E_{\pm}({\bf
k})=\epsilon({\bf k})\pm\sqrt{\sum_ad_a^2({\bf k})}$. When
$\sum_ad_a^2({\bf k})>0$ for all ${\bf k}$ in the BZ, the two
bands never touch each other. If we also require that ${\rm
max}_{\bf k} (E_{-}({\bf k})) < {\rm min}_{\bf k} (E_{+}({\bf
k}))$, so that the gap is not closed indirectly, then a gap always
exists between the two bands of the system. In the single particle
Hamiltonian $h({\bf k})$, the vector ${\bf d}({\bf k})$ acts as a
``Zeeman field" applied to a ``pseudospin" $\sigma_i$ of a two
level system. The occupied band satisfies $({\bf d}({\bf k})\cdot
{\bf \sigma}) \left|-,{\bf k}\right\rangle=-\left|{\bf d}({\bf
k})\right|\left|-,{\bf k}\right\rangle$, which thus corresponds to
the spinor with spin polarization in the $-{\bf d}({\bf k})$
direction. Thus the Berry's phase gained by $\left|-,{\bf
k}\right\rangle$ during an adiabatic evolution along some path $C$
in ${\bf k}$-space is equal to the Berry's phase a spin-$1/2$
particle gains during the adiabatic rotation of the magnetic field
along the path ${\bf d}(C).$ This is known to be half of the solid
angle subtended by ${\bf d}(C)$, as shown in Fig.\ref{berryphase}.
Consequently, the first Chern number $C_1$ is determined by the
winding number of ${\bf d}({\bf k})$ around the
origin\cite{volovik2003,qi2006}:
\begin{eqnarray}
C_1=\frac1{4\pi}\int dk_x\int dk_y\hat{\bf d}\cdot \frac{\partial
\hat{\bf d}}{\partial k_x}\times\frac{\partial \hat{\bf d}}{\partial
k_y}.\label{winding2band}
\end{eqnarray}
From the response equations we know that a non-zero $C_1$ implies
a quantized Hall response. The Hall effect can only occur in a
system with time-reversal symmetry breaking so if $C_1\neq 0$ then
time-reversal symmetry is broken. Historically, the first example
of such a two-band model with a non-zero Chern number was a
honeycomb lattice model with imaginary next-nearest-neighbor
hopping proposed by Haldane\cite{haldane1988}.

To be concrete, we shall study a particular two band model
introduced in Ref. \cite{qi2006}, which is given by
\begin{eqnarray}
h({\bf k})&=&(\sin k_x)\sigma_x+(\sin k_y)\sigma_y\nonumber\\
& &+\left(m+\cos k_x+\cos k_y\right)\sigma_z,\label{DiracH}
\end{eqnarray}
This Hamiltonian corresponds to the form (\ref{twobandH}) with
$\epsilon({\bf k})\equiv 0$ and $d({\bf k})=\left(\sin k_x,\sin
k_y,m+\cos k_x+\cos k_y\right)$. The Chern number of this system
is \cite{qi2006}
\begin{eqnarray}
C_1=\left\{\begin{array}{cc}1,&0<m<2\\-1,&-2<m<0\\0,&\text{otherwise}.\end{array}\right.
\end{eqnarray}
In the continuum limit, this model reduces to the $2+1$ dimensional
massive Dirac Hamiltonian $$h({\bf k})= k_x \sigma_x+ k_y\sigma_y+
(m+2)
\sigma_z=\left(\begin{array}{cc}m+2&k_x-ik_y\\k_x+ik_y&-m-2\end{array}\right).$$

In a real space, this model can be expressed in tight-binding form
as
\begin{eqnarray}
H&=&\sum_{n}\left[c_n^\dagger\frac{\sigma_z-i\sigma_x}2c_{n+\hat{x}}+c_n^\dagger\frac{\sigma_z-i\sigma_y}2c_{n+\hat{y}}+h.c.\right]\nonumber\\
& &+m\sum_n c_n^\dagger \sigma_zc_n\label{RSDiracH}
\end{eqnarray}
Physically, such a model describes the quantum anomalous Hall
effect realized with both strong spin-orbit coupling ($\sigma_{x}$
and $\sigma_y$ terms) and ferromagnetic polarization ($\sigma_z$
term). Initially this model was introduced for its simplicity in
Ref. \onlinecite{qi2006}, however, recently, it was shown that it
can be physically realized in ${\rm
Hg_{1-x}Mn_xTe/Cd_{1-x}Mn_xTe}$ quantum wells with a proper amount
of ${\rm Mn}$ spin polarization\cite{liu2008}.

\subsection{Dimensional reduction}\label{sec:dimreduction1d}

\begin{figure}[tbp]
\begin{center}
\includegraphics[width=3.5in] {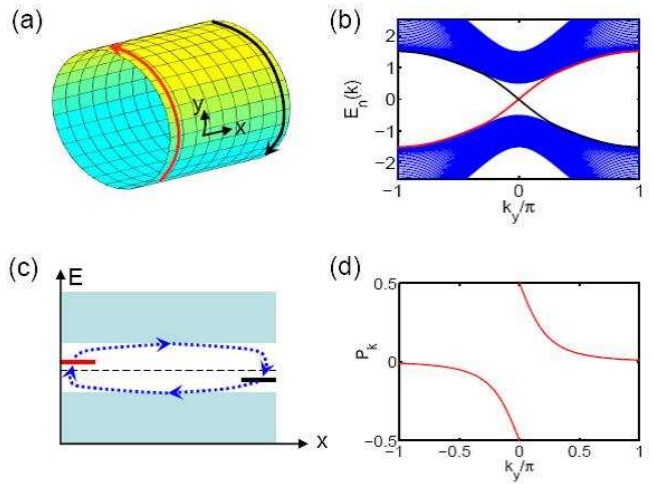}
\end{center}
\caption{(a) Illustration of a square lattice with cylindrical
geometry and the chiral edge states on the boundary. The
definition of $x$ and $y$ axis are also shown by black arrows. (b)
One-d energy spectrum of the model in Eq. (\ref{DiracH}) with
$m=-1.5$. The red and black line stands for the left and right
moving edge states, respectively, while the blue lines are bulk
energy levels. (c) Illustration of the edge states evolution for
$k_y=0\rightarrow 2\pi$. The arrow shows the motion of end states
in the space of center-of-mass position versus energy. (d)
Polarization of the one-d system versus $k_y$. (See text)}
\label{cylinder}
\end{figure}
To see how topological effects of $1+1$ dimensional insulators can
be derived from the first Chern number and the QH effect through
the procedure of dimensional reduction, we start by studying the
QH system on a cylinder. An essential consequence of the
nontrivial topology in the QH system is the existence of chiral
edge states. For the simplest case with the first Chern number
$C_1=1$, there is one branch of chiral fermions on each boundary.
These edge states can be solved for explicitly by diagonalizing
the Hamiltonian (\ref{RSDiracH}) in a cylindrical geometry. That
is, with periodic boundary conditions in the $y$-direction and
open boundary conditions in the $x$-direction, as shown in
Fig.\ref{cylinder} (a). Note that with this choice $k_y$ is still
a good quantum number. By defining the partial Fourier
transformation
\begin{eqnarray}
c_{k_y\alpha}(x)=\frac1{\sqrt{L_y}}\sum_{y}c_{\alpha}(x,y)e^{ik_y
y},\nonumber
\end{eqnarray}
with $(x,y)$ the coordinates of square lattice sites, the
Hamiltonian can be rewritten as
\begin{eqnarray}
H&=&\sum_{k_y,x}\left[c_{k_y}^\dagger(x)\frac{\sigma_z-i\sigma_x}2c_{k_y}(x+1)+h.c.\right]\nonumber\\
& &+\sum_{k_y,x}c_{k_y}^\dagger(x)\left[\sin
k_y\sigma_y+\left(m+\cos
k_y\right)\sigma_z\right]c_{k_y}(x)\nonumber\\
&\equiv &\sum_{k_y}H_{\rm 1D}(k_y).
\end{eqnarray}

In this way, the 2D system can be treated as  $L_y$ independent 1D
tight-binding chains, where $L_y$ is the period of the lattice in
the $y$-direction. The eigenvalues of the 1D Hamiltonian $H_{\rm
1D}(k_y)$ can be obtained numerically for each $k_y$, as shown in
Fig. \ref{cylinder} (b). An important property of the spectrum is
the presence of edge states, which lie in the bulk energy gap, and
are spatially localized at the two boundaries: $x=0,L_x.$  The
chiral nature of the edge states can be seen from their energy
spectrum. From Fig. \ref{cylinder} (b) we can see that the
velocity $v=\partial E/\partial k$ is always positive for the left
edge state and negative for the right one. The QH effect can be
easily understood in this edge state picture by Laughlin's gauge
argument\cite{laughlin1981}. Consider a constant electric field
$E_y$ in the $y$-direction, which can be chosen as
\begin{eqnarray}
A_y=-E_yt,~A_x=0.\nonumber
\end{eqnarray}
The Hamiltonian is written  $H=\sum_{k_y} H_{\rm 1D}(k_y+A_y)$ and
the current along the $x$-direction is given by
\begin{eqnarray}
J_x=\sum_{k_y}J_x(k_y)
\end{eqnarray}
with $J_x(k_y)$ the current of the 1D system. In this way, the
Hall response of the 2D system is determined by the current
response of the parameterized 1D systems $H_{\rm 1D}(q(t))$ to the
temporal change of the parameter $q(t)=k_y+A_y(t)$. The gauge
vector $A_y$ corresponds to a flux $\Phi=A_yL_y$ threading the
cylinder. During a time period $0\leq t\leq 2\pi/L_yE_y$, the flux
changes from $0$ to $2\pi$. The charge that flows through the
system during this time is given by
\begin{eqnarray}
\Delta Q&=&\int_0^{\Delta t} dt\sum_{k_y} J_x(k_y)\nonumber\\
&\equiv &\sum_{k_y}\left.\Delta P_x(k_y)\right|_{0}^{\Delta t}
\end{eqnarray}
with $\Delta t=2\pi/L_yE_y$. In the second equality we use the
relation between the current and charge polarization $P_x(k_y)$ of
the 1D systems $J_x(k_y)=dP_x(k_y)/dt$. In the adiabatic limit,
the 1D system stays in the ground state of $H_{\rm 1D}(q(t))$, so
that the change of polarization $\Delta P_x(k_y)$ is given by
$\Delta P_x(k_y)=P_x(k_y-2\pi/L_y)-P_x(k_y)$. Thus in the
$L_y\rightarrow \infty$ limit $\Delta Q$ can be written as
\begin{eqnarray}
\Delta Q=-\oint_0^{2\pi}dk_y\frac{\partial P_x(k_y)}{\partial
k_y}.
\end{eqnarray}
Therefore, the charge flow due to the Hall current generated by
the flux through the cylinder equals the charge flow through the
$1$-dimensional system $H_{\rm 1D}(k_y)$, when $k_y$ is cycled
adiabatically from $0$ to $2\pi$. From the QH response we know
$\Delta Q=\sigma_H\Delta t E_yL_y=2\pi \sigma_H$ is quantized as
an integer, which is easy to understand in the 1D picture. During
the adiabatic change of $k_y$ from $0$ to $2\pi$, the energy and
position of the edge states will change, as shown in
Fig.\ref{cylinder} (c). Since the edge state energy is always
increasing(decreasing) with $k_y$ for a state on the left (right)
boundary, the charge is always ``pumped" to the left for the
half-filled system, which leads to $\Delta Q=-1$ for each cycle.
This quantization can also be explicitly shown by calculating the
polarization $P_x(k_y)$, as shown in Fig.\ref{cylinder} (d), where
the jump of $P_x$ by one leads to $\Delta Q=-1$. In summary, we
have shown that the QH effect in the tight-binding model of Eq.
(\ref{DiracH}) can be mapped to an adiabatic pumping
effect\cite{thouless1983} by diagonalizing the system in one
direction and mapping the momentum $k$ to a parameter.

Such a dimensional reduction procedure is not restricted to
specific models, and can be generalized to any 2D insulators. For
any insulator with Hamiltonian (\ref{H2dk}), we can define the
corresponding 1D systems
\begin{eqnarray}
H_{\rm 1D}(\theta)=\sum_{k_x}c_{k_x\theta}^\dagger
h(k_x,\theta)c_{k_x\theta}
\end{eqnarray}
in which $\theta$ replaces the $y$-direction momentum $k_y$ and
effectively takes the place of $q(t).$ When
 $\theta$ is time-dependent, the current response can be
obtained by a similar Kubo formula to Eq. (\ref{kubo}), except that
the summation over all $(k_x,k_y)$ is replaced by that over only
$k_x$. More explicitly, such a linear response is defined as
\begin{eqnarray}
J_x(\theta)&=&G(\theta)\frac{d\theta}{dt}\label{kubo1d}\\
G(\theta)&=&\lim_{\omega\rightarrow
0}\frac{i}{\omega}Q(\omega+i\delta;\theta)\nonumber\\
Q(i\omega_n;\theta)&=&-\sum_{k_x,i\nu_m}{\rm tr}
\left(J_x(k_x;\theta)G_{\rm
1D}(k_x,i(\nu_m+\omega_n);\theta)\right.\nonumber\\
& &\left.\cdot\frac{\partial h(k_x;\theta)}{\partial \theta}G_{\rm
1D}(k_x,i\omega_n;\theta)\right)\frac1{L_x\beta}.\nonumber
\end{eqnarray}
Similar to Eq. (\ref{sigmaxy}) of the 2D case, the response
coefficient $G(k)$ can be expressed in terms of a Berry's phase
gauge field as
\begin{eqnarray}
G(\theta)&=&-\oint \frac{dk_x}{2\pi}
f_{x\theta}(k_x,\theta)\label{coefficient1d}\\
&=&\oint \frac{dk_x}{2\pi}\left(\frac{\partial a_x}{\partial
\theta}-\frac{\partial a_\theta}{\partial k_x}\right)\nonumber
\end{eqnarray}
with the sum rule
\begin{eqnarray}
\int G(\theta)d\theta=C_1\in \mathbb{Z}.\label{sumrule1d}
\end{eqnarray}
If we choose a proper gauge so that $a_\theta$ is always
single-valued, the expression of $G(\theta)$ can be further
simplified to
\begin{eqnarray}
G(\theta)=\frac{\partial}{\partial\theta}\left(\oint
\frac{dk_x}{2\pi} a_x(k_x,\theta)\right)\equiv \frac{\partial
P(\theta)}{\partial\theta}.\label{GP1d}
\end{eqnarray}
Physically, the loop integral
\begin{eqnarray}
P(\theta)=\oint dk_x a_x/2\pi\label{P1}
\end{eqnarray}
is nothing but the {\em charge polarization} of the 1D
system\cite{kingsmith1993,ortiz1994}, and the response equation
(\ref{kubo1d}) simply becomes  $J_x=\partial P/\partial t$. Since
the polarization $P$ is defined as the shift of the electron
center-of-mass position away from the lattice sites, it is only
well-defined modulo $1$. Consequently, the change $\Delta
P=P(\theta=2\pi)-P(\theta=0)$ through a period of adiabatic
evolution is an integer equal to $-C_1,$ and corresponds to the
charge pumped through the system. Such a relation between
quantized pumping and the first Chern number was shown by
Thouless\cite{thouless1983}.

Similar to the QH case, the current response can lead to a charge
density response, which can be determined by the charge
conservation condition. When the parameter $\theta$ has a smooth
spatial dependence $\theta=\theta(x,t)$, the response equation
(\ref{kubo1d}) still holds. From the continuity equation we obtain
\begin{eqnarray}
\frac{\partial \rho}{\partial t}&=&-\frac{\partial J_x}{\partial x}=
-\frac{\partial^2P(\theta)}{\partial x\partial t}\nonumber\\
\Rightarrow\rho&=&-\frac{\partial P(\theta)}{\partial
x}\label{charge1d}
\end{eqnarray}
in which $\rho$ is defined with respect to the background charge.
Similar to Eq. (\ref{response2d}), the density and current response
can be written together as
\begin{eqnarray}
j_\mu=-\epsilon_{\mu\nu}\frac{\partial P(\theta(x,t))}{\partial
x_\nu}\label{response1d}
\end{eqnarray}
where $\mu,\nu=0,1$ are time and space. It should be noted that
only differentiation with respect to $x,t$ appears in Eq.
(\ref{response1d}). This means, as expected, the current and
density response of the system do not depend on the
parametrization. In general, when the Hamiltonian has smooth space
and time dependence, the single particle Hamiltonian $h(k)$
becomes $h(k,x,t)\equiv h(k,\theta(x,t))$, which has the
eigenstates $\left|\alpha;k,x,t\right\rangle$ with $\alpha$ the
band index. Then relabelling $t,x,k$ as $q_A,~A=0,1,2$ we can
define the {\em phase space} Berry's phase gauge field
\begin{eqnarray}
{\cal A}_A&=&-i\sum_{\alpha}\left\langle \alpha
;q_A\right|\frac{\partial}{\partial
q_A}\left|\alpha ;q_A\right\rangle\nonumber\\
{\cal F}_{AB}&=&\partial_A {\cal A}_B-\partial_B {\cal A}_A
\end{eqnarray}
and the {\em phase space current}
\begin{eqnarray}
j_A^P=-\frac{1}{4\pi}\epsilon_{ABC}{\cal
F}_{BC}.\label{phasespace1d}
\end{eqnarray}
The physical current is obtained by integration over the wavevector
manifold:
\begin{eqnarray}
j_\mu=\int dk j_\mu^P=-\int\frac{dk}{2\pi}\epsilon^{\mu
2\nu}{\cal{F}}_{2\nu}
\end{eqnarray}
where $\mu,\nu=0,1.$ This recovers Eq. (\ref{response1d}). Note
that we could have also looked at the component $j_k=\int dk
j^{P}_k$ but this current does not have a physical interpretation.

Before moving to the next topic, we would like to apply this
formalism to the case of the Dirac model, which reproduces the
well-known result of fractional charge in the Su-Schrieffer-Heeger
(SSH) model\cite{su1979}, or equivalently the Jackiw-Rebbi
model\cite{jackiw1976}. To see this, consider the following
slightly different version of the tight-binding model
(\ref{DiracH}):
\begin{eqnarray}
h(k,\theta)&=&\sin k\sigma_x+\left(\cos
k-1\right)\sigma_z\nonumber\\
& &+m\left(\sin \theta \sigma_y+\cos\theta
\sigma_z\right)\label{Dirac1d}
\end{eqnarray}

\begin{figure}[tbp]

\begin{center}
\includegraphics[width=2.5in] {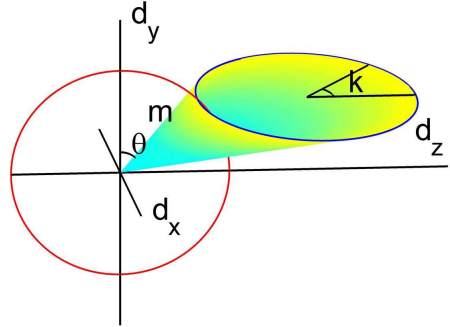}
\end{center}
\caption{Illustration of the ${\bf d}(k,\theta)$ vector for the 1D
Dirac model (\ref{Dirac1d}). The horizontal blue circle shows the
orbit of ${\bf d}(k)$ vector in the 3D space for $k\in[0,2\pi)$
with $\theta$ fixed. The red circle shows the track of the blue
circle under the variation of $\theta$. The cone shows the solid
angle $\Omega(\theta)$ surrounded by the ${\bf d}(k)$ curve, which
is $4\pi$ times the polarization $P(\theta)$.} \label{solidangle}
\end{figure}
\noindent with $m>0$. In the limit  $m\ll 1$, the Hamiltonian has
the continuum limit $h(k,\theta)\simeq k\sigma_x+m\left(\sin
\theta \sigma_y+\cos\theta \sigma_z\right)$, which is the
continuum Dirac model in $(1+1)$-d, with a real mass $m\cos\theta$
and an imaginary mass $m\sin\theta$. As discussed in Sec.
\ref{twobandmodel}, the polarization $\oint dk_x a_x/2\pi$ is
determined by the solid angle subtended by the curve ${\bf
d}(k)=\left(\sin k,m\sin\theta,m\cos\theta+\cos k-1\right)$, as
shown in Fig. \ref{solidangle}. In the limit $m\ll 1$ one can show
that the solid angle $\Omega(\theta)=2\theta$ so that
$P(\theta)\simeq\theta/2\pi$, in which case Eq. (\ref{response1d})
reproduces the Goldstone-Wilczek formula\cite{goldstone1981} :
\begin{eqnarray}
j_\mu=-\epsilon_{\mu\nu}\partial_\nu\theta.\label{GW1d}
\end{eqnarray}
Specifically, a charge
$Q=-\int_{-\infty}^{\infty}(d\theta/dx)(dx/2\pi)=-(\theta(+\infty)-\theta(-\infty))/2\pi$
is carried by a domain wall of the $\theta$ field. In particular,
for an anti-phase domain wall,
$\theta(+\infty)-\theta(-\infty)=\pi$, we obtain fractional charge
$q=1/2$. Our phase space formula (\ref{phasespace1d}) is a new
result, and it provides a generalization of the Goldstone-Wilczek
formula to the most general one-dimensional insulator.

\subsection{$Z_2$ classification of particle-hole
symmetric insulators in $(1+1)$-d}\label{sec:Z21d}

In the last subsection, we have shown how the first Chern number
of a Berry's phase gauge field appears in an adiabatic pumping
effect and the domain wall charge of one-dimensional insulators.
In these cases, an adiabatic spatial or temporal variation of the
single-particle Hamiltonian, through its parametric dependence on
$\theta(x,t)$, is required to define the Chern number. In other
words, the first Chern number is defined for {\em a parameterized
family} of Hamiltonians $h(k,x,t)$, rather than for a single 1D
Hamiltonian $h(k)$. In this subsection, we will show a different
application of the first Chern number, in which a $Z_2$
topological classification is obtained for particle-hole symmetric
insulators in 1D. Such a relation between Chern number and $Z_2$
topology can be easily generalized to the more interesting case of
second Chern number, where a similar $Z_2$ characterization is
obtained for TRI insulators, as will be shown in Sec.
\ref{sec:3dZ2} and \ref{sec:2dZ2}.

For a one-dimensional tight-binding Hamiltonian
$H=\sum_{mn}c_{m\alpha}^\dagger
h_{mn}^{\alpha\beta}(k)c_{n\beta}$, the particle-hole
transformation is defined by $c_{m\alpha}\rightarrow
C^{\alpha\beta}c_{m\beta}^\dagger$, where the {\em charge
conjugation matrix} $C$ satisfies $C^\dagger C=\mathbb{I}$ and
$C^*C=\mathbb{I}$. Under periodic boundary conditions the symmetry
requirement is
\begin{eqnarray}
H&=&\sum_{k}c_k^\dagger
h(k)c_k=\sum_{k}c_{-k} C^\dagger h(k)Cc_{-k}^\dagger\nonumber\\
&\Rightarrow & C^\dagger h(-k)C=-h^T(k).\label{Ccondition}
\end{eqnarray}

From Eq. (\ref{Ccondition}) it is straightforward to see the
symmetry of the energy spectrum: if $E$ is an eigenvalue of
$h(0)$, so is $-E$. Consequently, if the dimension of $h(k)$ is
odd, there must be at least one zero mode with $E=0$. Since the
chemical potential is constrained to vanish by the traceless
condition of $h$, such a particle-hole symmetric system cannot be
gapped unless the dimension of $h(k)$ is even. Since we are only
interested in the classification of  insulators, we will focus on
the case with $2N$ bands per lattice site.

Now consider two particle-hole symmetric insulators with
Hamiltonians $h_1(k)$ and $h_2(k)$, respectively. In general, a
continuous interpolation $h(k,\theta),~\theta\in[0,\pi]$ between
them can be defined so that
\begin{eqnarray}
h(k,0)=h_1(k),~h(k,\pi)=h_2(k)
\end{eqnarray}
Moreover, it is always possible to find a proper parametrization
so that $h(k,\theta)$ is gapped for all $\theta\in[0,\pi]$. In
other words, the topological space of all 1D insulating
Hamiltonians $h(k,\theta)$ is connected, which is a consequence of
the Wigner-Von Neumann theorem\cite{neumann1929}.

Suppose $h(k,\theta)$ is such a ``gapped interpolation" between
$h_1(k)$ and $h_2(k)$. In general, $h(k,\theta)$ for
$\theta\in(0,\pi)$ doesn't necessarily satisfy the particle-hole
symmetry. For $\theta\in[\pi,2\pi]$, define
\begin{eqnarray}
h(k,\theta)=-\left(C^{-1}h(-k,2\pi-\theta)C\right)^T.\label{Cparameterize}
\end{eqnarray}\noindent We choose this parameterization  so that if we replaced $\theta$ by a momentum wavevector then the corresponding higher
dimensional Hamiltonian would be particle-hole symmetric. Due to
the particle-hole symmetry of $h(k,\theta=0)$ and
$h(k,\theta=\pi)$, $h(k,\theta)$ is continuous for
$\theta\in[0,2\pi]$, and $h(k,2\pi)=h(k,0)$. Consequently, the
adiabatic evolution of $\theta$ from $0$ to $2\pi$ defines a cycle
of adiabatic pumping in $h(k,\theta)$, and a first Chern number
can be defined in the $(k,\theta)$ space. As discussed in Sec.
\ref{sec:dimreduction1d}, the Chern number $C[h(k,\theta)]$ can be
expressed as a winding number of the polarization
\begin{eqnarray}
C[h(k,\theta)]&=&\oint d\theta\frac{\partial
P(\theta)}{\partial\theta}\nonumber\\
P(\theta)&=&\oint \frac{dk}{2\pi}
\sum_{E_\alpha(k)<0}\left(-i\right)\left\langle
k,\theta;\alpha\right|\partial_k\left|k,\theta;\alpha\right\rangle\nonumber
\end{eqnarray}
where the summation is carried out over the occupied bands. In
general, two different parameterizations $h(k,\theta)$ and
$h'(k,\theta)$ can lead to different Chern numbers
$C[h(k,\theta)]\neq C[h'(k,\theta)]$. However, the symmetry
constraint in Eq. (\ref{Cparameterize}) guarantees that the two
Chern numbers always differ by an even integer:
$C[h(k,\theta)]-C[h'(k,\theta)]=2n,~n\in \mathbb{Z}$.

To prove this conclusion, we first study the behavior of
$P(\theta)$ under a particle-hole transformation. For an
eigenstate $\left|k,\theta;\alpha\right\rangle$ of the Hamiltonian
$h(k,\theta)$ with eigenvalue $E_\alpha(k,\theta)$, Eq.
(\ref{Cparameterize}) leads to
\begin{eqnarray}
h(-k,2\pi-\theta)C\left|k,\theta;\alpha\right\rangle^*=-E_\alpha(k)C\left|k,\theta;\alpha\right\rangle^*
\end{eqnarray}
in which $\left|k,\theta;\alpha\right\rangle^*$ is the complex
conjugate state:
$\left|k,\theta;\alpha\right\rangle^*=\sum_{m,\beta}\left(\left\langle
m,\beta\right|\left.k,\theta;\alpha\right\rangle\right)^*\left|m,\beta\right\rangle$
where $m,\beta$ are the position space lattice, and orbital index
respectively. Thus $C\left|k,\theta;\alpha\right\rangle^*\equiv
\left|-k,2\pi-\theta;\bar{\alpha}\right\rangle$ is an eigenstate
of $h(-k,2\pi-\theta)$ with energy
$E_{\bar{\alpha}}(k,2\pi-\theta)=-E_\alpha(k,\theta)$ and momentum
$-k$. Such a mapping between eigenstates of $h(k,\theta)$ and
$h(-k,2\pi-\theta)$ is one-to-one. Thus
\begin{eqnarray}
P(\theta)&=&\oint \frac{dk}{2\pi}
\sum_{E_\alpha(k)<0}\left(-i\right)\left\langle
k,\theta;\alpha\right|\partial_k\left|k,\theta;\alpha\right\rangle\nonumber\\
&=&\oint \frac{dk}{2\pi}
\sum_{E_{\bar{\alpha}}(-k)>0}\left(-i\right)\left(\left\langle
-k,2\pi-\theta;\bar{\alpha}\right|\right)^*\nonumber\\
& &\cdot\partial_k\left|-k,2\pi-\theta;\bar{\alpha}\right\rangle^*\nonumber\\
&=&-P(2\pi-\theta).\label{CsymmetryP}
\end{eqnarray}
Since $P(\theta)$ is only well-defined modulo $1$, the equality
(\ref{CsymmetryP}) actually means $P(\theta)+P(2\pi-\theta)=0{~\rm
mod}~1$. Consequently, for $\theta=0$ or $\pi$ we have
$2\pi-\theta=\theta~{\rm mod}~2\pi$, so that $P(\theta)=0$ or
$1/2$. In other words, the polarization $P$ is either $0$ or $1/2$
for any particle-hole symmetric insulator, which thus defines a
classification of particle-hole symmetric insulators. If two
systems have different $P$ value, they cannot be adiabatically
connected without breaking the particle-hole symmetry, because $P$
({\rm mod} $1$) is a continuous function during adiabatic
deformation, and a $P$ value other than $0$ and $1/2$ breaks
particle-hole symmetry. Though such an argument explains
physically why a $Z_2$ classification is defined for particle-hole
symmetric system, it is not so rigorous. As discussed in the
derivation from Eq. (\ref{coefficient1d}) to Eq. (\ref{GP1d}), the
definition $P(\theta)=\oint dka_k/2\pi$ relies on a proper gauge
choice. To avoid any gauge dependence, a more rigorous definition
of the $Z_2$ classification is shown below, which only involves
the gauge invariant variable $\partial P(\theta)/\partial \theta$
and Chern number $C_1$.

To begin with, the symmetry (\ref{CsymmetryP}) leads to
\begin{eqnarray}
\int_0^\pi
dP(\theta)&=&\int_{\pi}^{2\pi}dP(\theta),\label{CconstraintP}
\end{eqnarray}
which is independent of gauge choice since only the change of
$P(\theta)$ is involved. This equation shows that the change of
polarization during the first half and the second half of the closed
path $\theta\in[0,2\pi]$ are always the same.

\begin{figure}[tbp]
\begin{center}
\includegraphics[width=3in] {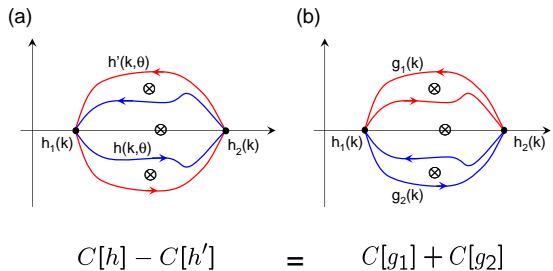}
\end{center}
\caption{Illustration of the interpolation between two
particle-hole symmetric Hamiltonians $h_1(k)$ and $h_2(k)$.}
\label{fig:Csymmetry}
\end{figure}

Now consider two different parameterizations $h(k,\theta)$ and
$h'(k,\theta)$, satisfying $h(k,0)=h'(k,0)=h_1(k)$,
$h(k,\pi)=h'(k,\pi)=h_2(k)$. Denoting the polarization $P(\theta)$
and $P'(\theta)$ corresponding to $h(k,\theta)$ and
$h'(k,\theta)$, respectively, the Chern number difference between
$h$ and $h'$ is given by
\begin{eqnarray}
C[h]-C[h']&=&\int_0^{2\pi}d\theta\left(\frac{\partial
P(\theta)}{\partial \theta}-\frac{\partial P'(\theta)}{\partial
\theta}\right).
\end{eqnarray}
Define the new interpolations $g_1(k,\theta)$ and $g_2(k,\theta)$
as
\begin{eqnarray}
g_1(k,\theta)&=&\left\{\begin{array}{cc}h(k,\theta),&\theta\in[0,\pi]\\h'(k,2\pi-\theta),&\theta\in[\pi,2\pi]\end{array}\right.\nonumber\\
g_2(k,\theta)&=&\left\{\begin{array}{cc}h'(k,2\pi-\theta),&\theta\in[0,\pi]\\h(k,\theta),&\theta\in[\pi,2\pi]\end{array}\right.\label{defg1g2}
\end{eqnarray}
$g_1(k,\theta)$ and $g_2(k,\theta)$ are obtained by recombination
of the two paths $h(k,\theta)$ and $h'(k,\theta)$, as shown in
Fig. \ref{fig:Csymmetry}. From the construction of $g_1$ and
$g_2$, it is straightforward to see that
\begin{eqnarray}
C[g_1]&=&\int_0^\pi d\theta\left(\frac{\partial P(\theta)}{\partial
\theta}-\frac{\partial P'(\theta)}{\partial
\theta}\right)\nonumber\\
C[g_2]&=&\int_\pi^{2\pi} d\theta\left(\frac{\partial
P(\theta)}{\partial \theta}-\frac{\partial P'(\theta)}{\partial
\theta}\right).
\end{eqnarray}
Thus $C[h]-C[h']=C[g_1]+C[g_2]$. On the other hand, from Eq.
(\ref{CconstraintP}) we know $C[g_1]=C[g_2]$, so that
$C[h]-C[h']=2C[g_1]$. Since $C[g_1]\in \mathbb{Z}$, we obtain that
$C[h]-C[h']$ is even for any two interpolations $h(k,\theta)$ and
$h'(k,\theta)$ between $h_1(k)$ and $h_2(k)$. Intuitively, such a
conclusion simply comes from the fact that the Chern number $C[h]$
and $C[h']$ can be different only if there are singularities between
these two paths, while the positions of the singularities in the
parameter space are always symmetric under particle-hole symmetry,
as shown in Fig. \ref{fig:Csymmetry}.

Based on the discussions above, we can define the {\em ``relative
Chern parity"} as
\begin{eqnarray}
 N_1[h_1(k),h_2(k)]=(-1)^{C[h(k,\theta)]}\text{,}\label{C1parity}
\end{eqnarray}
which is independent of the choice of interpolation $h(k,\theta)$,
but only determined by the Hamiltonians $h_1(k),~h_2(k)$.
Moreover, for any three particle-hole symmetric Hamiltonians
$h_1(k),~h_2(k),~h_3(k)$, it is easy to prove that the Chern
parity satisfies the following associative law:
\begin{eqnarray}
N_1[h_1(k),h_2(k)]N_1[h_2(k),h_3(k)]=N_1[h_1(k),h_3(k)].\nonumber\\
\label{associativity}
\end{eqnarray}
Consequently, $N_1[h_1(k),h_2(k)]=1$ defines an {\em equivalence
relation} between any two particle-hole symmetric Hamiltonians,
which thus classifies all the particle-hole symmetric insulators
into two classes. To define these two classes more explicitly, one
can define a ``vacuum" Hamiltonian as $h_0(k)\equiv h_0$, where
$h_0$ is an arbitrary matrix which does not depend on $k$ and which
satisfies the particle-hole symmetry constraint $C^\dagger
h_0C=-h_0^T$. Thus $h_0$ describes a totally local system, in which
there is no hopping between different sites. Taking such a trivial
system as a reference Hamiltonian, we can define
$N_1[h_0(k),h(k)]\equiv N_1[h(k)]$ as a $Z_2$ topological quantum
number of the Hamiltonian $h(k)$. All the Hamiltonians $h(k)$ with
$N_1[h_0(k),h(k)]=1$ are classified as $Z_2$ trivial, while those
with $N_1[h_0(k),h(k)]=-1$ are considered as $Z_2$ nontrivial.
(Again, this classification doesn't depend on the choice of
``vacuum" $h_0$, since any two vacua are equivalent.)

\begin{figure}[tbp]
\begin{center}
\includegraphics[width=3.5in] {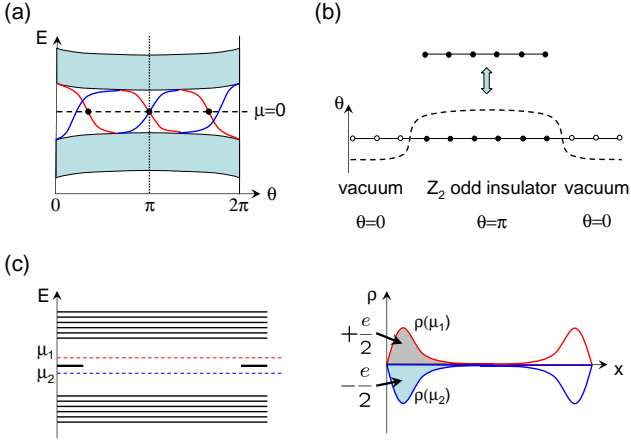}
\end{center}
\caption{(a) Schematic energy spectrum of a parameterized
Hamiltonian $h_{mn}(\theta)$ with open boundary conditions. The
red (blue) lines indicate the left (right) end states. The
$\theta$ values with zero-energy {\em left} edge states are marked
by the solid circles. (b) Illustration to show that the open
boundary of a $Z_2$ nontrivial insulator is equivalent to a domain
wall between $\theta=\pi$ (nontrivial) and $\theta=0$ (trivial
vacuum). (c) Illustration of the charge density distribution
corresponding to two different chemical potentials $\mu_1$ (red)
and $\mu_2$ (blue). The area below the curve $\rho(\mu_1)$ and
$\rho(\mu_2)$ is $+e/2$ and $-e/2$, respectively, which shows the
half charge confined on the boundary.} \label{Cedgestate}
\end{figure}

Despite its abstract form, such a topological characterization has a
direct physical consequence. For a $Z_2$ nontrivial Hamiltonian
$h_1(k)$, an interpolation $h(k,\theta)$ can be defined so that
$h(k,0)=h_0$, $h(k,\pi)=h_1(k)$, and the Chern number
$C[h(k,\theta)]$ is an odd integer. If we study the one-dimensional
system $h(k,\theta)$ with open boundary conditions, the tight
binding Hamiltonian can be rewritten in real space as
\begin{eqnarray}
h_{mn}(\theta)=\frac1{\sqrt{L}}\sum_ke^{ik(x_m-x_n)}h(k,\theta),~\forall
1\leq m,n\leq L.\nonumber
\end{eqnarray}
As discussed in Sec. \ref{sec:dimreduction1d}, there are mid-gap
end states in the energy spectrum of $h_{mn}(\theta)$ as a
consequence of the non-zero Chern number. When the Chern number
$C[h(k,\theta)]=2n-1,~n\in\mathbb{Z}$, there are values
$\theta^{L}_{s}\in[0,2\pi),~s=1,2,..2n-1$ for which the
Hamiltonian $h_{mn}(\theta_s)$ has zero energy localized states on
the left end of the 1D system, and the same number of
$\theta^{R}_{s}$ values where zero energy states are localized on
the right end, as shown in Fig. \ref{Cedgestate}. Due to the
particle-hole symmetry between $h_{mn}(\theta)$ and
$h_{mn}(2\pi-\theta)$,  zero levels always appear in pairs at
$\theta$ and $2\pi-\theta$. Consequently, when the Chern number is
odd, there must be a zero level at $\theta=0$ or $\theta=\pi$.
Since $\theta=0$ corresponds to a trivial insulator with flat
bands and no end states, the localized zero mode has to appear at
$\theta=\pi$. In other words, one zero energy localized state (or
an odd number of such states) is confined at each open boundary of
a $Z_2$ nontrivial particle-hole symmetric insulator.

The existence of a zero level leads to an important physical
consequence---a half charge on the boundary of the nontrivial
insulator. In a periodic system when the chemical potential
vanishes, the average electron density on each site is
$\bar{n}_m=\left\langle \sum_\alpha c_{m\alpha}^\dagger
c_{m\alpha}\right\rangle=N$ when there are $N$ bands filled. In an
open boundary system, define $\rho_m(\mu)=\left\langle \sum_\alpha
c_{m\alpha}^\dagger c_{m\alpha}\right\rangle_\mu-N$ to be the
density deviation with respect to $N$ on each site. Then
particle-hole symmetry leads to $\rho_m(\mu)=-\rho_m(-\mu)$. On
the other hand, when $\mu$ is in the bulk gap, the only difference
between $\mu$ and $-\mu$ is the filling of the zero levels
localized on each end $\left|0L\right\rangle$ and
$\left|0R\right\rangle$, so that
\begin{eqnarray}
\lim_{\mu\rightarrow
0^+}\left(\rho_m(\mu)-\rho_m(-\mu)\right)=\sum_\alpha\left|\left\langle
m\alpha|0L\right\rangle\right|^2\nonumber
\end{eqnarray}
for the sites $m$ that are far away enough from the right
boundary. Thus we have $\sum_m\rho_m(\mu\rightarrow 0^+)=1/2$
where the summation is done around the left boundary so that we do
not pick up a contribution from the other end. In summary, a
charge $e/2$ ($-e/2$) is localized on the boundary if the zero
level is vacant (occupied), as shown in Fig. \ref{Cedgestate}.

The existence of such a half charge can also be understood by
viewing the open boundary of a topologically nontrivial insulator
as a domain wall between the nontrivial insulator and the vacuum.
By defining the interpolation $h_{mn}(\theta)$, such a domain wall
is described by a spatial dependence of $\theta$ with
$\theta(x\rightarrow +\infty)=\pi,~\theta(x\rightarrow
-\infty)=0$. According to the response formula (\ref{charge1d}),
the charge carried by the domain wall is given by
\begin{eqnarray}
Q_d=e\int_{-\infty}^{+\infty}dx\frac{\partial
P(\theta(x))}{\partial x}=e\int_0^{\pi}dP(\theta).
\end{eqnarray}
By using Eq.. (\ref{CconstraintP}) we obtain
\begin{eqnarray}
Q_d=\frac e2\int_0^{2\pi}dP(\theta)=\frac e2C[h(k,\theta)].
\end{eqnarray}
It should be noted that an integer charge can always be added by
changing the filling of local states, which means $Q_d$ is only
fixed modulo $e$. Consequently, a $\pm e/2$ charge is carried by the
domain wall if and only if the Chern number is odd, i.e., when the
insulator is nontrivial.

\subsection{$Z_2$ classification of $(0+1)$-d particle-hole
symmetric insulators}\label{sec:0dZ2}

In the last subsection, we have shown how a $Z_2$ classification
of $(1+1)$-d particle-hole symmetric insulators is defined by
dimensional reduction from $(2+1)$-d systems. Such a dimensional
reduction can be repeated once more to study $(0+1)$-d systems,
that is, a single-site problem. In this subsection we will show
that a $Z_2$ classification of particle-hole symmetric
Hamiltonians in $(0+1)$-d is also obtained by dimensional
reduction. Although such a classification by itself is not as
interesting as the higher dimensional counterparts, it does
provide a simplest example of the ``dimensional reduction chain"
$(2+1)$-d$\rightarrow (1+1)$-d$\rightarrow (0+1)$-d, which can be
later generalized to its higher-dimensional counterpart
$(4+1)$-d$\rightarrow (3+1)$-d$\rightarrow (2+1)$-d. In other
words, the $Z_2$ classification of the $(0+1)$-d particle-hole
symmetric insulators can help us to understand the classification
of $(2+1)$-d TRI insulators as it is dimensionally reduced from
the $(4+1)$-d TRI insulator.

For a free, single-site fermion system with Hamiltonian matrix
$h$, the particle-hole symmetry is defined as
\begin{eqnarray}
C^\dagger hC=-h^T.\label{Ccondition0d}
\end{eqnarray}
Given any two particle-hole symmetric Hamiltonians $h_1$ and
$h_2$, we follow the same procedure as the last subsection and
define a continuous interpolation $h(\theta),~\theta\in[0,2\pi]$
satisfying
\begin{eqnarray}
h(0)=h_1,~h(\pi)=h_2,~C^\dagger
h(\theta)C=-h(2\pi-\theta)^T,\label{Cparameterize0d}
\end{eqnarray}
where $h(\theta)$ is gapped for all $\theta$. The Hamiltonian
$h(\theta)$ is the dimensional reduction of a $(1+1)$-d
Hamiltonian $h(k)$, with the wavevector $k$ replaced by the
parameter $\theta$. The constraint (\ref{Cparameterize0d}) is
identical to the particle-hole symmetry condition
(\ref{Ccondition}), so that $h(\theta)$ corresponds to a
particle-hole symmetric $(1+1)$-d insulator. As shown in last
subsection, $h(\theta)$ is classified by the value of the ``Chern
parity" $N_1[h(\theta)]$. If $N_1[h(\theta)]=-1$, no continuous
interpolation preserving particle-hole symmetry can be defined
between $h(\theta)$ and the vacuum Hamiltonian
$h(\theta)=h_0,\forall \theta\in[0,2\pi]$. To obtain the
classification of $(0+1)$-d Hamiltonians, consider two different
interpolations $h(\theta)$ and $h'(\theta)$ between $h_1$ and
$h_2$. According to the associative law (\ref{associativity}), we
know $N_1[h(\theta)]N_1[h'(\theta)]=N_1[h(\theta),h'(\theta)]$,
where $N_1[h(\theta),h'(\theta)]$ is the relative Chern parity
between two interpolations. In the following we will prove
$N_1[h(\theta),h'(\theta)]=1$ for any two interpolations $h$ and
$h'$ satisfying condition (\ref{Cparameterize0d}). As a result,
$N_1[h(\theta)]$ is independent of the choice of interpolation
between $h_1$ and $h_2$, so that $N_0[h_1,h_2]\equiv
N_1[h(\theta)]$ can be defined as a function of $h_1$ and $h_2$.
The $Z_2$ quantity $N_0$ defined for $(0+1)$-d Hamiltonians plays
exactly the same role as $N_1[h(k),h'(k)]$ in the $(1+1)$-d case,
from which a $Z_2$ classification can be defined.

To prove $N_1[h(\theta),h'(\theta)]=1$ for any two interpolations,
first define a continuous deformation $g(\theta,\varphi)$ between
$h(\theta)$ and $h'(\theta)$, which satisfies the conditions
below:
\begin{eqnarray}
g(\theta,\varphi=0)&=&h(\theta),~g(\theta,\varphi=\pi)=h'(\theta)\nonumber\\
g(0,\varphi)&=&h_1,~g(\pi,\varphi)=h_2\nonumber\\
C^\dagger g(\theta,\varphi)C&=&-g(2\pi-\theta,2\pi-\varphi)^T
.\label{gdefinition0d}
\end{eqnarray}
From the discussions in last subsection it is easy to confirm that
such a continuous interpolation is always possible, in which
$g(\theta,\varphi)$ is gapped for all $\theta$ and $\varphi$. In
the two-dimensional parameter space $\theta,\varphi$ one can
define the Berry phase gauge field and the first Chern number
$C_1[g(\theta,\varphi)]$. By the definition of the Chern parity,
we have $N_1[h(\theta),h'(\theta)]=(-1)^{C_1[g(\theta,\varphi)]}$.
However, the parameterized Hamiltonian $g(\theta,\varphi)$ can be
viewed in two different ways: it not only defines an interpolation
between $h(\theta)$ and $h'(\theta)$, but also defines an
interpolation between $g(0,\varphi)=h_1$ and $g(\pi,\varphi)=h_2$.
Since $g(0,\varphi)$ and $g(\pi,\varphi)$ are ``vacuum
Hamiltonians" without any $\varphi$ dependence, they have trivial
relative Chern parity, which means
$N_1[h(\theta),h'(\theta)]=N_1[g(0,\varphi),g(\pi,\varphi)]=N_1[h_1,h_2]=1$.

In conclusion, from the discussion above we have proved that any
two interpolations $h(\theta)$ and $h'(\theta)$ belong to the same
$Z_2$ class, so that the Chern parity $N_1[h(\theta)]$ only
depends on the end points $h_1$ and $h_2$. Consequently, the
quantity $N_0[h_1,h_2]\equiv N_1[h(\theta)]$ defines a relation
between each pair of particle-hole symmetric Hamiltonians $h_1$
and $h_2$. After picking a reference Hamiltonian $h_0$, one can
\emph{define} all the Hamiltonians with $N_0[h_0,h]=1$ as
``trivial" and $N_0[h_0,h]=-1$ as nontrivial. The main difference
between this classification and the one for $(1+1)$-d systems is
that there is no natural choice of the reference Hamiltonian
$h_0$. In other words, the names ``trivial" and ``non-trivial"
only have relative meaning in the $(0+1)$-d case. However, the
classification is still meaningful in the sense that any two
Hamiltonians with $N_0[h_1,h_2]=-1$ cannot be adiabatically
connected without breaking particle-hole symmetry. In other words,
the manifold of single-site particle-hole symmetric Hamiltonians
is disconnected, with at least two connected pieces.

\begin{figure}[tbp]
\begin{center}
\includegraphics[width=2.5in] {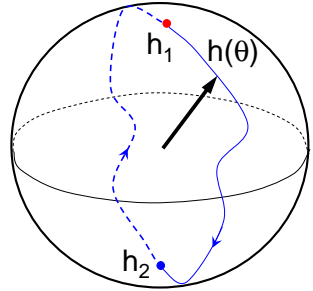}
\end{center}
\caption{Illustration of the $2\times 2$ single-site Hamiltonians.
Each point on the sphere represents an unit vector $\hat{\bf
d}=\vec{d}/|d|$, and the north and south poles correspond to the
particle-hole symmetric Hamiltonians $h_{1,2}=\pm \sigma^3$,
respectively. The blue path shows an interpolation between $h_1$
and $h_2$ satisfying the constraint (\ref{Cparameterize0d}), which
always encloses a solid angle $\Omega=2\pi$.} \label{twobytwo0d}
\end{figure}

As a simple example, we study  $2\times 2$ Hamiltonians. A general
$2\times 2$ single-site Hamiltonian can be decomposed as
\begin{eqnarray}
h=d_0\sigma^0+\sum_{a=1}^3d_a\sigma^a
\end{eqnarray}
where $\sigma^0=\mathbb{I}$ and $\sigma^{1,2,3}$ are the Pauli
matrices. When $C=\sigma^1$, particle-hole symmetry requires
$C^\dagger hC=-h^T$, from which we obtain $d_0=d_1=d_2=0$. Thus
$h=d_3\sigma^3$, in which $d_3\neq 0$ so as to make $h$ gapped.
Consequently, we can see that the two $Z_2$ classes are simply
$d_3>0$ and $d_3<0$. When an adiabatic interpolation
$h(\theta)=d_0(\theta)\sigma^0+\sum_ad_a(\theta)\sigma^a$ is
defined from $d_3>0$ to $d_3<0$, the spin vector $\vec{d}(\theta)$
has to rotate from the north pole to the south pole, and then
return along the image path determined by the particle-hole
symmetry (\ref{Cparameterize0d}), as shown in Fig.
\ref{twobytwo0d}. The topological quantum number $N_0[h_1,h_2]$ is
simply determined by the Berry's phase enclosed by the path
$d_a(\theta)$, which is $\pi$ when $h_1$ and $h_2$ are on
different poles, and $0$ otherwise. From this example we can
understand the $Z_2$ classification intuitively. In Sec.
\ref{sec:2dZ2} we show that the $Z_2$ classification of $(2+1)$-d
TRI insulators---the class that corresponds to the QSH effect---is
obtained as a direct analog of the $(0+1)$-d case discussed above.

\section{Second Chern number and its physical
consequences}\label{sec:secondChern}

In this section, we shall generalize the classification of the
$(2+1)$-d TRB topological insulator in terms of the first Chern
number and the $(2+1)$-d Chern-Simons theory to the classification
of the $(4+1)$-d TRI topological insulator in terms of the second
Chern number and the $(4+1)$-d Chern-Simons theory. We then
generalize the dimensional reduction chain $(2+1)$-d$\rightarrow
(1+1)$-d$\rightarrow (0+1)$-d to the case of $(4+1)$-d$\rightarrow
(3+1)$-d$\rightarrow (2+1)$-d for TRI insulators. Many novel
topological effects are predicted for the TRI topological
insulators in $(3+1)$-d and $(2+1)$-d.

\subsection{Second Chern number in $(4+1)$-d non-linear
response}\label{sec:secondtknn}

In this subsection, we will show how the second Chern number
appears as a non-linear response coefficient of $(4+1)$-d band
insulators in an external $U(1)$ gauge field, which is in exact
analogy with the first Chern number as the Hall conductance of a
$(2+1)$-d system. To describe such a non-linear response, it is
convenient to use the path integral formalism. The Hamiltonian of
a $(4+1)$-d insulator coupled to a $U(1)$ gauge field is written
as
\begin{eqnarray}
H[A]&=&\sum_{ m,n}\left(c_{m\alpha}^\dagger
h_{mn}^{\alpha\beta}e^{iA_{mn}}c_{n\beta}+h.c.\right)\nonumber\\ &
&+\sum_mA_{0m}c_{m\alpha}^\dagger c_{m\alpha}.
\end{eqnarray}
The effective action of gauge field $A^\mu$ is obtained by the
following path integral:
\begin{eqnarray}
e^{iS_{\rm eff}[A]}&=&\int
D[c]D[c^\dagger]e^{i\int d t\left[\sum_mc_{m\alpha}^\dagger\left(i\partial_t\right)c_{m\alpha}-H[A]\right]}\nonumber\\
&=&\det\left[{\left(i\partial_t-A_{0m}\right)\delta_{mn}^{\alpha\beta}-h_{mn}^{\alpha\beta}e^{iA_{mn}}}\right]
\end{eqnarray}
which determines the response of the fermionic system through the
equation
\begin{eqnarray}
j_\mu({\bf x})=\frac{\delta S_{\rm eff}[A]}{\delta A_\mu({\bf
x})}.\label{motioneqn4d}
\end{eqnarray}

\begin{figure}[t!]
\includegraphics[scale=0.4]{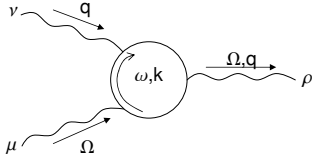}
\caption{The Feynman diagram that contributes to the topological
term (\ref{Seff4d}). The loop is a fermion propagator, and the wavy
lines are external legs corresponding to the gauge field.}
\label{Feynman4d}
\end{figure}

In the case of the $(2+1)$-d insulators, the effective action
$S_{\rm eff}$ contains a Chern-Simons term
$(C_1/4\pi)A_\mu\epsilon^{\mu\nu\tau}\partial_\nu A_\tau$ as shown
in Eq. (\ref{Seff2d}) of Sec. \ref{sec:response2d}, in which the
first Chern number $C_1$ appears as the coefficient. For the
$(4+1)$-d system, a similar topological term is in general present
in the effective action, which is the {\rm second Chern-Simons
term}:
\begin{eqnarray}
S_{\rm eff}=\frac{C_2}{24\pi^2}\int
d^4xdt\epsilon^{\mu\nu\rho\sigma\tau}A_\mu\partial_\nu
A_\rho\partial_\sigma A_\tau \label{Seff4d}
\end{eqnarray}
\noindent where $\mu,\nu,\rho,\sigma,\tau=0,1,2,3,4.$ As shown in
Refs. \onlinecite{niemi1983,golterman1993,volovik2002},  the
coefficient $C_2$ can be obtained by the one-loop Feynman diagram
in Fig. \ref{Feynman4d}, which can be expressed in the following
symmetric form:
\begin{widetext}
\begin{eqnarray}
C_2=-\frac{\pi^2}{ 15}\epsilon^{\mu\nu\rho\sigma\tau
}\int\frac{d^4kd\omega}{\left(2\pi\right)^5}{\rm
Tr}\left[\left(G\frac{\partial G^{-1}}{\partial
q^\mu}\right)\left(G\frac{\partial G^{-1}}{\partial
q^\nu}\right)\left(G\frac{\partial G^{-1}}{\partial
q^\rho}\right)\left(G\frac{\partial G^{-1}}{\partial
q^\sigma}\right)\left(G\frac{\partial G^{-1}}{\partial
q^\tau}\right)\right]\label{C2green}
\end{eqnarray}
\end{widetext}
in which $q^\mu=(\omega,k_1,k_2,k_3,k_4)$ is the frequency-momentum
vector, and $G(q^\mu)=\left[\omega+i\delta-h(k_i)\right]^{-1}$ is
the single-particle Green's function.

Now we are going to show the relation between $C_2$ defined in Eq.
(\ref{C2green}) and the non-abelian Berry's phase gauge field in
momentum space. To make the statement clear, we first write down
the conclusion:
\begin{itemize}
\item For any $(4+1)$-d band insulator with single particle
Hamiltonian $h({\bf k})$, the non-linear response coefficient
$C_2$ defined in Eq. (\ref{C2green}) is equal to the second Chern
number of the non-abelian Berry's phase gauge field in the BZ,
\emph{i.e.}:
\begin{eqnarray}
C_2&=&\frac1{32\pi^2}\int d^4k\epsilon^{ijk\ell}{\rm
tr}\left[f_{ij}f_{k\ell}\right]\label{2ndtknn}\\
\text{with~}f^{\alpha\beta}_{ij}&=&\partial_i
a^{\alpha\beta}_j-\partial_j
a^{\alpha\beta}_i+i\left[a_i,a_j\right]^{\alpha\beta},\nonumber\\
a_i^{\alpha\beta}({\bf k})&=&-i\left\langle \alpha,{\bf
k}\right|\frac{\partial}{\partial k_i }\left|\beta,{\bf
k}\right\rangle\nonumber
\end{eqnarray}
\end{itemize}
\noindent where $i,j,k,\ell=1,2,3,4.$

\begin{figure}[tbp]
\includegraphics[width=2.5in]{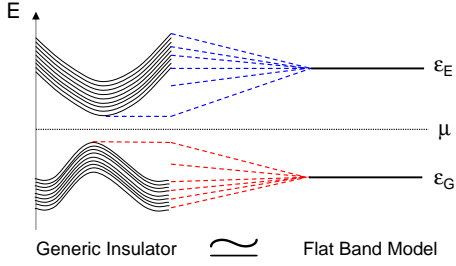}
\caption{Illustration showing that a band insulator with arbitrary
band structure $\epsilon_i(k)$ can be continuously deformed to a
flat band model with the same eigenstates. Since no level crossing
occurs at the Fermi level, the two Hamiltonians are topologically
equivalent.} \label{flatdeform}
\end{figure}

The index $\alpha$ in $a_i^{\alpha\beta}$ refers to the occupied
bands, therefore, for a general multi-band model,
$a_i^{\alpha\beta}$ is a non-abelian gauge field, and
$f^{\alpha\beta}_{ij}$ is the associated non-abelian field
strength. Here we sketch the basic idea of Eq. (\ref{2ndtknn}),
and leave the explicit derivation to Appendix \ref{app:C2green}.
The key point to simplify Eq. (\ref{C2green}) is noticing its {\em
topological invariance} \emph{i.e.} under any continuous
deformation of the Hamiltonian $h({\bf k})$, as long as no level
crossing occurs at the Fermi level, $C_2$ remains invariant.
Denote the eigenvalues of the single particle Hamiltonian $h({\bf
k})$ as $\epsilon_\alpha({\bf k}),\alpha=1,2,...,N$ with
$\epsilon_\alpha({\bf k})\leq \epsilon_{\alpha+1}({\bf k})$. When
$M$ bands are filled, one can always define a continuous
deformation of the energy spectrum so that $\epsilon_\alpha({\bf
k})\rightarrow \epsilon_G$ for $\alpha\leq M$ and
$\epsilon_\alpha({\bf k})\rightarrow \epsilon_E$ for $\alpha>M$
(with $\epsilon_E>\epsilon_G$), while all the corresponding
eigenstates $\left|\alpha,{\bf k}\right\rangle$ remain invariant.
In other words, each Hamiltonian $h({\bf k})$ can be continuously
deformed to some ``flat band" model, as shown in Fig.
\ref{flatdeform}. Since both Eq. (\ref{C2green}) and the second
Chern number in Eq. (\ref{2ndtknn}) are topologically invariant,
we only need to demonstrate Eq. (\ref{2ndtknn}) for the flat band
models, of which the Hamiltonians have the form
\begin{eqnarray}
h_0({\bf k})&=&\epsilon_G\sum_{1\leq \alpha\leq
M}\left|\alpha,{\bf k}\right\rangle \left\langle \alpha,{\bf
k}\right|+\epsilon_E\sum_{\beta>M}\left|\beta,{\bf k}\right\rangle
\left\langle \beta,{\bf k}\right|\nonumber\\
&\equiv &\epsilon_GP_G({\bf k})+\epsilon_EP_E({\bf
k}).\label{projectorH}
\end{eqnarray}
Here $P_G({\bf k})$ ($P_E({\bf k})$) is the projection operator to
the occupied (un-occupied) bands. Non-abelian gauge connections
can be defined in terms of these projection operators in a way
similar to Ref. \onlinecite{murakami2004}. Correspondingly, the
single particle Green's function can also be expressed by the
projection operators $P_G,~P_E$, and Eq. (\ref{2ndtknn}) can be
proved by straight-forward algebraic calculations, as shown in
Appendix \ref{app:C2green}.

In summary, we have shown that for any $(4+1)$-d band insulator,
there is a $(4+1)$-d Chern-Simons term (\ref{Seff4d}) in the
effective action of the external $U(1)$ gauge field, of which the
coefficient is the second Chern number of the non-abelian Berry
phase gauge field. Such a relation between Chern number and
Chern-Simons term in the effective action is an exact analogy of
the TKNN formula in $(2+1)$-d QH effect. By applying the equation
of motion (\ref{motioneqn4d}), we obtain
\begin{eqnarray}
j^\mu=\frac{C_2}{8\pi^2}\epsilon^{\mu\nu\rho\sigma\tau}\partial_\nu
A_\rho\partial_\sigma A_\tau\label{topomotioneqn4d}
\end{eqnarray}
which is the non-linear response to the external field $A_\mu$.
For example, consider a field configuration :
\begin{eqnarray}
A_x=0,~A_y=B_{z} x,~A_z=-E_zt,~A_w=A_t=0\label{EBchoice}
\end{eqnarray}
where $x,y,z,w$ are the spatial coordinates and $t$ is time. The
only non-vanishing components of the field curvature are
$F_{xy}=B_{z}$ and $F_{zt}=-E_z$, which according to Eq.
(\ref{topomotioneqn4d}) generates the current
\begin{eqnarray}
j_w=\frac{C_2}{4\pi^2}B_{z}E_z.\nonumber
\end{eqnarray}
If we integrate the equation above over the $x,y$ dimensions (with
periodic boundary conditions and assuming $E_z$ is does not depend
on $(x,y)$), we obtain
\begin{eqnarray}
\int dxdyj_w=\frac{C_2}{4\pi^2}\left(\int dxdy
B_{z}\right)E_z\equiv \frac{C_2N_{xy}}{2\pi}E_z
\end{eqnarray}
where $N_{xy}=\int dxdyB_{z}/2\pi$ is the number of flux quanta
through the $xy$ plane, which is always quantized to be an
integer. This is exactly the 4DQH effect proposed in Ref.
\onlinecite{zhang2001}. Thus, from this example we can understand
a physical consequence of the second Chern number: In a $(4+1)$-d
insulator with second Chern number $C_2$, a quantized Hall
conductance $C_2N_{xy}/2\pi$ in the $zw$ plane is induced by
magnetic field with flux $2\pi N_{xy}$ in the perpendicular ($xy$)
plane.

Similar to the $(2+1)$-d case, the physical consequences of the
second Chern number can also be understood better by studying the
surface states of an open-boundary system, which for the $(4+1)$-d
case is described by a $(3+1)$-d theory. In the next subsection we
will study an explicit example of a $(4+1)$-d topological
insulator, which helps us to improve our understanding of the
physical picture of the $(4+1)$-d topology; especially, after
dimensional reduction to the lower-dimensional physical systems.

\subsection{TRI topological insulators based on lattice Dirac models}\label{sec:Dirac5d}

In section \ref{twobandmodel}, we have shown that the model
introduced in Ref. \onlinecite{qi2006} realizes the fundamental
TRB topological insulator in $(2+1)$-d, and it reduces to the
Dirac model in the continuum limit. Generalizing this
construction, we propose the lattice Dirac model to be the
realization of the fundamental TRI topological insulator in
$(4+1)$-d. Such a model has also been studied in the field theory
literature\cite{golterman1993,creutz2001}. The continuum Dirac
model in $(4+1)$-d dimensions is expressed as
\begin{eqnarray}
H=\int
d^4x\left[\psi^\dagger(x)\Gamma^i\left(-i\partial_i\right)\psi(x)+m\psi^\dagger\Gamma^0\psi\right]
\end{eqnarray}
with $i=1,2,3,4$ the spatial dimensions, and
$\Gamma^{\mu},\mu=0,1,..,4$ the five Dirac matrices satisfying the
Clifford algebra
\begin{eqnarray}
\left\{\Gamma^\mu,\Gamma^\nu\right\}=2\delta_{\mu\nu}\mathbb{I}
\end{eqnarray}
with $\mathbb{I}$ the identity matrix\cite{footnoteGamma}.

The lattice (tight-binding) version of this model is written as
\begin{eqnarray}
H&=&\sum_{n,i}\left[\psi^\dagger_n\left(\frac
{c\Gamma^0-i\Gamma^i}2\right)\psi_{n+\hat{i}}+h.c.\right]\nonumber\\
& &+m\sum_{n}\psi^\dagger_n\Gamma^0\psi_n\label{Dirac5d}
\end{eqnarray}
or in  momentum space,
\begin{eqnarray}
H&=&\sum_{{\bf k}}\psi_{\bf k}^\dagger\left[ \sum_{i} \sin k_i
\Gamma^i +\left(m+c\sum_i\cos
k_i\right)\Gamma^0\right]\psi_{\bf k}.\label{Dirac5dk}\nonumber\\
\end{eqnarray}
Such a Hamiltonian can be written in the compact form
\begin{eqnarray}
H&=&\sum_{{\bf k}}\psi_{\bf k}^\dagger d_a({\bf k})\Gamma^a\psi_{\bf
k}\label{HGamma}
\end{eqnarray}
with \begin{eqnarray}d_a({\bf k})=\left(\left(m+c\sum_i\cos
k_i\right), \sin k_x, \sin k_y, \sin k_z, \sin
k_w\right)\nonumber\end{eqnarray} a five-dimensional vector.
Similar to the $(2+1)$-d two-band models we studied in Sec.
\ref{twobandmodel}, a single particle Hamiltonian with the form
$h({\bf k})=d_a({\bf k})\Gamma^a$ has two eigenvalues
$E_{\pm}({\bf k})=\pm\sqrt{\sum_ad_a^2({\bf k})}$, but with the
key difference that here both eigenvalues are doubly degenerate.
When $\sum_ad_a^2({\bf k})\equiv d^2({\bf k})$ is non-vanishing in
the whole BZ, the system is gapped at half-filling, with the two
bands with $E=E_{-}({\bf k})$ filled. Since there are two occupied
bands, an $SU(2)\times U(1)$ adiabatic connection can be
defined\cite{avron1988,demler1999,murakami2004}.  Starting from
the Hamiltonian (\ref{Dirac5dk}), one can determine the single
particle Green's function, and substituting it into the expression
for the second Chern number in Eq. (\ref{C2green}). We obtain
\begin{eqnarray}
C_2= \frac3{8\pi^2}\int
d^4k\epsilon^{abcde}\hat{d}_a\partial_x\hat{d}_b\partial_y\hat{d}_c\partial_z\hat{d}_d\partial_w\hat{d}_e\label{windingS4}
\end{eqnarray}
which is the winding number of the mapping $\hat{d}_a({\bf
k})\equiv d_a({\bf k})/\left|d({\bf k})\right|$ from the BZ $T^4$
to the sphere $S^4$ and $a,b,c,d,e=0,1,2,3,4.$ More details of
this calculation are presented in Appendix \ref{app:dirac4}.

Since the winding number (\ref{windingS4}) is equal to the second
Chern number of the Berry's phase gauge field, it is topologically
invariant. It is easy to calculate $C_2$ in the lattice Dirac
model (\ref{Dirac5d}). Considering the lattice Dirac model with a
fixed positive parameter $c$ and tunable mass term $m$, $C_2(m)$
as a function of $m$ can change only if the Hamiltonian is
gapless, {\em i.e.}, if ${\sum_ad_a^2({\bf k},m)}=0$ for some
${\bf k}$. It's easy to determine that $C_2(m)=0$ in the limit
$m\rightarrow +\infty$, since the unit vector $\hat{d}_a({\bf
k})\rightarrow (1,0,0,0,0)$ in that limit. Thus we only need to
study the change of $C_2(m)$ at each quantum critical points,
namely at critical values of $m$ where the system becomes gapless.

The solutions of equation ${\sum_ad_a^2({\bf k},m)}=0$ lead to five
critical values of $m$ and corresponding ${\bf k}$ points as listed
below:
\begin{eqnarray}
m&=&\left\{\begin{array}{cc}-4c,&{\bf k}=(0,0,0,0)\\-2c,&{\bf k}\in
P\left[(\pi,0,0,0)\right]\\
0,&{\bf k}\in P\left[(\pi,0,\pi,0)\right]\\
2c,&{\bf k}\in P\left[(\pi,\pi,\pi,0)\right]\\
4c,&{\bf k}=(\pi,\pi,\pi,\pi)\end{array}\right.
\end{eqnarray}
in which $P[{\bf k}]$ stands for the set of all the wavevectors
obtained from index permutations of wavevector ${\bf k}$. For
example, $P[(\pi,0,0,0)]$ consists of $(\pi,0,0,0)$,
$(0,\pi,0,0)$, $(0,0,\pi,0)$ and $(0,0,0,\pi)$. As an example, we
can study the change of $C_2(m)$ around the critical value
$m=-4c$. In the limit $m+4c\ll 2c$, the system has its minimal gap
at ${\bf k}={\bf 0}$, around which the $d_a({\bf k})$ vector has
the approximate form $d_a({\bf k})\simeq (\delta
m,k_x,k_y,k_z,k_w)+o(\left|k\right|)$, with $\delta m\equiv m+4c$.
Taking a cut-off $\Lambda\ll 2\pi$ in momentum space, one can
divide the expression (\ref{windingS4}) of $C_2$ into low-energy
and high-energy parts:
\begin{widetext}
\begin{eqnarray}
C_2=\frac3{8\pi^2}\left(\int_{\left|{\bf k}\right|\leq
\Lambda}d^4k+\int_{\left|{\bf k}\right|> \Lambda}d^4k\right)
\epsilon^{abcde}\hat{d}_a\partial_x\hat{d}_b\partial_y\hat{d}_c\partial_z\hat{d}_d\partial_w\hat{d}_e\equiv
C_2^{(1)}(\delta m,\Lambda)+C_2^{(2)}(\delta m,\Lambda).\nonumber
\end{eqnarray}
\end{widetext}
Since there is no level-crossing in the region $|{\bf
k}|>\Lambda$, the jump of $C_2$ at $\delta m=0$ can only come from
$C_2^{(1)}$. In the limit $\left|\delta m\right|<\Lambda\ll 2\pi$,
the continuum approximation of $d_a({\bf k})$ can be applied to
obtain
\begin{eqnarray}
C_2^{(1)}(\delta m,\Lambda)\simeq \frac3{8\pi^2}\int_{|{\bf k}|\leq
\Lambda}d^4k\frac{\delta m}{\left(\delta m^2+{\bf
k}^2\right)^{5/2}}\nonumber
\end{eqnarray}
which can be integrated and leads to
\begin{eqnarray}
\Delta {C_2}_{\delta m=0^-}^{\delta m=0^+}=\Delta
{C_2^{(1)}}_{\delta m=0^-}^{\delta m=0^+}=1.
\end{eqnarray}
From the analysis above we see that the change of the second Chern
number is determined only by the effective continuum model around
the level crossing wavevector(s). In this case the continuum model
is just the Dirac model. Similar analysis can be carried out at
the other critical $m$'s, which leads to the following values of
the second Chern number:
\begin{eqnarray}
C_2(m)=\left\{\begin{array}{cc}0,&m<-4c\text{ or
}m>4c\\1,&-4c<m<-2c\\-3,&-2c<m<0\\3,&0<m<2c\\-1,&2c<m<4c\end{array}.\right.
\end{eqnarray}
A more general formula is given in Ref. \onlinecite{golterman1993}.

\begin{figure}[tbp]
\includegraphics[width=2.5in]{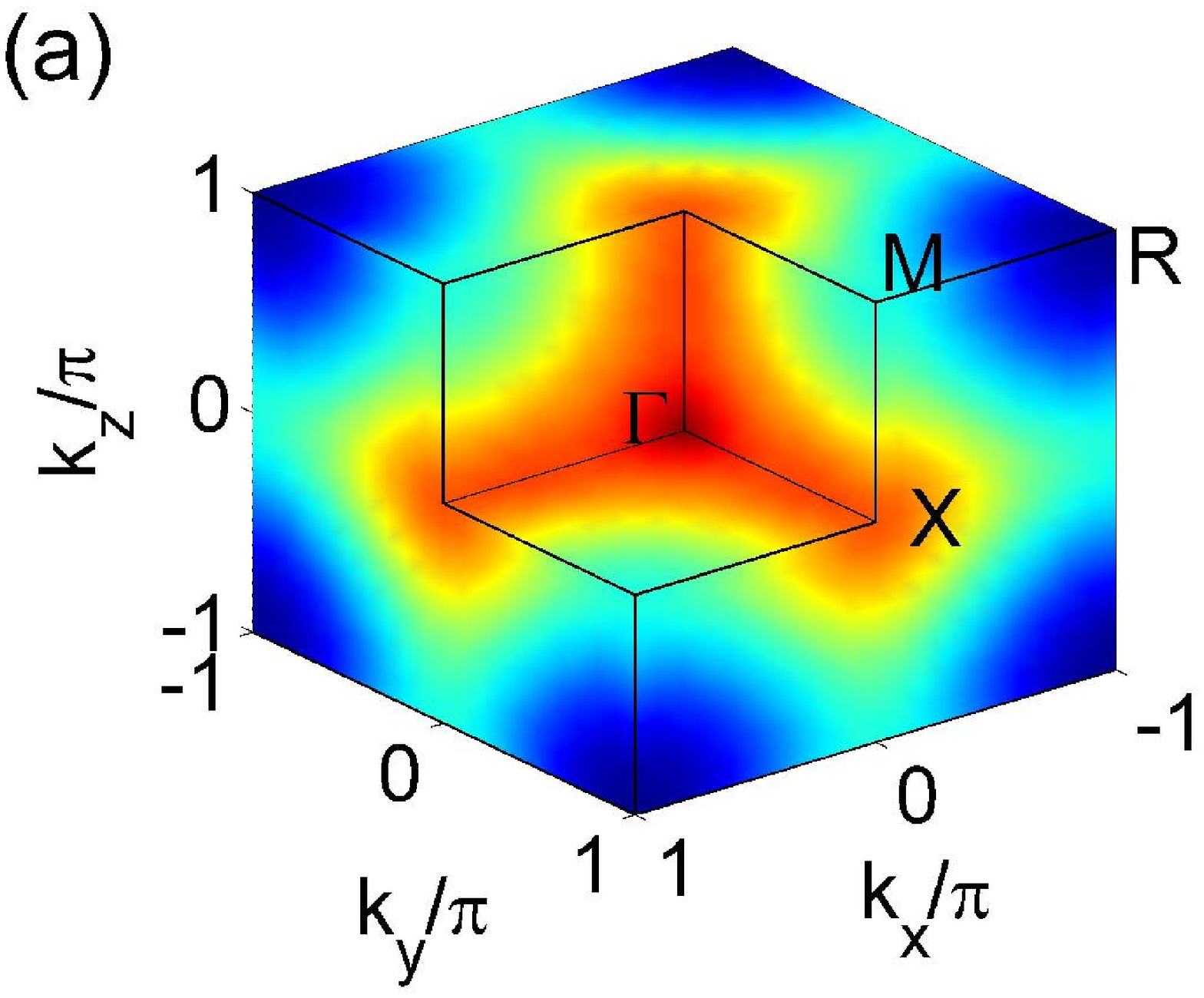}
\includegraphics[width=2.5in]{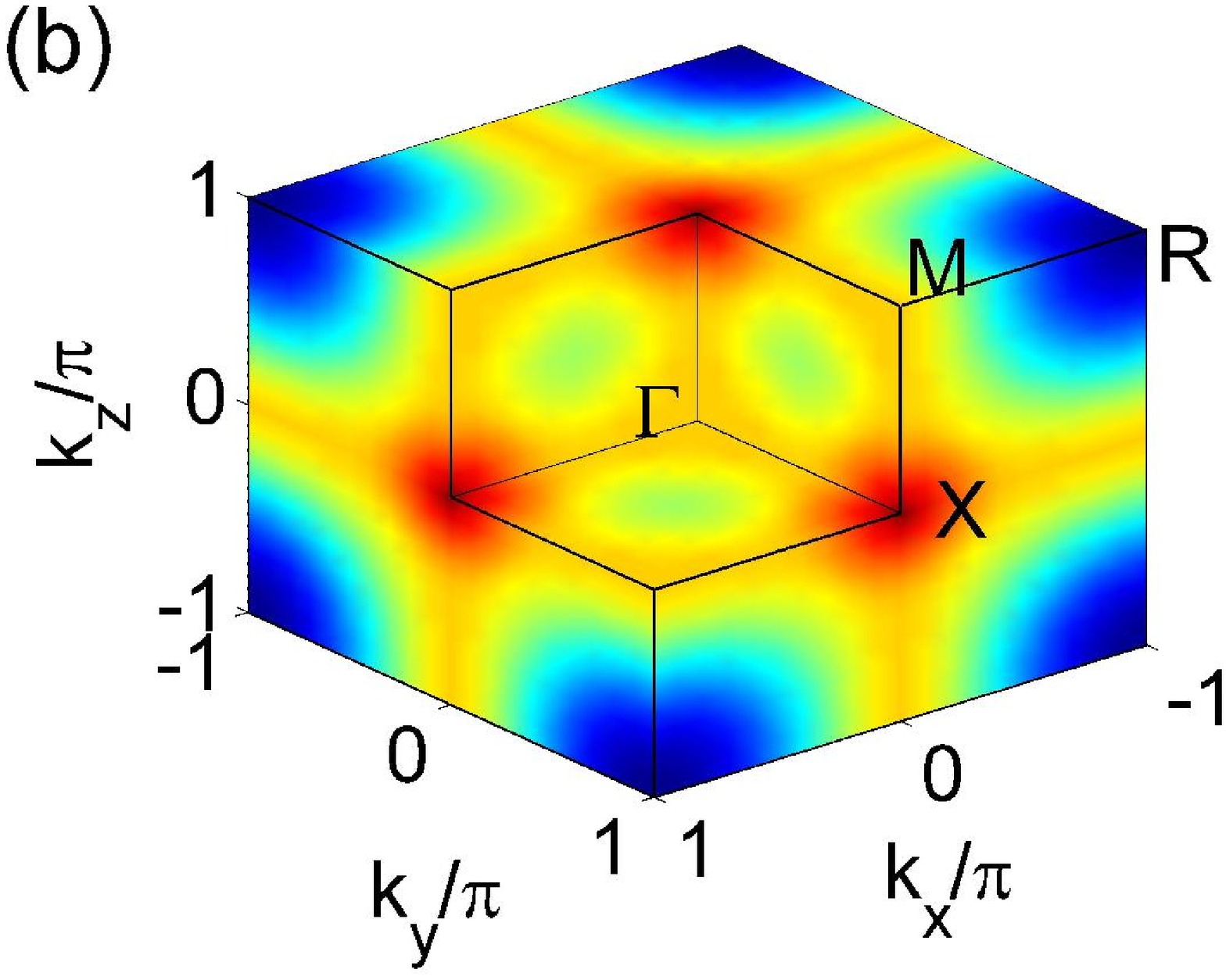}
\caption{Three dimensional energy spectrum of the surface states for
the parameters (a) $c=1,~m=-3$ and (b) $c=1,~m=-1$. The Dirac points
where energy gap vanishes are marked with deepest red color. For
$m=-3$ there is one Dirac point at $\Gamma$ point while for $m=-1$
there are three of them at $X$ points.} \label{fig:surface4d}
\end{figure}

\begin{figure}[tbp]
\includegraphics[width=2.5in]{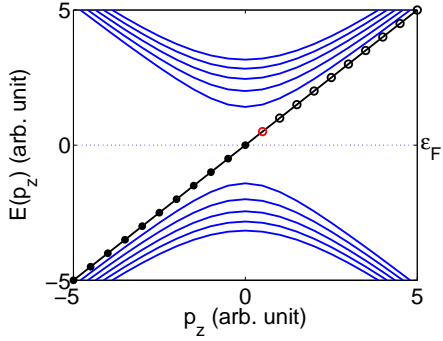}
\caption{Illustration of the surface Landau level spectrum given
by Eq. (\ref{LLspectrum}). Each level in the figure is $N_{xy}$
fold degenerate. The solid circles are the occupied states of the
zeroth Landau level, and the red open circle shows the states that
are filled when the gauge vector potential $A_z$ is shifted
adiabatically from $0$ to $2\pi/L_z$.} \label{fig:anomaly4d}
\end{figure}

After obtaining the second Chern number, we can study the surface
states of the topologically nontrivial phases of this model. In
the same way as in Sec. \ref{sec:dimreduction1d}, we can take open
boundary conditions for one dimension, say, $w$, and periodic
boundary conditions for all other dimensions, so that
$k_x,k_y,k_z$ are still good quantum numbers. The Hamiltonian is
transformed to a sum of 1D tight-binding models:
\begin{eqnarray}
H&=&\sum_{\vec{k},w}\left[\psi_{\vec{k}}^\dagger(w)\left(\frac{c\Gamma^0-i\Gamma^4}2\right)
\psi_{\vec{k}}(w+1)+h.c.\right]\nonumber\\
&+&\sum_{\vec{k},w}\psi_{\vec{k}}^\dagger(w)\left[\sin
k_i\Gamma^i+\left(m+c\sum_i\cos
k_i\right)\Gamma^0\right]\psi_{\vec{k}}(w)\nonumber\\
\end{eqnarray}
in which $\vec{k}=(k_x,k_y,k_z),$  $i=1,2,3$, and $w=1,2,..,L$ are
the $w$ coordinates of lattice sites. The single-particle energy
spectrum can be obtained as $E_\alpha(\vec{k}),~\alpha=1,2,..4L$,
among which the mid-gap surface states are found when $C_2\neq 0$,
as shown in Fig. \ref{fig:surface4d}. When the Chern number is
$C_2$, there are $|C_2|$ branches of gapless surface states with
linear dispersion, so that the low energy effective theory is
described by $|C_2|$ flavors of chiral fermions\cite{creutz2001}:
\begin{eqnarray}
H={\rm
sgn}(C_2)\int\frac{d^3p}{\left(2\pi\right)^3}\sum_{i=1}^{|C_2|}v_{i}\psi^\dagger_i(\vec{p})\vec{\bf\sigma}\cdot\vec{\bf
p}\psi_i(\vec{p}).
\end{eqnarray}
The factor ${\rm sgn}(C_2)$ ensures that the chirality of the
surface states is determined by the sign of the Chern number. From
such a surface theory we can obtain a more physical understanding
of the nonlinear response equation (\ref{topomotioneqn4d}) to an
external $U(1)$ gauge field. Taking the same gauge field
configuration as in Eq. (\ref{EBchoice}), the non-vanishing
components of the field curvature are $F_{xy}=B_{z}$ and
$F_{zt}=-E_z$. Consequently, the $(3+1)$-d surface states are
coupled to a magnetic field ${\bf B}=B_{z}\hat{\bf z}$ and an
electric field ${\bf E}=E_z\hat{\bf z}$. For simplicity, consider
the system with $-4c<m<-2c$ and $C_2=1$, in which the surface
theory is a single chiral fermion with the single particle
Hamiltonian
$$h=v\vec{\bf\sigma}\cdot\left(\vec{p}+\vec{A}\right)=v\sigma_xp_x+v\sigma_y\left(p_y+B_{z}x\right)+v\sigma_z\left(p_z-E_zt\right).$$
If $E_z$ is small enough so that the time-dependence of
$A_z(t)=-E_zt$ can be treated adiabatically, the single particle
energy spectrum can be solved for a fixed $A_z$ as
\begin{eqnarray}
E_{n\pm}(p_z)&=&\pm v\sqrt{(p_z+A_z)^2+2n|B_{z}|},~n=1,2,...\nonumber\\
E_0(p_z)&=&v(p_z+A_z){\rm sgn}(B_{z}).\label{LLspectrum}
\end{eqnarray}
When the size of the surface is taken as $L_x\times L_y\times L_z$
with periodic boundary conditions, each Landau level has the
degeneracy $N_{xy}=L_xL_yB_{z}/2\pi$. Similar to Laughlin's gauge
argument for QH edge states\cite{laughlin1983}, the effect of an
infinitesimal electric field $E_z$ can be obtained by
adiabatically shifting the momentum $p_z\rightarrow p_z+E_zt$. As
shown in Fig. \ref{fig:anomaly4d}, from the time $t=0$ to
$t=T\equiv 2\pi/L_zE_z$, the momentum is shifted as
$p_z\rightarrow p_z+2\pi/L_z$, so that the net electron number of
the surface 3D system increases by $N_{xy}$. In other words, a
``generalized Hall current" $I_w$ must be flowing towards the $w$
direction:
\begin{eqnarray}
I_w=\frac {N_{xy}}T=\frac{L_xL_yL_zB_{z}E_{z}}{4\pi^2}.\nonumber
\end{eqnarray}
This ``generalized Hall current" is the key property of the 4DQH
effect studied in Ref. \onlinecite{zhang2001}.

In terms of current density we obtain $j_w=B_{z}E_z/4\pi^2$, which
is consistent with the result of Eq. (\ref{topomotioneqn4d})
discussed in the last subsection. More generally the current
density $j_w$ can be written as
\begin{eqnarray}
j_w=C_2\frac{\bf E\cdot
B}{4\pi^2}=\frac{C_2}{32\pi^2}\epsilon^{\mu\nu\sigma\tau}F_{\mu\nu}F_{\sigma\tau}\label{anomaly4d}
\end{eqnarray}
which is the {\em chiral anomaly} equation of  massless $(3+1)$-d
Dirac fermions\cite{adler1969,bell1969}. Since the gapless states
on the 3D edge of the 4D lattice Dirac model are chiral fermions,
the current $I_w$ carries away chiral charge, leading to the
non-conservation of chirality on the 3D edge.

\section{Dimensional reduction to
$(3+1)$-d TRI insulators}\label{sec:dimreduction3d}

As shown in Sec. \ref{sec:dimreduction1d}, one can start from a
$(2+1)$-d TRB topological insulator described by a Hamiltonian
$h(k_x,k_y)$, and perform the procedure of dimensional reduction
by replacing $k_y$ by a parameter $\theta$. The same dimensional
reduction procedure can be carried out for the $(4+1)$-d TRI
insulator with a non-vanishing second Chern number. From this
procedure, one obtains the topological effective theory of
insulators in $(3+1)$-d and $(2+1)$-d. Specifically, for TRI
insulators a general $Z_2$ topological classification is defined.
Compared to the earlier proposals of the $Z_2$ topological
invariant\cite{kane2005A,fu2006,fu2007a,fu2007b,moore2007,roy2006a,roy2006b,roy2006c},
our approach provides a direct relationship between the
topological quantum number and the physically measurable
topological response of the corresponding system. We discuss a
number of theoretical predictions, including the TME effect, and
propose experimental settings where these topological effects can
be measured in realistic materials.

\subsection{Effective action of $(3+1)$-d
insulators}\label{sec:eff3d}

To perform the dimensional reduction explicitly, in the following
we show the derivation for the $(4+1)$-d Dirac model
(\ref{Dirac5d}). However, each step of the derivation is
applicable to any other insulator model, so the conclusion is
completely generic.

The Hamiltonian of Dirac model (\ref{Dirac5d}) coupled to an
external $U(1)$ gauge field is given by
\begin{eqnarray}
H[A]&=&\sum_{n,i}\left[\psi^\dagger_n\left(\frac
{c\Gamma^0-i\Gamma^i}2\right)e^{iA_{n,n+\hat{i}}}\psi_{n+\hat{i}}+h.c.\right]\nonumber\\
& &+m\sum_{n}\psi^\dagger_n\Gamma^0\psi_n .\label{Dirac5dgauge}
\end{eqnarray}
Now consider a special ``Landau"-gauge configuration satisfying
$A_{n,n+\hat{i}}=A_{n+\hat{w},n+\hat{w}+\hat{i}},~\forall n$,
which is translationally invariant in the $w$ direction. Thus,
under periodic boundary conditions the $w$-direction momentum
$k_w$ is a good quantum number, and the Hamiltonian can be
rewritten as
\begin{eqnarray}
H[A]&=&\sum_{k_w,\vec{x},s}\left[\psi^\dagger_{\vec{x},k_w}\left(\frac
{c\Gamma^0-i\Gamma^s}2\right)e^{iA_{\vec{x},\vec{x}+\hat{s}}}\psi_{\vec{x}+\hat{s},k_w}+h.c.\right]\nonumber\\
&
&+\sum_{k_w,\vec{x},s}\psi^\dagger_{\vec{x},k_w}\left[\sin\left(k_w+A_{\vec{x}4}\right)\Gamma^4\right.\nonumber\\
& &\left.+\left(m+c\cos
\left(k_w+A_{\vec{x}4}\right)\right)\Gamma^0\right]\psi_{\vec{x},k_w}\nonumber
\end{eqnarray}
where $\vec{x}$ stands for the three-dimensional coordinates,
$A_{\vec{x}4}\equiv A_{\vec{x},\vec{x}+\hat{w}},$ and $s=1,2,3$
stands for the $x,y,z$ directions. In this expression, the states
with different $k_w$ decouple from each other, and the $(4+1)$-d
Hamiltonian $H[A]$ reduces to a series of $(3+1)$-d Hamiltonians.
Pick one of these $(3+1)$-d Hamiltonians with fixed $k_w$ and
rename $k_w+A_{\vec{x}4}=\theta_{\vec{x}}$, we obtain the
$(3+1)$-d model
\begin{eqnarray}
H_{\rm
3D}[A,\theta]&=&\sum_{\vec{x},s}\left[\psi^\dagger_{\vec{x}}\left(\frac
{c\Gamma^0-i\Gamma^s}2\right)e^{iA_{\vec{x},\vec{x}+\hat{s}}}\psi_{\vec{x}+\hat{s}}+h.c.\right]\nonumber\\
&
&+\sum_{\vec{x},s}\psi^\dagger_{\vec{x}}\left[\sin\theta_{\vec{x}}\Gamma^4+\left(m+c\cos
\theta_{\vec{x}}\right)\Gamma^0\right]\psi_{\vec{x}}\nonumber\\
\label{Dirac3d}
\end{eqnarray}
which describes a band insulator coupled to an electromagnetic
field $A_{\vec{x},\vec{x}+\hat{s}}$ and an adiabatic parameter
field $\theta_{\vec{x}}$.

Due to its construction, the response of the model (\ref{Dirac3d})
to $A_{\vec{x},\vec{x}+\hat{s}}$ and $\theta_{\vec{x}}$ fields is
closely related to the response of the $(4+1)$-d Dirac model
(\ref{Dirac5d}) to the $U(1)$ gauge field. To study the response
properties of the $(3+1)$-d system, the effective action $S_{\rm
3D}[A,\theta]$ can be defined as
\begin{eqnarray}
\exp^{iS_{\rm 3D}[A,\theta]}&=&\int D[\psi]D[\bar{\psi}]e^{i\int
dt\left[\sum_{\vec{x}}\bar{\psi}_{\vec{x}}
\left(i\partial_\tau-A_{\vec{x}0}\right)\psi_{\vec{x}}-H[A,\theta]\right]}.\nonumber
\end{eqnarray}
A Taylor expansion of $S_{\rm 3D}$ can be carried out around the
field configuration $A_s(\vec{x},t)\equiv
0,~\theta(\vec{x},t)\equiv\theta_0$, which contains a non-linear
response term directly derived from the $(4+1)$-d Chern-Simons
action (\ref{Seff4d}):
\begin{eqnarray}
S_{\rm 3D}=\frac{G_3(\theta_0)}{4\pi}\int
d^3xdt\epsilon^{\mu\nu\sigma\tau}\delta\theta\partial_\mu
A_\nu\partial_\sigma A_\tau .\label{Seff3d}
\end{eqnarray}
Compared to the Eq. (\ref{Seff4d}), the field
$\delta\theta(\vec{x},t)=\theta(\vec{x},t)-\theta_0$ plays the
role of $A_{4}$, and the coefficient $G_3(\theta_0)$ is determined
by the same Feynman diagram (\ref{Feynman4d}), but evaluated for
the three-dimensional Hamiltonian (\ref{Dirac3d}). Consequently,
$G_3(\theta_0)$ can be calculated and is equal to Eq.
(\ref{C2green}), but without the integration over $k_w$:
\begin{widetext}
\begin{eqnarray}
G_3(\theta_0)=-\frac{\pi}{6}\int
\frac{d^3kd\omega}{\left(2\pi\right)^4}{\rm
Tr}\epsilon^{\mu\nu\sigma\tau}\left[\left(G\frac{\partial
G^{-1}}{\partial q^\mu}\right)\left(G\frac{\partial
G^{-1}}{\partial q^\nu}\right)\left(G\frac{\partial
G^{-1}}{\partial q^\sigma}\right)\left(G\frac{\partial
G^{-1}}{\partial q^\tau}\right)\left(G\frac{\partial
G^{-1}}{\partial \theta_0}\right)\right] \label{correlation3d}
\end{eqnarray}
\end{widetext}\noindent where  $q^{\mu}=\left(\omega,k_x,~k_y,~k_z\right).$
Due to the same calculation as  Sec. \ref{sec:secondtknn} and
Appendix \ref{app:C2green}, $G_3(\theta_0)$ is determined from the
Berry phase curvature as
\begin{eqnarray}
G_3(\theta_0)&=&\frac1{8\pi^2}\int d^3k \epsilon^{ijk}{\rm
tr}\left[f_{\theta i}f_{jk}\right],\label{G3Berrycurvature}
\end{eqnarray}
in which  the Berry phase gauge field is defined in the
four-dimensional space $\left(k_x,~k_y,~k_z,~\theta_0\right)$,
{\em i.e.}, $a_i^{\alpha\beta}=-i\left\langle
\vec{k},\theta_0;\alpha\right|\left(\partial/\partial
k_i\right)\left|\vec{k},\theta_0;\beta\right\rangle$ and
$a_\theta^{\alpha\beta}=-i\left\langle
\vec{k},\theta_0;\alpha\right|\left(\partial/\partial
\theta_0\right)\left|\vec{k},\theta_0;\beta\right\rangle$.
Compared to the second Chern number (\ref{2ndtknn}), we know that
$G_3(\theta_0)$ satisfies the sum rule
\begin{eqnarray}
\int G_3(\theta_0)d\theta_0=C_2\in\mathbb{Z}\label{sumrule3d},
\end{eqnarray}
which is in exact analogy with the sum rule of the pumping
coefficient $G_1(\theta)$ in Eq. (\ref{sumrule1d}) of the
$(1+1)$-d system. Recall that $G_1(\theta)$ can be expressed as
$\partial P_1(\theta)/\partial\theta$, where $P_1(\theta)$ is
simply the charge polarization. In comparison, a generalized
polarization $P_3(\theta_0)$ can also be defined in $(3+1)$-d so
that $G_3(\theta_0)=\partial P_3(\theta_0)/\partial\theta_0$. (
Recently, a similar quantity has also been considered in Ref.
\onlinecite{xiao2007} from the point of view of semiclassical
particle dynamics.) The conventional electric polarization ${\bf
P}$ couples linearly to the external electric field ${\bf E}$, and
the magnetic polarization ${\bf M}$ couples linearly to the
magnetic field ${\bf B}$, however, as we shall show, $P_3$ is a
pseudo-scalar which couples non-linearly to the external
electromagnetic field combination ${\bf E} \cdot {\bf B}$. For
this reason, we coin the term ``magneto-electric polarization" for
$P_3$. To obtain $P_3(\theta_0)$, one needs to introduce the
non-Abelian Chern-Simons term:
\begin{eqnarray}
\mathcal{K}^{A}=\frac1{16\pi^2}\epsilon^{ABCD}{\rm
Tr}\left[\left(f_{BC}-\frac13\left[a_B,a_C\right]\right)\cdot
a_D\right],\label{CS3form}
\end{eqnarray}
which is a vector in the four-dimensional parameter space
$q=(k_x,k_y,k_z,\theta_0)$ and $A,B,C,D=x,y,z,\theta.$
$\mathcal{K}^A$ satisfies
\begin{eqnarray}
\partial_A
\mathcal{K}^A&=&\frac1{32\pi^2}\epsilon^{ABCD}{\rm
tr}\left[f_{AB}f_{CD}\right]\nonumber\\
\Rightarrow G_3(\theta_0)&=&\int
d^3k\partial_A\mathcal{K}^A.\nonumber
\end{eqnarray}
When the second Chern number is nonzero, there is an obstruction
to the definition of $a_A$, which implies that $\mathcal{K}_A$
cannot be a single-valued continuous function in the whole
parameter space. However, in an appropriate gauge choice,
$\mathcal{K}^i,i=x,y,z$ can be single-valued, so that
$G_3(\theta_0)=\int d^3k\partial_\theta \mathcal{K}^\theta\equiv
\partial P_3(\theta_0)/\partial \theta_0$, with
\begin{eqnarray}
P_3(\theta_0)&=&\int
d^3k\mathcal{K}^\theta\nonumber\\
&=&\frac1{16\pi^2}\int d^3k\epsilon^{\theta ijk}{\rm
Tr}\left[\left(f_{ij}-\frac13\left[a_i,a_j\right]\right)\cdot
a_k\right]. \nonumber\\\label{P3}
\end{eqnarray}\noindent Thus, $P_3(\theta_0)$ is given by the
integral of the non-Abelian Chern-Simons  $3$-form over momentum
space. This is analogous to the charge polarization defined as the
integral of the adiabatic connection $1$-form over a path in
momentum space.

As is well-known, the three-dimensional integration of the
Chern-Simons term is only gauge-invariant modulo an integer. Under
a gauge transformation $a_i\rightarrow u^{-1}
a_iu-iu^{-1}\partial_i u$ ($u\in U(M)$ when $M$ bands are
occupied), the change of $P_3$ is
$$\Delta P_3=\frac i{24\pi^2}\int d^3k\epsilon^{\theta ijk}{\rm
Tr}\left[\left(u^{-1}\partial_iu\right)\left(u^{-1}\partial_ju\right)\left(u^{-1}\partial_ku\right)\right],$$
which is an integer. Thus $P_3(\theta_0)$, just like
$P_1(\theta)$, is only defined modulo $1$, and its change during a
variation of $\theta_0$ from $0$ to $2\pi$ is well-defined, and
given by $C_2$.

The effective action (\ref{Seff3d}) can be further simplified by
introducing $G_3=\partial P_3/\partial\theta$. Integration by
parts of $S_{\rm 3D}$ leads to $$S_{\rm 3D}=\frac1{4\pi}\int
d^3xdt\epsilon^{\mu\nu\sigma\tau}A_\mu(\partial
P_3/\partial\theta)\partial_\nu\delta\theta\partial_\sigma
A_\tau.$$ $(\partial P_3/\partial\theta)\partial_\nu\delta\theta$
can be written as $\partial_\nu P_3$, where
$P_3(\vec{x},t)=P_3(\theta(\vec{x},t))$ has space-time dependence
determined by the $\theta$ field. Such an expression is only
meaningful when the space-time dependence of $\theta$ field is
smooth and adiabatic, so that locally $\theta$ can still be
considered as a parameter. In summary, the effective action is
finally written as
\begin{eqnarray}
S_{\rm 3D}=\frac1{4\pi}\int d^3xdt\epsilon^{\mu\nu\sigma\tau}
P_3(x,t) \partial_\mu A_\nu\partial_\sigma A_\tau
.\label{Seff3dP3}
\end{eqnarray}
This effective topological action for the $(3+1)$-d insulator is
one of the central results of this paper. As we shall see later,
many physical consequences can be directly derived from it. It
should be emphasized that this effective action is well-defined
for an arbitrary $(3+1)$-d insulator Hamiltonian
$h(\vec{k},\vec{x},t)$ in which the dependence on $\vec{x},t$ is
adiabatic. We obtained this effective theory by the dimensional
reduction from a $(4+1)$-d system; and we presented it this way
since we believe that this derivation is both elegant and
unifying. However, for readers who are not interested in the
relationship to higher dimensional physics, a self-contained
derivation can also be carried out directly in $(3+1)$-d, as we
explained earlier, by integrating out the fermions in the presence
of the $A_\mu(x,t)$ and the $\theta(x,t)$ external fields.

This effective action is known in the field theory literature as
axion electrodynamics\cite{huang1985,wilczek1987,lee1987}, where
the adiabatic field $P_3$ plays the role of the axion
field\cite{peccei1977,sikivie1984}. When the $P_3$ field becomes a
constant parameter independent of space and time, this effective
action is referred to as the topological term for the $\theta$
vacuum\cite{callan1976,jackiw1976b}.  The axion field has not yet
been experimentally identified, and it remains as a deep mystery
in particle physics. Our work shows that the same physics can
occur in a condensed matter system, where the adiabatic ``axion"
field $P_3(x,t)$ has a direct physical interpretation and can be
accessed and controlled experimentally.

From the discussion above it is clear that 3D TRI topological
insulators realize a non-trivial solitonic background $\theta$
field. In Ref. \onlinecite{fu2007a} the authors suggest several
candidate materials which could be 3D topological insulators.
These 3D materials are topologically non-trivial because of band
inversion mechanism similar to that of the HgTe quantum
wells\cite{bernevig2006d}. Ref. \onlinecite{dai2007} closely
studied the strained, bulk HgTe. We will keep this system in mind
since it has a simple physical interpretation, and its essential
physics can be described by the Dirac model presented earlier. We
can consider the trivial vacuum outside the material to have a
constant axion field with the value $\theta=0$ and the interior of
a 3D topological insulator to have a $\theta=\pi$ background
field. The value $\theta=\pi$ does not violate time-reversal (or
CP in high-energy language). HgTe is a zero-gap semiconductor and
has no topologically protected features. However, when strained,
the system develops a bulk insulating gap between the p-wave
light-hole ``conduction band" and the p-wave heavy-hole ``valence
band" around the $\Gamma$-point. To study the topological features
we must also include the s-wave band which in a conventional
material like GaAs would be a conduction band. Because of the
strong spin-orbit coupling in HgTe the band structure is actually
inverted the s-wave band becomes a valence band. For a moment we
will ignore the heavy-hole band and only consider the light-hole
and s-wave band\cite{dai2007}. The effective Hamiltonian of these
two bands is a massive Dirac Hamiltonian, but with a negative
mass. The negative mass indicates a phase shift of $\pi$ in the
vacuum angle $\theta$ from its original unshifted value in the
trivial vacuum. The axion domain wall structure at the surface of
the topological insulator traps fermion zero modes which are
simply the topologically protected surface states. If we include
the effects of the heavy-hole band the dispersion of the bulk
bands and surface states are quantitatively modified. However,  as
long as the crystal is strained enough to maintain the bulk gap
the topological phenomena will be unaffected and the boundary of
the 3D topological insulator can still be described as an axion
domain wall. Thus, this material in condensed matter physics
provide a direct realization of axion electrodynamics.

\subsection{Physical
Consequences of the Effective Action $S_{3D}$}\label{sec:phys3d}

In this subsection we present the general physical consequences of
the effective topological action (\ref{Seff3dP3}) for $(3+1)$-d
insulators coupled to a $P_3$ polarization, and in subsection
\ref{sec:3dZ2} we focus on its consequences for TRI insulators.
Since the effective action is quadratic in $A_\mu$, it describes a
linear response to the external electromagnetic fields which
depends on the spatial and temporal gradients of $P_3$. Taking a
variation of $S_{\rm 3D}[A,\theta]$ we obtain the response
equation:
\begin{eqnarray}
j^\mu=\frac{1}{2\pi}\epsilon^{\mu\nu\sigma\tau}\partial_\nu
P_3\partial_\sigma A_\tau .\label{motioneqn3d}
\end{eqnarray}
The physical consequences Eq. (\ref{motioneqn3d}) can be
understood by studying the following two cases.

{\bf (1) Hall effect induced by spatial gradient of P$_3$}.

\begin{figure}[tbp]
\includegraphics[width=3.5in]{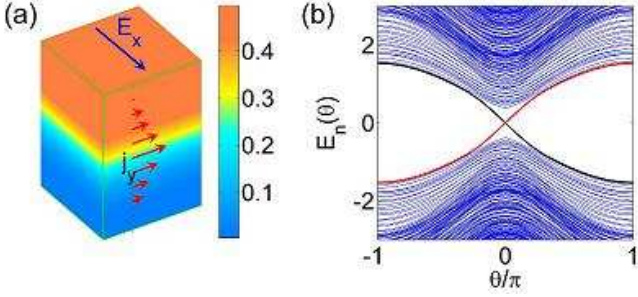}
\caption{(a) Illustration of the Hall effect induced by a spatial
gradient of $P_3$. The colors represent different values of $P_3$,
which decrease from $0$ at the bottom to $1/2$ on top. The blue
arrow shows the direction of a uniform electric field $E_x$ and
the white arrows show the Hall current density induced, given by
the formula $j_y=(\partial_zP_3/2\pi)E_x$. (b) Energy spectrum of
the $(3+1)$-d lattice Dirac model (\ref{Dirac3d}) in a magnetic
field $B_z$ towards the $z$ direction. The boundary conditions are
periodic in the $x$ and $y$-directions and open on the
$z$-direction. The red and black curves show the surface states on
the top and bottom surfaces, respectively, each of which is a
$B_zL_xL_y/2\pi$ fold degenerate Landau Level. The parameters of
model (\ref{Dirac3d}) are chosen to be $m=-3,~c=1$.}
\label{3dschematic}
\end{figure}

Consider a system in which $P_3=P_3(z)$ only depends on  $z$. For
example, this can be realized by the lattice Dirac model
(\ref{Dirac3d}) with $\theta=\theta(z)$. (This type of domain wall
has also been considered in Ref. \onlinecite{fradkin1986}). In
this case Eq. (\ref{motioneqn3d}) becomes
\begin{eqnarray}
j^\mu=\frac{\partial_zP_3}{2\pi}\epsilon^{\mu\mu\rho}\partial_\nu
A_\rho,~\mu,\nu,\rho=t,x,y\nonumber
\end{eqnarray}
which describes a QH effect in the $xy$ plane with the Hall
conductivity $\sigma_{xy}=\partial_zP_3/2\pi$, as shown in Fig.
\ref{3dschematic} (a). For a uniform electric field $E_x$ in the
$x$-direction, the Hall current density is
$j_y=(\partial_zP_3/2\pi)E_x$. Thus the integration over $z$ in a
finite range gives the 2D current density in the $xy$ plane:
$$J_y^{\rm 2D}=\int_{z_1}^{z_2} dz j_y=\frac{1}{2\pi}\left(\int_{z_1}^{z_2}
dP_3\right)E_x.
$$
In other words, the net Hall conductance of the region $z_1\leq
z\leq z_2$ is
\begin{eqnarray}
\sigma^{\rm 2D}_{xy}=\int_{z_1}^{z_2}dP_3/2\pi,\label{sigmaHP3}
\end{eqnarray}
which only depends on the change of $P_3$ in this region, and is
not sensitive to any details of the function $P_3(z)$. Analogously
in the $(1+1)-d$ case, if we perform the spatial integration of
Eq. (\ref{charge1d}), we obtain the total charge induced by the
charge polarization $P$:
\begin{eqnarray}
Q=-\int_{z_1}^{z_2}dP/2\pi.
\end{eqnarray}
By comparing these two equations, we see that the relation between
$P_3$ and Hall conductance in $(3+1)$-d insulators is the same as
the relation between charge polarization $P$ and the total charge
in the $(1+1)$-d case. As a specific case, a domain wall between
two homogeneous materials with different $P_3$ will carry Hall
conductance $\sigma_H=\Delta P_3/2\pi$, while the fractional
charge carried by a domain wall in $(1+1)$-d is given by
$Q=-\Delta P/2\pi$.

{\bf (2) Topological Magneto-electric effect(TME) induced by
temporal gradient of P$_3$.}

When $P_3=P_3(t)$ is spatially uniform, but time-dependent, Eq.
(\ref{motioneqn3d}) becomes
\begin{eqnarray}
j^i=-\frac{\partial_tP_3}{2\pi}\epsilon^{ijk}\partial_jA_k,~i,j,k=x,y,z
.\nonumber
\end{eqnarray}
In other words, we have
\begin{eqnarray}
\vec{j}=-\frac{\partial_tP_3}{2\pi}\vec{B} .\label{MEeffect}
\end{eqnarray}
Since the charge polarization $\vec{P}$ satisfies
$\vec{j}=\partial_t\vec{P}$, in a static uniform magnetic field $B$
we have $\partial_t\vec{P}=-\partial_t\left(P_3\vec{B}/2\pi\right)$,
so that
\begin{eqnarray}
\vec{P}=-\frac{\vec{B}}{2\pi}\left(P_3+{\rm
const.}\right).\label{MEeffect2}
\end{eqnarray}
Such an equation describes the charge polarization induced by a
magnetic field, which is a magneto-electric effect. Compared to
similar effects in multiferroic
materials\cite{hur2004,Eerenstein2006}, the magneto-electric
effect obtained here is of topological origin, and only determined
by the magneto-electric polarization $P_3$.

Similar to the $(1+1)$-d adiabatic pumping effect, the response
Eq. (\ref{MEeffect}) can be understood in a surface state picture.
For example, consider the lattice Dirac model (\ref{Dirac3d}) with
periodic boundary conditions in the $x,y$ directions and open
boundary conditions in the $z$-direction. In the presence of a
static magnetic field $B_z$ in the $z$-direction, the single
particle energy spectrum $E_n(\theta)$ can be solved for at a
fixed $\theta$ value. As shown in Fig. \ref{3dschematic} (b),
mid-gap states appear for generic $\theta$, which are localized on
the $(2+1)$-d boundary. It should be noticed that each state is
$N$-fold degenerate where $N=B_zL_xL_y/2\pi$ is the Landau level
degeneracy. In the lattice Dirac model, when $-4c<m<-2c$ so that
$C_2=\int_{\theta=0}^{\theta=2\pi} dP_3=1$, we find that during a
period $\theta=0\rightarrow 2\pi$, $N$ degenerate surface states
on the bottom boundary sink below fermi level and $N$ states on
the top float up. Consequently, when $\theta$ is adiabatically
tuned from $0$ to $2\pi$, there are $N$ electrons pumped from the
top surface to the bottom one, which is in consistent with the
result of Eq. (\ref{MEeffect}):$$\Delta Q=\int dt\int dxdy
j_z=-\frac{\int_0^{2\pi} dP_3}{2\pi}B_zL_xL_y=-NC_2.$$

Just like the relation between $(2+1)$-d QH edge states and the
mid-gap end states in the $(1+1)$-d pumping effect, there is a
direct relationship between the $(3+1)$-d surface states of the
$(4+1)$-d lattice Dirac model and the adiabatic pumping discussed
above. As discussed in Sec. \ref{sec:Dirac5d}, the surface theory
of a $(4+1)$-d lattice Dirac model with nontrivial $C_2$ is a
$(3+1)$-d chiral fermion. As shown in Fig. \ref{fig:anomaly4d},
the energy spectrum in a magnetic field $B_z$ has a chiral
dependence on the wavevector $p_z$. During the dimensional
reduction procedure, $p_z$ (in the notation of Sec.
\ref{sec:Dirac5d}) is replaced by the parameter $\theta$, so that
the chiral energy spectrum $E(p_z)$ changes to the ``chiral"
$\theta$ dependence of $E_n(\theta)$ in Fig. \ref{3dschematic}
(b). In other words, the adiabatic pumping in the magnetic field
in the $(3+1)$-d system is a dimensionally reduced version of the
chiral anomaly on the surface of a $(4+1)$-d topological
insulator.

The TME leads to a striking consequence if magnetic monopoles are
present. For a uniform $P_3$, Eq. (\ref{MEeffect}) leads to
$$\nabla\cdot
\vec{j}=-\frac{\partial_tP_3}{2\pi}\nabla\cdot\vec{B}.$$ Suppose
we consider a compact $U(1)$ electromagnetic field on a lattice,
where the monopole density $\rho_m=\nabla\cdot \vec{B}/2\pi$ can
be non-vanishing, then we obtain
\begin{eqnarray}
\partial_t\rho_e=\left(\partial_tP_3\right)\rho_m.\label{monopolecharge}
\end{eqnarray}
Therefore, when $P_3$ is adiabatically changed from zero to
$\Theta/2\pi$, the magnetic monopole will acquire a charge of
\begin{eqnarray}
Q_e=\frac{\Theta}{2\pi} Q_m.
\end{eqnarray}
Such a relation was first derived by Witten in the context of the
topological term obtained from QCD\cite{witten1979}.

\subsection{$Z_2$ topological
classification of time-reversal invariant
insulators}\label{sec:3dZ2}

In Sec. \ref{sec:Z21d} we have seen how a $Z_2$ topological
classification is obtained for $(1+1)$-d particle-hole symmetric
insulators. The key point for that case is to show that any
interpolation between two particle-hole symmetric insulators
$h_1(k)$ and $h_2(k)$ carries the same parity of Chern number, so
that the ``relative Chern parity" is well-defined for each two
Hamiltonians with particle-hole symmetry. In this section, we will
show that the same approach can be applied to $(3+1)$-d
insulators, where the time-reversal symmetry plays the same role
as particle-hole symmetry does in $(1+1)$-d.

For a Hamiltonian $H=\sum_{m,n}c_{m\alpha}^\dagger
h_{mn}^{\alpha\beta}c_{n\beta}$, the time-reversal transformation
is an anti-unitary operation defined by $c_{m\alpha}\rightarrow
T^{\alpha\beta}c_{m\beta}$, where the {\em time-reversal} matrix
$T$ satisfies $T^\dagger T=\mathbb{I}$ and $T^*T=-\mathbb{I}$. In
$\vec{k}$-space  time-reversal symmetry requires
\begin{eqnarray}
T^\dagger h(-\vec{k})T=h^T(\vec{k}).\label{Tcondition}
\end{eqnarray}
The condition $T^*T=-\mathbb{I}$ is essential, and leads to
Kramers's degeneracy. Now we will follow the same approach as
Sec. \ref{sec:Z21d} and show how to define a $Z_2$ invariant for
the TRI insulators in $(3+1)$-d. For any two TRI band insulators
$h_1(\vec{k})$ and $h_2(\vec{k})$, an interpolation
$h(\vec{k},\theta)$ can be defined, satisfying
\begin{eqnarray}
h(\vec{k},0)=h_1(\vec{k}),~h(\vec{k},\pi)=h_2(\vec{k})\nonumber\\
T^\dagger
h(-\vec{k},-\theta)T=h^T(\vec{k},\theta),\label{Tparametrization}
\end{eqnarray}
and $h(\vec{k},\theta)$ is gapped for any $\theta\in[0,2\pi]$.
Since the interpolation is periodic in $\theta$, a second Chern
number $C_2[h(\vec{k},\theta)]$ of the Berry phase gauge field can
be defined in the $(\vec{k},\theta)$ space. In the same way as in
Sec. \ref{sec:Z21d}, we will demonstrate below that
$C_2[h(\vec{k},\theta)]-C_2[h'(\vec{k},\theta)]=0~{\rm mod}~2$ for
any two interpolations $h$ and $h'$. First of all, two new
interpolations $g_{1,2}(\vec{k},\theta)$ can be defined by Eq.
(\ref{defg1g2}), which we repeat here for convenience:
\begin{eqnarray}
g_1(k,\theta)&=&\left\{\begin{array}{cc}h(k,\theta),&\theta\in[0,\pi]\\h'(k,2\pi-\theta),&\theta\in[\pi,2\pi]\end{array}\right.\nonumber\\
g_2(k,\theta)&=&\left\{\begin{array}{cc}h'(k,2\pi-\theta),&\theta\in[0,\pi]\\h(k,\theta),&\theta\in[\pi,2\pi]\end{array}\right.
.\nonumber
\end{eqnarray}
By their definition, $g_1$ and $g_2$ satisfy
$C_2[h]-C_2[h']=C_2[g_1]+C_2[g_2]$ and $T^\dagger
g_1(-\vec{k},-\theta)T=g_2^T(\vec{k},\theta)$. To demonstrate
$C_2[g_1]=C_2[g_2]$, consider an eigenstate
$\left|\vec{k},\theta;\alpha\right\rangle_1$ of
$g_1(\vec{k},\theta)$ with eigenvalue $E_\alpha(\vec{k},\theta)$.
We have
\begin{eqnarray}
g_2^T(-\vec{k},-\theta)T^\dagger\left|\vec{k},\theta;\alpha\right\rangle_1&=&T^\dagger
g_1(\vec{k},\theta)\left|\vec{k},\theta;\alpha\right\rangle_1\nonumber\\
&=&E_\alpha(\vec{k},\theta)T^\dagger\left|\vec{k},\theta;\alpha\right\rangle_1\nonumber\\
\Rightarrow
g_2(-\vec{k},-\theta)T^T\left(\left|\vec{k},\theta;\alpha\right\rangle_1\right)^*&=&E_\alpha(\vec{k},\theta)T^T\left(\left|\vec{k},\theta;\alpha\right\rangle_1\right)^*
.\nonumber
\end{eqnarray}
Thus $T^T(|\vec{k},\theta;\alpha\rangle_1)^*$ is an eigenstate of
$g_2(-\vec{k},-\theta)$ with the same eigenvalue
$E_\alpha(\vec{k},\theta)$. Expand over the eigenstates
$|-\vec{k},-\theta,\beta\rangle_2$ of $g_2(-\vec{k},-\theta)$, we
have
\begin{eqnarray}
T^T\left(\left|\vec{k},\theta;\alpha\right\rangle_1\right)^*=\sum_\beta
U_{\alpha\beta}(\vec{k},\theta)\left|-\vec{k},-\theta;\beta\right\rangle_2
.
\end{eqnarray}
Consequently the Berry phase gauge vector of the $g_1$ and $g_2$
systems satisfies
\begin{widetext}
\begin{eqnarray}
a^{\alpha\beta}_{1j}(\vec{k},\theta)&=&-i\left\langle
\vec{k},\theta;\alpha\right|\partial_j\left|\vec{k},\theta;\beta\right\rangle_1=-i\left[\sum_{\gamma,\delta}{U^*_{\alpha\gamma}}\left\langle
-\vec{k},-\theta;\gamma\right|\partial_j
\left(U_{\beta\delta}\left|
-\vec{k},-\theta;\delta\right\rangle_2\right)\right]^*\nonumber\\
&=&\sum_{\gamma,\delta}U_{\alpha\gamma}a^{\gamma\delta
*}_{2j}(-\vec{k},-\theta){(U^\dagger)}_{\delta\beta}-i\sum_\gamma
U_{\alpha\gamma}(\vec{k},\theta)\partial_j
U_{\beta\gamma}^*(\vec{k},\theta).\label{TsymmetryA}
\end{eqnarray}
\end{widetext}
In other words, $a_{1j}^{\alpha\beta}(\vec{k},\theta)$ is equal to
$a_{2j}^{\alpha\beta}(-\vec{k},-\theta)$ up to a gauge
transformation. Consequently, the Berry phase curvature satisfies
$f_{1ij}^{\alpha\beta}(\vec{k},\theta)=U_{\alpha\gamma}f_{2ij}^{\gamma\delta
*}(-\vec{k},-\theta){(U^\dagger)}_{\delta\beta}$, which thus leads
to $C_2[g_1(\vec{k},\theta)]=C_2[g_2(\vec{k},\theta)]$. In
summary, we have proved
$C_2[h(\vec{k},\theta)]-C_2[h'(\vec{k},\theta)]=2C_2[g(\vec{k},\theta)]=0~{\rm
mod}~2$ for any two symmetric interpolations $h$ and $h'$. Thus
the ``relative second Chern parity"
$$N_3[h_1(\vec{k}),h_2(\vec{k})]=(-1)^{C_2[h(\vec{k},\theta)]}$$ is
well-defined for any two time-reversal invariant $(3+1)$-d
insulators, independent on the choice of interpolation. In the
same way as in $(1+1)$-d, a vacuum Hamiltonian $h_0(\vec{k})\equiv
h_0,~\forall \vec{k}$ can be defined as a reference. All the
Hamiltonians with $N_3[h_0,h]=-1$ are called $Z_2$ nontrivial,
while those with $N_3[h_0,h]=1$ are trivial.

Similar to the $(1+1)$-d case, there is a more intuitive, but less
rigorous, way to define the $Z_2$ invariant $N_3$. Through the
derivation of Eq. (\ref{TsymmetryA}) one can see that for a TRI
Hamiltonian satisfying Eq. (\ref{Tcondition}), the Berry's phase
gauge potential satisfies
$a_{i}(\vec{k})=Ua_i(-\vec{k})U^\dagger-iU\partial_i U^\dagger$,
so that the magneto-electric polarization $P_3$ satisfies
$$2P_3=\frac i{24\pi^2}\int d^3k\epsilon^{ijk}{\rm Tr}\left[\left(U\partial_iU^\dagger\right)
\left(U\partial_jU^\dagger\right)\left(U\partial_kU^\dagger\right)\right]\in\mathbb{Z}.$$
Consequently, there are only two inequivalent, TRI values of
$P_3$, which are $P_3=0$ and $P_3=1/2$. For two Hamiltonians $h_1$
and $h_2$, the second Chern number
$C_2[h(\vec{k},\theta)]=2\left(P_3[h_2]-P_3[h_1]\right)~{\rm
mod}~2$, so the difference of $P_3$ determines the relative Chern
parity $N_3[h_1,h_2]$ by
$N_3[h_1,h_2]=(-1)^{2(P_3[h_1]-P_3[h_2])}$. Since the trivial
Hamiltonian $h_0$ obviously has $P_3=0$, we know that all the
Hamiltonians with $P_3=1/2$ are topologically non-trivial, while
those with $P_3=0$ are trivial.

\begin{figure}[tbp]
\includegraphics[width=3.5in]{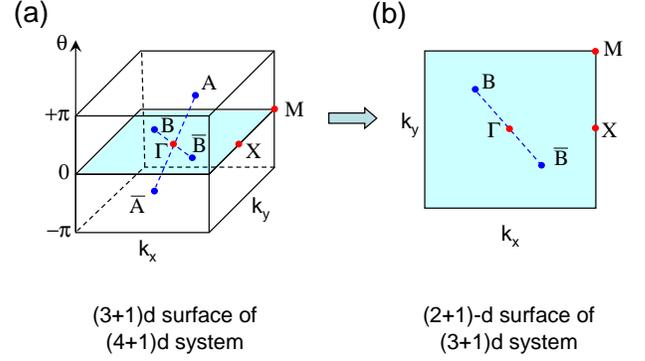}
\caption{Illustration of the nodal points in the surface state
energy spectrum of $(4+1)$-d and $(3+1)$-d insulators. (a) The
nodal points in the $(k_x,k_y,\theta)$ BZ, for a $z={\rm const.}$
surface of the $(4+1)$-d system. The red (blue) points stand for
nodal points at time-reversal symmetric (asymmetric) wavevectors.
The dashed lines are a guide for the eyes. There are two pairs of
asymmetric nodal points and three symmetric points in this
example, which correspond to a bulk second Chern number $C_2=7$.
(b) The nodal points in the $(k_x,k_y)$ BZ, for a $z={\rm const.}$
surface of the $(3+1)$-d system. According to the dimensional
reduction procedure (see text) the 2D surface energy spectrum is
given by the $\theta=0$ slice of the 3D surface spectrum in (a).
Since $\theta=\pi$ corresponds to a vacuum Hamiltonian $h_0$, no
nodal points exist in that plane. Consequently, the number of
nodal points in the 2D BZ ($5$ in this example) has the same
parity as $C_2$ in the $(4+1)$-d system.} \label{fig:surface4dT}
\end{figure}

Once the $Z_2$ classification is obtained, the physical
consequences of this topological quantum number can be studied by
the effective theory (\ref{Seff3dP3}), as has been done in the
last subsection. In the $(1+1)$-d case, we have shown that a
zero-energy localized state exists at each open boundary of a
$Z_2$ nontrivial particle-hole symmetric insulator, which leads to
a half charge $Q_d=e/2(~{\rm mod}~e)$ confined on the boundary.
Similarly, the nontrivial $(3+1)$-d insulators also have
topologically protected surface states. The easiest way to study
the surface physics of the $(3+1)$-d insulator is again by
dimensional reduction. As discussed above, for any
three-dimensional Hamiltonian $h_1(\vec{k})$, an interpolation
$h(\vec{k},\theta)$ can be defined between $h_1$ and the ``vacuum
Hamiltonian" $h_0$. If we interpret $\theta$ as the fourth
momentum, $h(\vec{k},\theta)$ defines a $(4+1)$-d band insulator.
Moreover, the constraint Eq. (\ref{Tparametrization}) on
$h(\vec{k},\theta)$ requires time-reversal symmetry for the
corresponding $(4+1)$-d system. The Hamiltonian
$h(\vec{k},\theta)$ can be written in a real space form and then
defined on a four-dimensional lattice with open boundary
conditions in the $z$-direction and periodic boundary conditions
for all the other directions. As discussed in Sec.
\ref{sec:Dirac5d}, there will be $|C_2[h]|$ flavors of $(3+1)$-d
chiral fermions on the surface when the second Chern number
$C_2[h]$ is nonzero. In other words, in the 3D BZ of the surface
states there are $|C_2[h]|$ nodal points
$\left(k_{xn},k_{yn},\theta_n\right),~n=1,..,|C_2[h]|$ where the
energy spectrum $E_n(k_x,k_y,\theta)$ is gapless and disperses
linearly as a Dirac cone. From time-reversal symmetry it is easy
to prove that the energy spectrum is identical for
$(k_x,k_y,\theta)$ and $(-k_x,-k_y,-\theta)$. Consequently, if
$\left(k_{x},k_{y},\theta\right)$ is a nodal point, so is
$\left(-k_{x},-k_{y},-\theta\right)$. In other words,
time-reversal symmetry requires the chiral fermions to appear in
pairs, except for the ones at time-reversal symmetric points, as
shown in Fig. \ref{fig:surface4dT} (a). Thus, when the second
Chern number $C_2[h]$ is odd, there must be an odd number of Dirac
cones at the $8$ symmetric points in the 3D BZ. Actually, the
$(4+1)$-d lattice Dirac model (\ref{Dirac5dk}) provides an example
of TRI insulators with nontrivial second Chern number, since one
can define $\Gamma^0$ to be time-reversal even and
$\Gamma^{1,2,3,4}$ to be odd, as in conventional relativistic
quantum mechanics\cite{footnoteGamma2}. As shown in Fig.
\ref{fig:surface4d}, all the nodal points of the surface states
are located at the symmetric points $\Gamma$, $M$, $R$ or $X$.

 Now we return to the surface of $(3+1)$-d insulator. Since
$h_1({\vec k})=h(\vec{k},0),~h_0=h(\vec{k},\pi)$ by definition of
the interpolation, the surface energy spectra of $h_1$ and $h_0$
are given by the $\theta=0$ and $\theta=\pi$ slices of the 3D
surface spectrum. Since all $8$ time-reversal symmetric points
($\Gamma$, $X$ and $M$) are at $\theta=0$ or $\theta=\pi$, we know
that the net number of Dirac cones on the surface energy spectrum
of $h_1$ and $h_0$ is odd (even) when $C_2[h(\vec{k},\theta)]$ is
odd (even). However, $h_0$ is defined as the vacuum Hamiltonian,
which is totally local without any hopping between different
sites. Thus, there cannot be any mid-gap surface states for $h_0$.
Consequently, the number of 2D Dirac cones in the surface state
spectrum of $h_1$ must be odd (even) when $C_2[h]$ is odd (even).
Since the parity of $C_2[h]$ determines the $Z_2$ invariant
$N_3[h_1]$, we finally reach the conclusion that there must be an
odd (even) number of $(2+1)$-d gapless Dirac fermions confined on
the surface of a $(3+1)$-d nontrivial (trivial) topological
insulator.

Compared to earlier works on $Z_2$ invariants and surface states
in $(3+1)$-d, one can see that the $Z_2$ nontrivial topological
insulator defined here corresponds to the ``strong topological
insulator" of Ref. \onlinecite{fu2007b}. The present approach has
the advantage of (i) demonstrating the bulk-edge relationship more
explicitly, (ii) clarifying the connection between the second
Chern number and the $Z_2$ topological number and (iii) naturally
providing the effective theory that describes the physically
measurable topological response properties of the system. The
``weak topological insulators" defined in Ref.
\onlinecite{fu2007b} are not included in the present approach,
since these $Z_2$ invariants actually correspond to topological
properties of $(2+1)$-d insulators (QSH insulators, as will be
discussed in next section), just as the QH effect in $(3+1)$-d
systems\cite{halperin1987} still corresponds to a first Chern
number, but defined in a 2D projection of the 3D BZ.

\subsection{Physical properties of $Z_2$-nontrivial
insulators}\label{sec:3dphysT} In the last subsection we have
defined the $Z_2$ topological quantum number for the $(3+1)$-d TRI
insulators, and discussed the gapless Dirac fermions on the
surface of a non-trivial insulator. Now we will study the physical
response properties of the non-trivial insulators. Since the
non-trivial insulator has a magneto-electric polarization
$P_3=1/2{~\rm mod~}1$, according to Eq. (\ref{Seff3dP3}) the
effective action of the bulk system should be
\begin{eqnarray}
S_{\rm 3D}=\frac{2n+1}{8\pi}\int
d^3xdt\epsilon^{\mu\nu\sigma\tau}\partial_\mu A_\nu
\partial_\sigma A_\tau .\label{Seff3dTI}
\end{eqnarray}
in which $n=P_3-1/2\in\mathbb{Z}$ is the integer part of $P_3$.
Under time-reversal symmetry, the term
$\epsilon^{\mu\nu\sigma\tau}\partial_\mu A_\nu\partial_\sigma
A_\tau=2{\bf E\cdot B}$ is odd, so that for general $P_3$, the
effective action (\ref{Seff3dP3}) breaks time-reversal symmetry.
However, when the space-time manifold is closed ({\em i.e.}, with
periodic boundary conditions in the spatial and temporal
dimensions), the term $\int
d^3xdt\epsilon^{\mu\nu\sigma\tau}\partial_\mu A_\nu\partial_\sigma
A_\tau$ is quantized to be $8\pi^2m,~m\in\mathbb{Z}$.
Consequently, $S_{\rm 3D}=(2n+1)m\pi$ so that the action
$e^{iS_{\rm 3D}}=e^{im\pi}=(-1)^m$ is time-reversal invariant and
is independent of  $n,$ the integer part of $P_3$. This
time-reversal property of the effective action is consistent with
that of $P_3$ discussed in the last subsection. Thus, in a closed
space-time manifold, the topological action (\ref{Seff3dTI}) is a
consistent effective theory of the $Z_2$-nontrivial TRI
insulators, and the integer part of $P_3$ is not a physical
quantity. However, the case is different when the system has open
boundaries. On a space-time manifold with boundary, the value of
$S_{\rm 3D}$ is not quantized, which thus breaks time-reversal
symmetry even for $P_3=1/2$ or $0$. In this case, the integer part
$n$ of $P_3$ does enter the action $e^{iS_{\rm 3D}}$, and becomes
physical; its value depends on the quantitative details of the
boundary.

To understand the physics in the open boundary system, we study a
semi-infinite nontrivial insulator occupying the space $z\leq 0.$
Since the vacuum, which fills  $z>0,$ is effectively a  trivial
insulator (with an infinitely large gap), the effective action
(\ref{Seff3dP3}) can be written in the whole of $\mathbb{R}^3$ as
\begin{eqnarray}
S_{\rm 3D}=\frac1{4\pi}\int
d^3xdt\epsilon^{\mu\nu\sigma\tau}A_\mu\partial_\nu
P_3\partial_\sigma A_\tau\nonumber
\end{eqnarray}
Since $P_3=1/2~{\rm mod}~1$ for $z<0$ and $P_3=0~{\rm mod}~1$ for
$z>0$, we have $$\partial_zP_3=(n+1/2)\delta(z),$$ where
$n\in\mathbb{Z}$ depends on the non-topological details of the
surface, as will be studied later. In this case, the effective
action is reduced to a $(2+1)$-d Chern-Simons term on the surface:
\begin{eqnarray}
S_{\rm surf}=-\frac{2n+1}{8\pi}\int
dxdydt\epsilon^{3\mu\nu\rho}A_\mu\partial_\nu
A_\rho.\label{surfaceCS}
\end{eqnarray}
This is consistent with the observation in Sec. \ref{sec:phys3d}
that a domain wall of $P_3$ carries a QH effect. The Hall
conductance of such a surface of a $Z_2$ nontrivial insulator is
thus $\sigma_H=(n+1/2)/2\pi$, which is quantized as a half odd
integer times the quanta $e^2/h$. On the other hand, from the
discussion in the last subsection we know that there are always an
odd number of $(2+1)$-d Dirac fermions living on the surface of a
nontrivial insulator. Thus the half QH effect on the surface can
be easily understood by the parity anomaly of massless Dirac
fermions\cite{redlich1984}. Here we need to be more careful. The
Hall conductance carried by a Dirac fermion is well-defined only
when the fermion mass is non-vanishing, so that a gap is opened.
With the continuum Hamiltonian
$H=k_x\sigma^x+k_y\sigma^y+m\sigma^z$, the Berry phase curvature
can be calculated as in Eq. (\ref{winding2band}). The ${\bf d}$
vector is given by ${\bf d}=(k_x,k_y,m)$, which has a meron-type
configuration in  $k_x,k_y$ space, and thus carries a Hall
conductance\cite{redlich1984}  $$\sigma_H=\frac1{4\pi}{\rm
sgn}(m)\; (=\frac{e^2}{2h}{\rm sgn}(m)).$$ Now consider the
surface of a topological insulator with $2n+1$  gapless Dirac
cones. From the discussion in the last subsection we know that
without breaking time-reversal symmetry, at least one of these
Dirac cones cannot be gapped. Now consider a perturbation that
breaks time-reversal symmetry, \emph{i.e.} a term which assigns a
mass $m_i, i=1,2,..2n+1$ to each Dirac cone and induces a net Hall
conductance $\sigma_H=\sum_{i=1}^{2n+1}{\rm sgn}(m_i)/4\pi$. Since
$\sum_{i=1}^{2n+1}{\rm sgn}(m_i)$ is an odd integer, the Hall
conductance we obtain is consistent with the surface Chern-Simons
theory (\ref{surfaceCS}). From this discussion we can understand
that the effective action (\ref{surfaceCS}) describes a surface
with time-reversal symmetry breaking, though the bulk system
remains time-reversal invariant. The bulk topology requires there
to be a $1/2$ quanta in the Hall conductance, and the surface
time-reversal symmetry breaking term determines the integer part
$n$. This is an exact analog of the half charge on an end of the
$(1+1)$-d particle-hole symmetric insulator. As shown in Fig.
\ref{Cedgestate}, whether the localized state on the end of a
nontrivial insulator is filled or vacant can only be determined by
choosing a chemical potential $\mu_1>0$ or $\mu_2<0$, or
equivalently, by breaking the particle-hole symmetry around the
boundary. The charge localized at one end of the insulator chain
is $(n+1/2)e$, in which the integer part $n$ depends on the
symmetry breaking term on the surface, but the $1/2$ part is
guaranteed by the bulk topology.

\begin{figure}[tbp]
\includegraphics[width=3in]{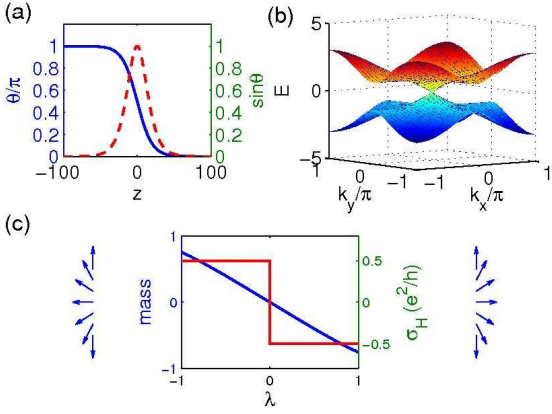}
\caption{(a) Domain wall configuration $\theta(z)$ (blue line) and
corresponding $\sin\theta(z)$ (red dashed line) as defined by Eq.
(\ref{thetadomain}) with $\xi=20$. (b) Dispersion relation of the
two-dimensional bands trapped on the domain wall for the TRI
Hamiltonian $h_0$. (c) The mass of surface Dirac fermion (blue
line) and the half-quantized Hall conductance carried by the
domain wall (red line) in the Hamiltonian $h(\lambda)=h_0+\lambda
h_1$ as a function of $\lambda$. The arrows on the left and right
side show schematically the rotation of angle $\theta$ across the
domain wall for $\lambda<0$ and $\lambda>0$, respectively.}
\label{fig:domainDirac}
\end{figure}

To show such a surface state picture more explicitly, again the
lattice Dirac model can be taken as an example. Consider the
$(3+1)$-d lattice Dirac model (\ref{Dirac3d}) with a domain wall
configuration of the $\theta(\vec{x})$ field given by
\begin{eqnarray}
\theta(\vec{x})=\theta(z)=\frac{\pi}2\left[1-\tanh(z/4\xi)\right],\label{thetadomain}
\end{eqnarray}
which has the asymptotic behavior $\theta(z\rightarrow
-\infty)=\pi$, $\theta(z\rightarrow +\infty)=0$. The domain wall
width is $\xi$, as shown in Fig. \ref{fig:domainDirac} (a). With
periodic boundary conditions in the $x$ and $y$-directions, the
Hamiltonian can be block diagonalized:
\begin{eqnarray}
H&=&\sum_{z,k_x,k_y}\left[\psi_{k_xk_y}^\dagger(z)\left(\frac{c\Gamma^0-i\Gamma^3}2\right)
\psi_{k_xk_y}(z+1)+h.c.\right]\nonumber\\
&
+&\sum_{z,k_x,k_y}\psi_{k_xk_y}^\dagger(z)\left[\left(m+c\cos\theta(z)+c\cos
k_x\right.\right.\nonumber\\
& &\left.+c\cos k_y\right)\Gamma^0 \left.+\sin k_x\Gamma^1+\sin
k_y\Gamma^2\right]\psi_{k_xk_y}(z)\nonumber\\
&+&\sum_{z,k_x,k_y}\psi_{k_xk_y}^\dagger(z)\sin
\theta(z)\Gamma^4\psi_{k_xk_y}(z)\equiv
H_0+H_1.\label{Dirac3dDomainH}
\end{eqnarray}
Under a time-reversal transformation, $\Gamma^{0}$ is even and
$\Gamma^{1,2,3,4}$ are odd. Thus the only time-reversal odd term
in Hamiltonian (\ref{Dirac3dDomainH}) is the last term, which is
localized around the boundary due to the factor $\sin\theta(z)$
(See Fig. (\ref{fig:domainDirac}) (a)). Hereby we denote
$H=H_0+H_1$ with $H_1$ the last term and $H_0$ all the other TRI
terms, and define $h_{0},h_{1}$ as the single particle Hamiltonian
corresponding to $H_0,H_1$, respectively. Then the Hamiltonian
$H_0$ describes a {\em time-reversal invariant} interface between
two insulators $\theta=0$ and $\theta=\pi$. For $-4c<m<-2c$, the
parameterized Hamiltonian $H(\theta),~\theta\in[0,2\pi]$ has a
Chern number $C_2=1$, so that the $\theta=0$ and $\theta=\pi$
system has relative Chern parity $-1$. It's easy to show that
Hamiltonian (\ref{Dirac3d}) for $\theta=0,~-4c<m<-2c$ is
adiabatically connected to the $m\rightarrow -\infty$ limit. Thus
we know that $\theta=0$ and $\theta=\pi$ correspond to $Z_2$
trivial and nontrivial insulators, respectively. Consequently, on
the domain wall at $z=0$ there should be an odd number of gapless
Dirac cones for $H_0$. As shown in Fig. \ref{fig:domainDirac} (b),
numerical diagonalization of $h_0$ shows one single Dirac cone at
$(k_x,k_y)=(0,0)$ on the surface. To understand the effect of
$h_1$ term, notice that $\left\{\Gamma^4,h_0\right\}=0$, with
$\{~\}$ being the anti-commutator. The effective Hamiltonian of
the Dirac cone can always be written as $h_{\rm
surf}=k_x\sigma_x+k_y\sigma_y$ in a proper basis, and it should
also anti-commute with $\Gamma^4$ since the bulk Hamiltonian does.
Since the only term that anti-commutes with $h_{\rm surf}$ in the
$2\times 2$ Hilbert space is $\sigma_z$, we know that the effect
of $\Gamma^4$ term is to induce a mass term $m\sigma_z$ in the
lattice Dirac model. More accurately, the amplitude and the sign
of $m$ can be determined by standard perturbation theory. Given
the two zero-energy surface states
$\left|k=0,\alpha\right\rangle,~\alpha=1,2$, the representation of
the matrices $\sigma_x$ and $\sigma_y$ in the effective theory
$h_{\rm surf}$ can be determined by
$$\sigma^i_{\alpha\beta}=\left\langle
k=0,\alpha\right|\left.\frac{\partial h_0}{\partial
k_i}\right|_{k=0}\left|k=0,\beta\right\rangle,~i=x,y.$$ Then the
$\sigma^z$ is given by $\sigma^z=-i\sigma^x\sigma^y$, so that the
mass $m$ is determined by
$$m=\frac12\sum_{\alpha\beta}\sigma^z_{\alpha\beta}\left\langle
k=0,\beta\right|h_1\left|k=0,\alpha\right\rangle.$$ If we consider
the parameterized Hamiltonian $h=h_0+\lambda h_1$, then the mass
of the surface Dirac fermion is proportional to $\lambda$ for
$\lambda\rightarrow 0$. As shown in Fig. \ref{fig:domainDirac}
(c), the mass is positive for $\lambda<0$, which leads to a
surface Hall conductance $\sigma_H=-{\rm sgn}(\lambda)/4\pi$. On
the other hand, the surface Hall conductance can also be
calculated by the effective theory through Eq. (\ref{sigmaHP3}).
For $\lambda=1$, the phase field $\theta$ winds from $\pi$ to $0$,
which leads to
$\sigma_H=\int_{-\infty}^{+\infty}dP_3(z)/2\pi=\int_\pi^0dP_3(\theta)/2\pi=-1/4\pi$
(since $\int_0^\pi dP_3=C_2/2=1/2$). The Hamiltonian for
$\lambda=-1$ can be considered to be the same lattice Dirac
Hamiltonian $H(\theta)$ with $\theta(z)$ replaced by $-\theta(z).$
This keeps $h_0$ invariant but reverses the sign of $h_1$.
Consequently, the winding of the $\theta$ field is from $-\pi$ to
$0$, which leads to a Hall conductance
$\sigma_H=\int_{-\pi}^0dP_3(\theta)/2=1/4\pi$. The winding of
$\theta$ in the two cases is shown schematically in Fig.
\ref{fig:domainDirac} (c).

In summary, from this example we learn that the effect of a
time-reversal symmetry breaking term on the surface is to assign a
mass to the Dirac fermions which determines  the winding direction
of $P_3$ through the domain wall. Once each Dirac cone on the
surface gains a mass, the whole system is gapped and the Berry
phase curvature is well-defined, so that the winding number of
$P_3$ through the domain wall is determined. Physically, the
time-reversal symmetry breaking term on the surface can come from
magnetic fields or magnetic moments localized on the surface; it
could also arise from the spontaneous breaking of time reversal
symmetry on the surface due to interactions. Once such a
``T-breaking surface field" (denoted by $M$) is applied, the
effective action (\ref{Seff3dP3}) is well-defined for open
boundaries, and describes the electromagnetic response of the
$Z_2$ nontrivial insulator. Actually, the T-breaking field should
be considered to be an external field applied to the TRI system,
such that the topological action (\ref{Seff3dP3}) describes a
nonlinear response of the system to the combination of $M$ and
electromagnetic field $A_\mu$. For a $Z_2$ nontrivial insulator
occupying a spatial region $\mathcal{V}$ with boundary $\partial
\mathcal{V}$, the spatial gradient of $P_3$ is given by
\begin{eqnarray}
\nabla P_3(\vec{x})=\left(g[M(\vec{x})]+\frac12\right)\int_{\partial
\mathcal{V}}d\hat{\bf
n}(\vec{y})\delta^3\left(\vec{x}-\vec{y}\right)
\end{eqnarray}
where $g[M(\vec{x})]\in\mathbb{Z}$ is the integer part of the
winding number determined by the T-breaking field $M(\vec{x})$, and
$\hat{\bf n}$ is the normal vector of the surface. Under such a
configuration of $\nabla P_3$, the effective action (\ref{Seff3dP3})
is reduced to the surface Chern-Simons action
\begin{eqnarray}
S_{\rm
surf}=\frac1{4\pi}\int_{\partial\mathcal{V}}d\hat{n}_\mu\left(g[M(\vec{x})]+\frac12\right)\epsilon^{\mu\nu\sigma\tau}A_\nu\partial_\sigma
A_\tau .\label{surfaceCSgeneral}
\end{eqnarray}
Since $g[M(\vec{x})]$ can only take discrete values, in general
the surface of a nontrivial insulator consists of several domains
with different Hall conductance. To obtain more realistic
predictions of the effective theory (\ref{surfaceCSgeneral}), in
the rest of this subsection we will study a specific case---the
interface between a ferromagnetic insulator and a $Z_2$ nontrivial
insulator, where the surface time-reversal symmetry breaking is
generated by the magnetization of the FM material. Several
specific experimental proposals will be discussed.

{\bf (1) Magnetization-induced QH effect.}

\begin{figure}[tbp]
\includegraphics[width=3.5in]{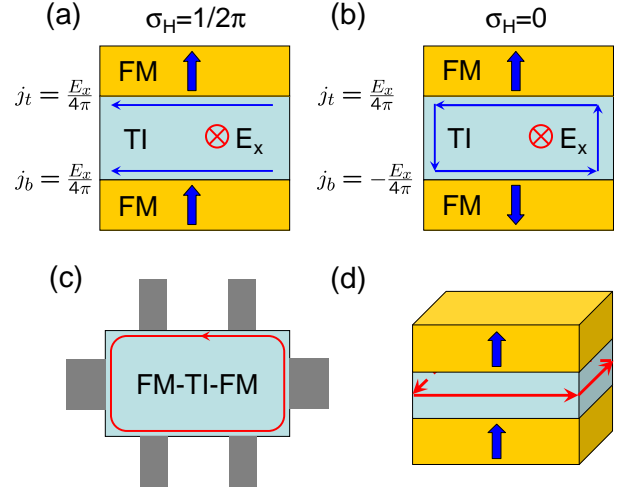}
\caption{Illustration of the QH effect in ferromagnet-topological
insulator heterostructure. (a) and (b) The electric field $E_x$
(with direction into the paper) and the induced Hall current $j_t$
and $j_b$ for parallel and antiparallel magnetization,
respectively. In (b) the Hall current on the two surfaces are
opposite and form a circulating current. (c) A top-down view of
the device for Hall measurement. The grey regions are leads that
contact to the surface of the topological insulator. (d) For the
case of parallel magnetization, the chiral edge states are trapped
on the side surfaces of the topological insulator. These carry the
quantized Hall current.} \label{fig:domainqhe}
\end{figure}

Consider the ferromagnet-topological insulator heterostructure
shown in Fig. \ref{fig:domainqhe}. The magnetization of the two FM
layers can be parallel or antiparallel, and the standard
six-terminal measurement can be performed to measure the in-plane
Hall conductance. The net Hall conductance is given by the
summation of the contributions of the top and bottom surfaces.
When the topological insulator is uniform, an outward pointing
magnetization vector, ({\em i.e.}, towards the direction of the
surface normal vector $\hat{\bf n}$), will have the same effect,
no matter to which surface it is applied. Suppose the Hall current
on the top surface induced by an electric field ${\bf E}=E_x{\bf
\hat{x}}$ is ${\bf j}_t=\hat{\bf n}_t\times {\bf E}/4\pi$, then on
the bottom surface the same formula applies, such that ${\bf
j}_b=\hat{\bf n}_b\times{\bf E}/4\pi$. Since $\hat{\bf
n}_t=-\hat{\bf n}_b=\hat{\bf z}$, the current ${\bf j}_t=-{\bf
j}_b$, as shown in Fig. \ref{fig:domainqhe}. Consequently, the
antiparallel magnetization leads to a vanishing net Hall
conductance, while the parallel magnetization leads to
$\sigma_H=e^2/h$.

Just like the usual integer QH effect, the quantized Hall
conductance here is carried by chiral edge states. To understand
the edge state picture, notice that for parallel magnetizations in
Fig. \ref{fig:domainqhe} (a), the magnetization vector is outward
pointing at the top surface and inward pointing at the bottom
surface. Although the Hall conductance of the two surfaces are the
same in the global $x,y,z$ basis, they are opposite in the {\em
local} basis defined with respect to the normal vector $\hat{\bf
n}$. In other words, the integer $g[M(\vec{x})]$ in the surface
Chern-Simons theory (\ref{surfaceCSgeneral}) is $0$ for the top
surface and $-1$ for the bottom surface. Consequently, the side
surface is a domain wall between two different QH regions with
Hall conductances that differ by one quantum. Just like a domain
wall between $\nu=0$ and $\nu=1$ regions in the usual QH system,
such a domain wall will trap a chiral fermi liquid, which in this
experimental proposal is responsible for the net Hall effect. It
should be noticed that the side surface is 2D, so that generically
there are also other non-chiral propagating modes on the side
surface, besides the one branch of chiral edge states. However,
the existence of these non-chiral states does not change the
stability of the chiral edge state, since there is always one more
right mover than left mover. The stability of the QH effect is
still protected by the ``bulk" gap, which is the
magnetization-induced gap $E_M$ in this case. Thus we will expect
to observe this QH effect under the following two requirements:
(i) temperature $k_BT\ll E_M$; (ii) the chemical potential on the
top and bottom surfaces remains in the gap induced by the applied
magnetization.

We would like to point out that this experimental proposal
provides a direct demonstration of the half QH effect on the
surface of a topological insulator. If the $\sigma_H=e^2/h$
measured for parallel magnetization were contributed by one
surface, then the magnetization flip of the other surface would
have no effect on the net Hall conductance. Thus if an $e^2/h$
Hall conductance is observed for parallel magnetization, and the
magnetization flip of either magnet leads to vanishing Hall
conductance, one can conclude that the Hall conductance is
contributed equally by the two surfaces.

{\bf (2) Topological Magneto-Electric Effect (TME).}

\begin{figure}[tbp]
\includegraphics[width=3.3in]{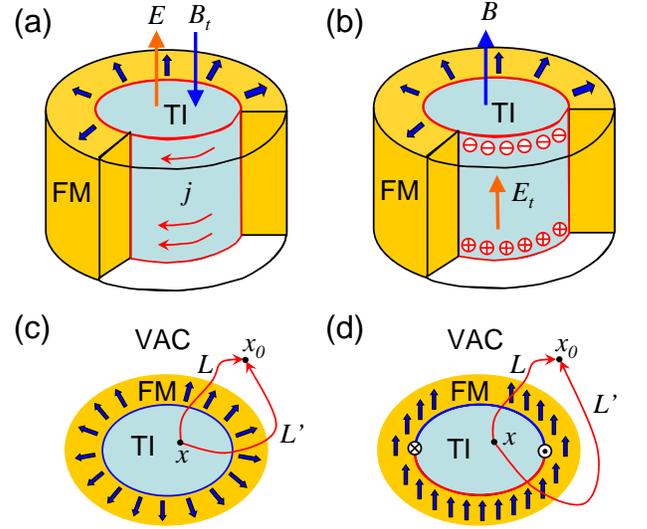}
\caption{(a) Illustration of the magnetic field induced by an
electric field in a cylinder geometry. The magnetization of the FM
layer points outward from the side surface of the TI, and a
circulating current is induced by the electric field. (b)
Illustration of the electric field induced by a perpendicular
magnetic field in the same geometry. ``$\oplus$" and ``$\ominus$"
indicate the positive and negative charge induced by the magnetic
field on the top and bottom surface, respectively. (c)
Illustration of a topological insulator with fully gapped surface
states induced by a hedgehog magnetization configuration. $P_3$ of
the topological insulator is well-defined since the integral along
any two pathes $L$ and $L'$ gives the same $\Delta
P_3=P_3(\vec{x})-P_3(\vec{x}_0)$. (d) Illustration of a
topological insulator with parallel magnetization on the surface.
A one-d domain wall between two-d surface regions with different
Hall conductance (as shown by red and blue) is induced, which
carries chiral edge states as shown by $\otimes$ and $\odot$. In
this case the $P_3$ of the topological insulator cannot be
determined, since two different pathes $L$ and $L'$ lead to
different $\Delta P_3$.} \label{fig:MEeffectTI}
\end{figure}

As has been discussed in Sec. \ref{sec:phys3d}, a TME effect is
induced by $P_3$, which is described by Eqs. (\ref{MEeffect}) and
(\ref{MEeffect2}). Now we consider the realization of this effect
in a nontrivial topological insulator. Similar to the surface QH
effect, a magnetization (or any other time-reversal symmetry
breaking term) is necessary to determine the integer part of
$P_3$. Consider the FM-TI-FM structure in Fig. \ref{fig:domainqhe}
(b). With antiparallel magnetizations, the current induced by an
electric field $E_x$ on the top and bottom surfaces flows in
opposite directions. If we consider an isolated system rather than
the Hall bar with leads as discussed above, a circulating current
is formed, which induces a magnetic field parallel or
anti-parallel to the electric field. However, in the geometry
shown in Fig. (\ref{fig:domainqhe}) (b), dissipation occurs when
the circulating current flows on the gapless side surface and the
adiabatic condition of the TME effect is violated. To obtain the
TME effect, a T-breaking gap for the side surface is necessary.
This is satisfied in the cylindrical geometry shown in Fig.
\ref{fig:MEeffectTI} (a). With a magnetization pointing out of the
cylinder's surface, the surface is gapped and has a fixed Hall
conductance $\sigma_H=(n+\frac12)e^2/h$. In an electric field
parallel to the cylinder as shown in Fig. \ref{fig:MEeffectTI}
(a), a tangential circulating current is induced, with the
strength $j_t=\sigma_HE$. The magnetic field generated by such a
current in the topological insulator can be obtained by solving
Maxwell's equations:
\begin{eqnarray}
{\bf B}_t=-\frac{4\pi}c\sigma_H^t{\bf E}=- (2n+1) \frac{e^2}{\hbar
c}{\bf E}\label{HEresponse}
\end{eqnarray}
in which CGS units are reintroduced. From this formula we can see
that the magnetic field induced by an electric field is proportional
to the electric field, where the response coefficient is quantized
in odd multiples of the fine structure constant. When the
magnetization of the side surface is reversed, the magnetic field
induced is also reversed, as is expected from time-reversal
symmetry. Combined with the conventional, non-topological response,
we obtain ${\bf B}= {\bf H}+4\pi{\bf M} - (2n+1) \frac{e^2}{\hbar
c}{\bf E}$, or
\begin{eqnarray}
{\bf H}= {\bf B}-4\pi{\bf M}+(2n+1) \frac{e^2}{\hbar c}{\bf
E}\label{FullHEresponse}
\end{eqnarray}

Similar induction between electric and magnetic fields also occurs
when a magnetic field is applied. Consider the process of applying
a magnetic field $B$ parallel to the cylinder as shown in Fig.
\ref{fig:MEeffectTI} (b). When the magnetic field is turned on
from zero, a circulating electric field parallel to the side
surface is generated, which then induces a Hall current $j\propto
dB/dt$ parallel or anti-parallel to the magnetic field.
Consequently, a charge density proportional to $B$ is accumulated
on the top and bottom surfaces, so that a magnetic field induces
an electric field parallel to it. By solving Maxwell equations,
such a topological contribution to the electric field is obtained
as
\begin{eqnarray}
{\bf E}_t=(2n+1)\frac{e^2}{\hbar c}{\bf B} . \label{DBresponse}
\end{eqnarray}
Combined with the conventional, non-topological response, we
obtain:
\begin{eqnarray}
{\bf D}= {\bf E}+4\pi {\bf P}- (2n+1) \frac{e^2}{\hbar c}{\bf
B}\label{FullDBresponse}
\end{eqnarray}
The conventional Maxwell's equations, supplemented by the
constituent relations (\ref{FullDBresponse}) and
(\ref{FullHEresponse}) give the complete description of the
electrodynamics of the 3D topological insulators.

An alternative description is to use the conventional constituent
relations ${\bf D}={\bf E}+4\pi{\bf P}$ and ${\bf H}={\bf B}-4\pi
{\bf M}$, and work with a set of Maxwell's equations modified by
the topological term. The total action of the electro-magnetic
field including the topological term is given by (\ref{Seff3dP3}):
\begin{eqnarray}
S_{\rm tot}&=&S_{\rm Maxwell}+S_{\rm topo}\nonumber\\
&=&\int d^3xdt\left[\frac1{16\pi}F_{\mu\nu}F^{\mu\nu}+\frac12F_{\mu\nu}\mathcal{P}^{\mu\nu}-\frac1c j^\mu A_\mu\right]\nonumber\\
& &+\frac{\alpha}{16\pi}\int
d^3xdtP_3\epsilon^{\mu\nu\sigma\tau}F_{\mu\nu}F_{\sigma\tau}\label{Stot3d}
\end{eqnarray}
in which $\alpha\equiv e^2/\hbar c$ is the fine structure
constant, and $\mathcal{P}^{0i}=P^i$ and
$\mathcal{P}^{ij}=\epsilon^{ijk}M_k$ are the electric and magnetic
polarization vectors, respectively. The equations of motion are
obtained by variation of the action over $A^\mu$ as
\begin{eqnarray}
\frac1{4\pi}\partial_\nu F^{\mu\nu}+\partial_\nu
\mathcal{P}^{\mu\nu}+\frac{\alpha}{4\pi}\epsilon^{\mu\nu\sigma\tau}\partial_\nu
(P_3F_{\sigma\tau})=\frac1c j^\mu .
\end{eqnarray}
These equations can also be written in the more familiar component
form as
\begin{eqnarray}
& & \nabla\cdot {\bf D}=4\pi\rho+2\alpha (\nabla P_3\cdot{\bf
B})\nonumber\\
& & \nabla\times{\bf H}-\frac1c\frac{\partial{\bf D}}{\partial t}
=\frac{4\pi}c{\bf j}-2\alpha\left( (\nabla P_3\times {\bf
E})+\frac1c\left(\partial_tP_3\right){\bf B}\right)\nonumber\\
& & \nabla\times {\bf E} + \frac1c\frac{\partial {\bf B}}{\partial
t}= 0\nonumber\\
& & \nabla\cdot{\bf B}=0\label{topoMaxwell}
\end{eqnarray}
where ${\bf D}={\bf E}+4\pi{\bf P}$ and ${\bf H}={\bf B}-4\pi {\bf
M}$ include only the non-topological contributions. These are the
equations of motion of axion
electrodynamics\cite{huang1985,wilczek1987,lee1987}. By shifting
the topological terms to the left-hand side, and redefining ${\bf
D}$ and ${\bf H}$ according to (\ref{FullDBresponse}) and
(\ref{FullHEresponse}), and taking $P_3=n+\frac{1}{2}$, we recover
the conventional Maxwell's equations, but with modified
constituent relations, thus demonstrating the equivalence to the
formulation given above. The quantization of the TME effect in odd
units of the fine structure constant is a deeply profound
quantization phenomenon in condensed matter physics. The flux
quantization inside a superconductor determines the fundamental
constant $hc/e$, while the quantization of the Hall resistance
determines the fundamental constant $h/e^2$. To date, there has
been no other known quantization phenomenon in units of the
dimensionless fine structure constant $\alpha=e^2/\hbar c$.

It should be emphasized that the value of $P_3$ in the topological
insulator can only be determined when a magnetization is applied
to open a gap on the surface. As shown in Fig.
\ref{fig:MEeffectTI} (c), by defining a path $L$ from a reference
point $\vec{x}_0$ deep in the vacuum, $P_3$ can be determined by
$P_3(\vec{x})={\int_{\vec{x}_0}^{\vec{x}}}_L \vec{dl}\cdot\nabla
P_3$. However, this definition only applies when the result does
not depend on the choice of path. If a magnetic ``shell" covered
the surface of the topological insulator, with the magnetization
outgoing everywhere on the interface, then the change of $P_3$
across the interface is the same for different points on the
surface, such that the bulk $P_3$ is well-defined without
dependence on the choice of path. In this case, Eqs.
(\ref{FullHEresponse}) and (\ref{FullDBresponse}) are
well-defined, and the integer part of $P_3$ can change if the
magnetization direction is reversed. On the other hand, when there
are domain walls on the surface, the integer part of $P_3$ is not
well-defined in the bulk of the topological insulator, and Eqs.
(\ref{FullHEresponse}) and (\ref{FullDBresponse}) do not apply, as
shown in Fig. \ref{fig:MEeffectTI} (d). Physically, the failure of
Eqs. (\ref{FullHEresponse}) and (\ref{FullDBresponse}) is simply
due to the existence of a QH edge current on the domain wall,
which requires the more general Maxwell equations Eqs.
(\ref{topoMaxwell}) including the contribution of the current.
This analysis also provides a new picture of the surface QH
effect, that is, the QH effect on the surface is carried by the
chiral edge states living on vortex rings of the $P_3$ field. It
is only when there are no vortex rings of $P_3$ on the surface,
that the surface is fully gapped and the electro-magnetic response
is simply given by Eqs. (\ref{FullHEresponse}) and
(\ref{FullDBresponse}).

{\bf (3) Low-frequency Faraday rotation.}

\begin{figure}[tbp]
\includegraphics[width=3in]{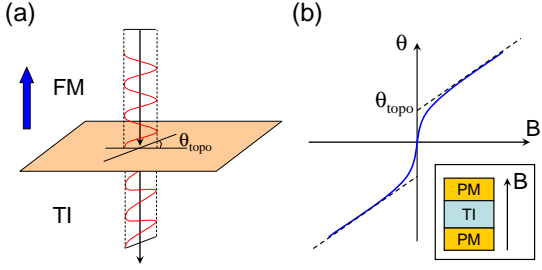}
\caption{(a) Illustration of the Faraday rotation $\theta_{\rm
topo}$ on the interface between a ferromagnet (or equivalently, a
paramagnet in magnetic field) and a topological insulator. (b)
Illustration of the total Faraday rotation angle $\theta$ as a
function of magnetic field $B$ in the sandwich structure as shown
in the inset. The zero-field extrapolation of $\theta(B)$
determines the topological term $\theta_{\rm topo}$.}
\label{fig:Faraday}
\end{figure}

The TME effect can be experimentally observed in the settings
discussed above, by applying an electric field through a
capacitor, and measuring the magnetic field by a SQUID device.
Alternatively, we consider the experiment of Faraday or Kerr
rotation. The modified Maxwell equations (\ref{topoMaxwell}) can
be applied to another phenomenon---photon propagation in the
system\cite{huang1985}. It should be noted that the effective
theory (\ref{Stot3d}) only applies in the low-energy limit $E\ll
E_g$, where $E_g$ is the gap of the surface state. Thus, to detect
the topological phenomena we should study the low frequency
photons with $\omega\ll E_g/\hbar$. Consider a FM-TI interface at
$z=0$, as shown in Fig. \ref{fig:Faraday}. Normally incident,
linearly-polarized light can be written as:
\begin{eqnarray}
{\bf A}(z,t)=\left\{\begin{array}{cc}{\bf a}e^{i(-kz-\omega t)}+{\bf
b}e^{i(kz-\omega t)},&z>0\\{\bf c}e^{i(-k'z-\omega
t)},&z<0\end{array}\right.
\end{eqnarray}
in which $k=\omega/v$ and $k'=\omega/v'$ are the wavevectors of
the photon in the $z>0$ and $z<0$ region, respectively. The
$\nabla P_3$ terms in Eq. (\ref{topoMaxwell}) contribute a
non-conventional boundary condition at $z=0$. Define $\nabla
P_3=\Delta {\bf \hat{z}}\delta(z)$ (with
$\Delta-1/2\in\mathbb{Z}$), the boundary conditions are given by
\begin{eqnarray}
{\bf a}+{\bf b}&=&{\bf c}\nonumber\\
{\bf \hat{z}}\times\left[k\left(-{\bf a}+{\bf b}\right)/\mu+k'{\bf
c}/\mu'\right]&=&-\frac{2\alpha\Delta\omega}c{\bf c}\nonumber
\end{eqnarray}
in which the dimensionless constants $\epsilon,\epsilon'$ and
$\mu,\mu'$ are the permittivity and permeability of the $z>0$ and
$z<0$ materials, respectively. Denote $a_\pm=a_x\pm ia_y$ and the
same for $b_\pm,~c_\pm$, the equations above lead to
\begin{eqnarray}
a_+&=&\frac12\left[1+\frac{k'/\mu'-2i\alpha\Delta\omega/c}{k/\mu}\right]c_+
.\nonumber
\end{eqnarray}
Consequently, when the incident wave ${\bf a}$ is linearly
polarized, the transmission wave ${\bf c}$ is also linearly
polarized, with the polarization plane rotated by an angle
\begin{eqnarray}
\theta_{\rm
topo}=\arctan\frac{2\alpha\Delta}{\sqrt{\epsilon/\mu}+\sqrt{\epsilon'/\mu'}}.\label{topoFaraday}
\end{eqnarray}
 Here we always assume that the
magnetization of the FM material is perpendicular to the $xy$
plane, so that ${\bf H}=\mu{\bf B}$ holds for in-plane magnetic
fields. In the simplest case $\mu,\mu'\simeq
1,~\epsilon,\epsilon'\simeq 1$ and $\Delta=1/2$, we get
$\theta\simeq \alpha\simeq 7.3\times 10^{-3}{\rm rad}$.

Since the ferromagnetic material itself also induces a Faraday
rotation, its necessary to distinguish these two contributions in
order to measure the topological contribution (\ref{topoFaraday}).
Replace the FM layers by paramagnetic materials with large
susceptibilities, and apply an external magnetic field to polarize
them. In this case the magnetization is proportional to magnetic
field, such that the Faraday rotation contributed by the bulk is
also proportional to magnetic field. The net Faraday rotation is
given by $\theta=\theta_{\rm topo}^{(t)}+\theta_{\rm
topo}^{(b)}+\theta_{\rm bulk}$, which has the following dependence
on the magnetic field:
\begin{eqnarray}
\theta(B)&=&uB+2{\rm
sgn}(B)\arctan\frac{\alpha}{\sqrt{\epsilon/\mu}+\sqrt{\epsilon'/\mu'}}.
\end{eqnarray}
Consequently, the topological contribution can be obtained by
measuring $\theta(B)$ at different applied magnetic fields and
extracting the linear extrapolation of $\theta(B)$ as
$B\rightarrow 0^+$.

Experimentally, the main difficulty of measuring such an effect
comes from the low-frequency constraint $\omega\ll E_g/\hbar$. For
a typical value $E_g=10{\rm meV}$ we get $f\equiv\omega/2\pi\ll
2.4{\rm THz}$, which is in the far infrared or microwave region.
In principle, it is possible to find a topological insulator with
a larger gap which can support an accurate measurement of Faraday
rotation. Similar proposals as above can also be worked out for
the rotation of reflected wave (Kerr effect).

\section{Dimensional reduction to
$(2+1)$-d}\label{sec:dimreduction2d}

By carrying out the same dimensional reduction procedure once
more, we can obtain the topological effective theory for TRI
$(2+1)$-d insulators. Additionally a $Z_2$ classification can be
defined for $(2+1)$-d TRI insulators, which is in exact analogy to
the $Z_2$ classification of $(0+1)$-d particle-hole symmetric
insulators. We will show that the $(2+1)$-d $Z_2$ nontrivial phase
corresponds to the QSH insulator proposed recently
\cite{kane2005A,bernevig2006a,bernevig2006d,koenig2007}, and study
the physical consequences of the effective theory.

\subsection{Effective action of $(2+1)$-d insulators}\label{sec:eff2d}

In Sec. \ref{sec:eff3d} we have seen how a $(3+1)$-d insulator
with a parameter field $\theta(\vec{x},t)$ is related to a
$(4+1)$-d insulator through dimensional reduction. In the same
way, two parameter fields can be defined to obtain the dimensional
reduction from $(4+1)$-d to $(2+1)$-d. In the following we will
still take the lattice Dirac model as a canonical example to show
the dimensional reduction procedure and derive the effective
theory.

Starting from the lattice Dirac model (\ref{Dirac5dgauge}) and
choosing a special gauge vector configuration satisfying
$A_{n,n+\hat{i}}=A_{n+\hat{w},n+\hat{w}+\hat{i}}=A_{n+\hat{z},n+\hat{z}+\hat{i}}$,
(so that the gauge vector is homogeneous along $z$ and $w$) we
obtain the Hamiltonian
\begin{widetext}
\begin{eqnarray}
H[A]&=&\sum_{k_z,k_w,{\bf x}}\sum_{s=1,2}\left[\psi^\dagger_{{\bf
x};k_z,k_w}\left(\frac {c\Gamma^0-i\Gamma^s}2\right)e^{iA_{{\bf
x},{\bf x}+\hat{s}}}\psi_{{\bf
x}+\hat{s};k_z,k_w}+h.c.\right]+\sum_{k_z,k_w,{\bf
x}}\sum_{s=1,2}\psi^\dagger_{{\bf
x};k_z,k_w}\nonumber\\
& &\cdot\left[\sin\left(k_z+A_{{\bf
x}3}\right)\Gamma^3+\sin\left(k_w+A_{{\bf
x}4}\right)\Gamma^4+\left(m+c\cos\left(k_z+A_{{\bf x}3}\right)+c\cos
\left(k_w+A_{{\bf x}4}\right)\right)\Gamma^0\right]\psi_{{\bf
x};k_z,k_w}\nonumber
\end{eqnarray}
\end{widetext}
in which ${\bf x}=(x,y)$ is the two-dimensional coordinate. As in
the  $(3+1)$-d case, the gauge fields in the $z$ and $w$
directions can be replaced by parameter fields $(k_z+A_{{\bf
x}3})\rightarrow \theta_{\bf x}$ and $(k_w+A_{{\bf
x}4})\rightarrow \varphi_{\bf x}$, resulting in the parameterized
family of $(2+1)$-d Hamiltonians:
\begin{eqnarray}
H_{\rm 2D}[A,\theta,\varphi]&=&\sum_{{\bf
x},s}\left[\psi^\dagger_{{\bf x}}\left(\frac
{c\Gamma^0-i\Gamma^s}2\right)e^{iA_{\vec{x},\vec{x}+\hat{s}}}\psi_{{\bf x}+\hat{s}}+h.c.\right]\nonumber\\
& &+\sum_{{\bf x},s}\psi^\dagger_{{\bf x}}\left[\sin\theta_{{\bf
x}}\Gamma^3+\sin\varphi_{\bf
x}\Gamma^4\right.\nonumber\\
& &+\left.\left(m+c\cos \theta_{{\bf x}}+c\cos\varphi_{\bf
x}\right)\Gamma^0\right]\psi_{{\bf x}}.\label{Dirac2d}
\end{eqnarray}

By integrating out the fermion fields and expanding the resulting
effective action around
$A_s=0,~\theta=\theta_0,~\varphi=\varphi_0$, the same nonlinear
term shown in the Feynman diagram in Fig. \ref{Feynman4d} leads to
the topological term
\begin{eqnarray}
S_{\rm 2D}=\frac{G_2(\theta_0,\varphi_0)}{2\pi}\int
d^2xdt\epsilon^{\mu\nu\rho}A_\mu\partial_\nu\delta\theta\partial_\rho\delta\varphi\label{Seff2dthetaphi}
\end{eqnarray}
in which the coefficient $G_2(\theta_0,\varphi_0)$ is determined
by the same correlation function as Eq. (\ref{C2green}), but
without the integrations over $k_z,k_w$:
\begin{widetext}
\begin{eqnarray}
G_2(\theta_0,\varphi_0)&=&\frac{2\pi}{3}\int
\frac{d^2kd\omega}{\left(2\pi\right)^3}{\rm
Tr}\epsilon^{\mu\nu\rho}\left[\left(G\frac{\partial
G^{-1}}{\partial q^\mu}\right)\left(G\frac{\partial
G^{-1}}{\partial q^\nu}\right)\left(G\frac{\partial
G^{-1}}{\partial q^\rho}\right)\left(G\frac{\partial
G^{-1}}{\partial \theta_0}\right)\left(G\frac{\partial
G^{-1}}{\partial
\varphi_0}\right)\right]\nonumber\\
&=&\frac1{4\pi}\int d^2k\epsilon^{ij}{\rm
Tr}\left[2f_{i\theta}f_{j\varphi}-f_{ij}f_{\theta\varphi}\right]\label{correlation2d}
\end{eqnarray}
\end{widetext}
in which $\mu,\nu,\rho=0,1,2,\;$ $i,j=1,2,\;$
$q^{\mu}=(\omega,k_x,k_y)$, and the Berry curvature is defined in
the four-dimensional parameter space $(k_x,k_y,\theta,\varphi)$.
The coefficient $G_{2}\left(\theta_0,\varphi_0\right)$ satisfies
the sum rule
\begin{eqnarray}
\int G_2\left(\theta_0,\varphi_0\right)d\theta_0d\varphi_0=2\pi
C_2.\label{sumrule2d}
\end{eqnarray}
To simplify the expression further, the Chern-Simons form
$\mathcal{K}^A$ in Eq. (\ref{CS3form}) can be introduced again.
Here $A$ runs over $k_x,k_y,\theta,\varphi$, and
$G_2(\theta_0,\varphi_0)$ can be written in terms of
$\mathcal{K}^A$ as
$$G_2\left(\theta_0,\varphi_0\right)=-2\pi\int
d^2k\left(\partial_x \mathcal{K}^x+\partial_y
\mathcal{K}^y+\partial_\theta\mathcal{K}^\theta+\partial_\varphi\mathcal{K}^\varphi\right).$$
Similar to the $(3+1)$-d case, the momentum derivative terms
$\partial_{(x,y)}\mathcal{K}^{(x,y)}$ lead to vanishing
contributions if $\mathcal{K}^{(x,y)}$ is single-valued, in which
case $G_2$ can be expressed as
\begin{eqnarray}
G_2\left(\theta_0,\varphi_0\right)=\partial_\theta
\Omega_\varphi-\partial_\varphi\Omega_\theta,\nonumber
\end{eqnarray}
with
\begin{eqnarray}
\Omega_\varphi&=&-2\pi\int
d^2k\mathcal{K}^\theta,~\Omega_\theta=2\pi\int
d^2k\mathcal{K}^\varphi.\label{A2d}
\end{eqnarray}
Notice that
\begin{eqnarray}
-\mathcal{K}^\theta&=&-\frac1{16\pi^2}\epsilon^{ij}{\rm
Tr}\left[\left(f_{ij}-\frac13\left[a_i,a_j\right]\right)\cdot
a_\varphi\right.\nonumber\\
&
&-\left.2\left(f_{i\varphi}-\frac13\left[a_i,a_\varphi\right]\right)\cdot
a_j\right],\nonumber
\end{eqnarray}
and similarly for $\mathcal{K}^\varphi$. We know that the vector
$\Omega=\left(\Omega_{\theta},\Omega_{\varphi}\right)$ has the
correct transformation properties of a gauge vector potential
under the coordinate transformations of the parameter space
$(\theta,\varphi)$, and also under gauge transformations of the
wave functions. Consequently, when the parameters
$\theta=\theta(x^\mu)$ and $\varphi=\varphi(x^\mu)$ have smooth
dependence on space-time coordinates, an effective gauge vector
potential $\Omega_\mu$ can be defined in  $(2+1)$-d
\emph{space-time} as
\begin{eqnarray}
\Omega_\mu\equiv\Omega_\theta\partial_\mu\delta\theta+\Omega_\varphi\partial_\mu\delta\varphi,
\label{pullback}
\end{eqnarray} the gauge curvature of which is
related to $G_2$ as
\begin{eqnarray}
\partial_\mu
\Omega_\nu-\partial_\nu\Omega_\mu&=&\left(\partial_\theta\Omega_\varphi-\partial_\varphi\Omega_\theta\right)
\left(\partial_\mu\delta\theta\partial_\nu\delta\varphi-\partial_\nu\delta\theta\partial_\mu\delta\varphi\right)\nonumber\\
&=&G_2\left(\partial_\mu\delta\theta\partial_\nu\delta\varphi-\partial_\nu\delta\theta\partial_\mu\delta\varphi\right).\label{pullback2}
\end{eqnarray}
Mathematically, $G_2$ is a density of second Chern form in the 2D
parameter space $(\theta,\phi)$, and
$(\theta,\varphi)=(\theta(x^\mu),\varphi(x^\mu))$ defines a smooth
map from the $(2+1)$-d space-time manifold to the 2D parameter
space. The curvature of the gauge potential $\Omega_\mu$ is the
pullback of $G_2$ to  $(2+1)$-d space-time. By making use of Eq.
(\ref{pullback}) the effective action (\ref{Seff2dthetaphi}) can
be rewritten in a parameter-independent form:
\begin{eqnarray}
S_{\rm 2D}=\frac1{2\pi}\int
d^2xdt\epsilon^{\mu\nu\tau}A_\mu\partial_\nu\Omega_{\tau}.\label{Seff2dG2}
\end{eqnarray}

The physical consequences of the effective theory (\ref{Seff2dG2})
can be understood by studying the response equation:
\begin{eqnarray}
j^\mu=\frac1{2\pi}\epsilon^{\mu\nu\rho}\partial_\nu\Omega_{\rho}.\label{response2dQSH}
\end{eqnarray}
As will be shown in the next subsection, Eq. (\ref{response2dQSH})
is the fundamental response equation for the QSH effect, which
takes the form similar to the fundamental response equation for QH
effect (\ref{response2d}), with the replacement of the external
gauge field by a effective Berry's phase gauge field. In this
sense, our formalism provides a unifying theory for both effects.
This type of relationship between different types of topological
insulators will be discussed in more detail in section
\ref{sec:unifiedCS}.

It is worth to noting that the response equation
(\ref{response2dQSH}) can be expressed in an explicit form for the
Dirac model (\ref{Dirac2d}). According to Eq. (\ref{windingS4}),
the momentum-space second Chern number of the $(4+1)$-d Dirac
model $h({\bf k})=\sum_ad_a({\bf k})\Gamma^a$ is equal to the
winding number of $\hat{\bf d}({\bf k})$ on the unit sphere $S^4$.
Correspondingly, the Hamiltonian of the $(2+1)$-d Dirac model
(\ref{Dirac2d}) with constant $\theta$ and $\varphi$ has the form
 $h({\bf k},\theta,\varphi)=\sum_ad_a({\bf
k},\theta,\varphi)\Gamma^a$, so the correlation function $G_2$
defined in Eq. (\ref{correlation2d}) can be obtained as
\begin{eqnarray}
G_2(\theta,\varphi)= \frac3{4\pi}\int
d^2k\epsilon^{abcde}\frac{{d}_a\partial_{k_x}{d}_b\partial_{k_y}{d}_c\partial_\theta{d}_d\partial_\varphi{d}_e}{|{\bf
d}({\bf k},\theta,\varphi)|^5}.\nonumber
\end{eqnarray}
Thus the curvature of effective gauge vector potential $\Omega_\mu$
is expressed as
\begin{eqnarray}
\partial_\mu \Omega_\nu-\partial_\nu\Omega_\mu=
{3\epsilon^{abcde}}\int
\frac{d^2k}{4\pi}\frac{{d}_a\partial_{k_x}{d}_b\partial_{k_y}{d}_c\partial_\mu{d}_d\partial_\nu{d}_e}{|{\bf
d}({\bf k},\theta,\varphi)|^5}.\nonumber\\
\label{G2Dirac}
\end{eqnarray}
Now consider a slightly different version of lattice Dirac model
given by
\begin{eqnarray}
h({\bf k},{\bf n})&=&\sin k_x\Gamma^1+\sin k_y\Gamma^2+\left(\cos
k_x+\cos
k_y-2\right)\Gamma^0\nonumber\\
& &+m\sum_{a=0,3,4}\hat{n}_a\Gamma^{a},\nonumber
\end{eqnarray}
in which $m>0$ and $\hat{\bf n}=(\hat{n}_0,\hat{n}_3,\hat{n}_4)$
is a 3D unit vector. For such a model the ${\bf d}$ vector can be
decomposed as
\begin{eqnarray}
{\bf d}({\bf k},\theta,\varphi)={\bf d}_0({\bf
k})+\left(\begin{array}{c}0\\0\\m\hat{\bf
n}\end{array}\right),\nonumber
\end{eqnarray}
with ${\bf d}_0({\bf k})=(\sin k_x,\sin k_y,0,0,\cos k_x+\cos
k_y-2)$. In the limit $m\ll 2$, the Hamiltonian has the continuum
limit $h({\bf k},{\bf \hat{n}})\simeq
\sum_{a=1,2}k_a\Gamma^a+\sum_{b=0,3,4}m\hat{n}_b\Gamma^b$, which
is the continuum $4\times 4$ Dirac model with three possible mass
terms. In this limit the integral over ${\bf k}$ in Eq.
(\ref{G2Dirac}) can be explicitly carried out, leading to the
following expression:
\begin{eqnarray}
\partial_\mu \Omega_\nu-\partial_\nu\Omega_\mu=\frac12\hat{\bf
n}\cdot\partial_\mu\hat{\bf n}\times \partial_\nu \hat{\bf
n},\nonumber
\end{eqnarray}
which is the {\em skyrmion density} of the unit vector
${\bf\hat{n}}$. Combined with Eq. (\ref{response2dQSH}) we obtain
the response equation for the Dirac model in the continuum limit:
\begin{eqnarray}
j^\mu=\frac1{8\pi}\epsilon^{\mu\nu\tau}\hat{\bf
n}\cdot\partial_\mu\hat{\bf n}\times \partial_\nu \hat{\bf
n}.\label{ResponseSkyrmion}
\end{eqnarray}
Eq. (\ref{ResponseSkyrmion}) describes a topological response
where the charge density and current are equal to the skyrmion
density and current, respectively. Such an equation can be
considered as a $(2+1)$-d version of Goldstone-Wilczek formula
(\ref{GW1d}), which has been studied extensively in the
literature\cite{goldstone1981,jaroszewicz1984,abanov2001,chamon2007,grover2008}.
Thus, through the discussion above we have shown that the
topological response formula(\ref{ResponseSkyrmion}) of $(2+1)$-d
Dirac fermions is a special example of the generic response
equation (\ref{response2dQSH}).

\begin{figure}[tbp]
\includegraphics[width=3.3in]{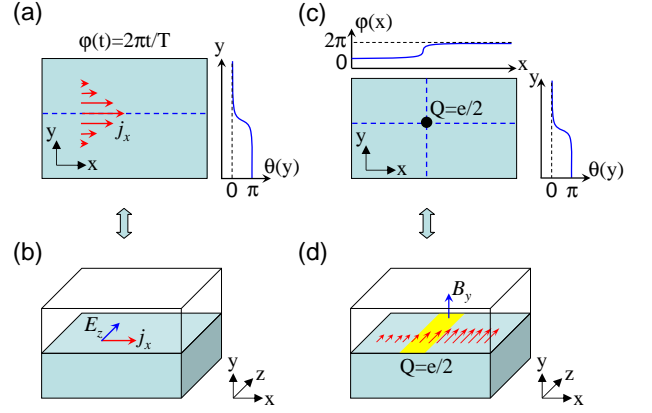}
\caption{(a) Illustration of the charge pumping effect in the
$(2+1)$-d lattice Dirac model with spatial dependent
$\theta=\theta(y)$ and time-dependent $\varphi=\varphi(t)$. This
effect is the dimensional reduction of the domain-wall QH effect
in the $(3+1)$-d system as shown in Fig. (b), in which
$\partial_t\varphi$ plays the role of electric field $E_z$. (c)
Illustration of the half charge trapped at the crossing point of
the $\theta$ and $\varphi$ domain walls. This effect is the
dimensional reduction of the charge trapped on the domain wall by
a magnetic field, as shown in Fig. (d). $\varphi(x)$ corresponds
to the gauge vector potential $A_z$ as shown by red arrows, the
curvature of which leads to a magnetic field $B_y$ in the yellow
region with net flux $2\pi$. Due to the half quantized Hall
conductance of the $\theta$ domain wall, such a magnetic flux
induces a half charge.} \label{fig:domain2dQSH}
\end{figure}

To understand the physics described by  Eq. (\ref{response2dQSH}),
consider the lattice Dirac model in Eq. (\ref{Dirac2d}) with an
adiabatic time-evolution of $\varphi(t)=2\pi t/T$, and a spatial
domain wall configuration of $\theta( \vec{x})$:
$$\theta (y)=\frac{\pi}2\left[1+\tanh\left(\frac{y}\xi\right)\right],$$ as
shown in Fig. \ref{fig:domain2dQSH} (a). According to Eq.
(\ref{response2dQSH}), the charge current along the domain wall is
given by
$$j_x=\frac1{2\pi}\left(\partial_y\Omega_t-\partial_t\Omega_y\right).$$
When the parameter $\varphi$ evolves adiabatically from $0$ to
$2\pi$, the net charge flowing across the line $x=0$ is given by
$\int dtI_x=\int
dtdy(\partial_y\Omega_t-\partial_t\Omega_y)/2\pi=\int_{0}^\pi
d\theta
\int_0^{2\pi}d\varphi(\partial_\theta\Omega_\varphi-\partial_\varphi\Omega_\theta)/2\pi$,
which is the integration of the second Chern form ${\rm
Tr}\left[\epsilon^{ABCD}f_{AB}f_{CD}\right]/32\pi^2$ over the
parameter range $\theta\in[0,\pi],~\varphi\in[0,2\pi]$, where
$A,B,C,D=k_x,k_y,\theta,\varphi.$ According to the discussion in
the $(3+1)$-d case, we know that a magneto-electric polarization
$P_3$ can be defined as
$$P_3(\theta)=\int d^2kd\varphi\mathcal{K}^\theta=-\int d\varphi
\Omega_\varphi/2\pi,$$ which implies $\int dt I_x=-\int_0^\pi
d\theta
\partial_\theta P_3(\theta)$. For $-4c<m<-2c$, we have
$\int_0^\pi dP_3(\theta)=1/2$, corresponding to the pumped charge
$\Delta Q=\int dtI_x=1/2$. In summary, the physical consequence of
the topological response equation (\ref{response2dQSH}) is to
induce a topological charge pumping effect during the adiabatic
evolution of $\varphi$, in which the charge pumped in one period
is proportional to the spatial gradient of the magneto-electric
polarization $P_3$ defined in $(k_x,k_y,\varphi)$ space.
Specifically, a charge $e/2$ is pumped along each
$\Delta\theta=\pi$ domain wall of $\theta$ when $\varphi$ evolves
from $0$ to $2\pi$.

Such a charge pumping effect can also be viewed as the dimensional
reduction of the half QH effect on the $\theta$ domain wall of
$(3+1)$-d lattice Dirac model, which has been studied in Sec.
\ref{sec:phys3d}. This dimensional reduction procedure is in exact
analogy with the usual reduction from the $(2+1)$-d QH effect to
$(1+1)$-d quantized pumping effect studied in Sec.
\ref{sec:dimreduction1d}. Similar to the latter case, a fractional
charge effect can also be proposed in $(2+1)$-d according to Eq.
(\ref{response2dQSH}). To show this effect, one can consider the
same $\theta$ domain wall as shown above, and a $2\pi$ domain wall
of $\varphi$ along the $y$-direction $\varphi({\bf
x})=\pi\left[1+\tanh(x/\xi)\right]$, as shown in Fig.
\ref{fig:domain2dQSH} (b). The charge density is given by
$\rho=\left({\partial_x
\Omega_y-\partial_y\Omega_x}\right)/{2\pi}$. By integrating over
the $x$-direction we obtain $\int \Omega_xdx=\int \Omega_\varphi
d\varphi=-2\pi P_3(\theta)$, such that $\rho_{\rm 1D}=\int
dx\rho=\partial_y P_3(\theta)$ and $\int dy\rho_{\rm 1D}=1/2$.
Thus, a half charge is localized at the crossing of $\theta$ and
$\varphi$ domain walls\cite{qi2007}. Such a fractional charge
effect can also be understood through the dimensional reduction
from $(3+1)$-d. The spatial dependence of $\varphi(x)$ corresponds
to the spatial dependence of $k_z-A_z(x)$, which describes a
magnetic field perpendicular to the 2D domain wall in $(3+1)$-d
system. When $\varphi(x)$ has a $2\pi$ domain wall, the net flux
of the corresponding magnetic field is $2\pi$, which thus induces
a half charge as shown in Fig. \ref{fig:domain2dQSH} (b).

In summary, we have studied the physical consequences of the
topological effective action (\ref{Seff2dG2}) in a spatially
and/or temporally inhomogeneous insulator. In the rest of this
section we will show how to define a $Z_2$ topological invariant
in $(2+1)$-d TRI insulators and study the physical properties of
the $Z_2$ nontrivial phase---QSH phase---by applying the effective
theory (\ref{Seff2dG2}).

\subsection{$Z_2$ topological classification of TRI insulators}\label{sec:2dZ2}

In Sec. \ref{sec:Z21d} and \ref{sec:0dZ2} we have shown how a
$Z_2$ classification of particle-hole invariant insulators can be
defined in both $(1+1)$-d and $(0+1)$-d through dimensional
reduction from the $(2+1)$-d QH effect. The second-Chern-class
analogy of the $(2+1)$-d QH effect is the $(4+1)$-d QH effect
{\cite{zhang2001} described by the Chern-Simons theory
(\ref{Seff4d}), which then leads to the $Z_2$ classification of
TRI insulators in $(3+1)$-d, as shown in Sec. \ref{sec:3dZ2}.
Following this line of reasoning, it is straightforward to see
that a $Z_2$ classification can be defined for $(2+1)$-d TRI
insulators, as an analog of $(0+1)$-d particle-hole symmetric
insulators. In this subsection we will sketch the demonstration of
such a topological classification without going into  detail since
the derivation here is exactly parallel to that in Sec.
\ref{sec:0dZ2}.

First of all, for two TRI $(2+1)$-d insulators $h_1({\bf
k}),~h_2({\bf k})$ an adiabatic interpolation $h({\bf k},\theta)$
can be defined, satisfying
\begin{eqnarray}
h({\bf k},0)&=&h_1,~h({\bf k},\pi)=h_2\nonumber\\
T^\dagger h(-{\bf k},-\theta)T&=&h^T({\bf
k},\theta),\label{InterpQSH}
\end{eqnarray}
Since $h({\bf k},\theta)$ corresponds to the Hamiltonian of a
$(3+1)$-d TRI insulator, a $Z_2$ topological quantity $N_3[h({\bf
k},\theta)]=\pm 1$ can be defined as shown in Sec. \ref{sec:3dZ2}.
The key point to defining a $Z_2$ invariant for the $(2+1)$-d
Hamiltonians $h_1,h_2$ is to demonstrate the independence of
$N_3[h({\bf k},\theta)]$ on the choice of $h({\bf k},\theta)$.
Consider two different parameterizations $h({\bf k},\theta)$ and
$h'({\bf k},\theta)$. An interpolation $g({\bf k},\theta,\varphi)$
can be defined between them  which satisfies
\begin{eqnarray}
g({\bf k},\theta,0)&=&h({\bf k},\theta),~g({\bf
k},\theta,\pi)=h'({\bf
k},\theta)\nonumber\\
g({\bf k},0,\varphi)&=&h_1({\bf k}),~g({\bf k},\pi,\varphi)=h_2({\bf k})\nonumber\\
g^T({\bf k},\theta,\varphi)&=&T^\dagger g(-{\bf
k},-\theta,-\varphi)T .\nonumber
\end{eqnarray}
$g({\bf k},\theta,\varphi)$ corresponds to a $(4+1)$-d insulator
Hamiltonian, for which a second Chern number $C_2[g]$ is defined.
By its definition, the ``second Chern parity" $N_3$ of $h({\bf
k},\theta)$ and $h'({\bf k},\theta)$ satisfies
$N_3[h]N_3[h']=(-1)^{C_2[g]}$. At the same time, $g({\bf
k},\theta,\varphi)$ can also be considered as an interpolation
between $\theta=0$ and $\theta=\pi$ systems, {\em i.e.}, between
$g({\bf k},0,\varphi)\equiv h_1({\bf k})$ and $g({\bf
k},\pi,\varphi)\equiv h_2({\bf k})$. Since $h_{1,2}({\bf k})$ are
both independent of $\varphi$, the $\varphi$-component of the
Berry's phase gauge field vanishes for $g({\bf k},0,\varphi)$ and
$g({\bf k},\pi,\varphi)$. Consequently, it can be shown that
$(-1)^{C_2[g]}=N_3[g({\bf k},0,\varphi)]=N_3[g({\bf
k},\pi,\varphi)]=1$, so that $N_3[h]N_3[h']=1$ for any two
interpolations $h$ and $h'$. Thus, we have shown that the $Z_2$
quantity $N_2[h_1({\bf k}),h_2({\bf k})]\equiv N_3[h({\bf
k},\theta)]$ only depends on the $(2+1)$-d Hamiltonians $h_1$ and
$h_2$. By defining a constant Hamiltonian $h_0({\bf k})=h_0$ as
reference, all $(2+1)$-d TRI insulators are classified by the
value of $N_2[h_0,h({\bf k})]$. An insulator with $N_2[h_0,h]=-1$
cannot be adiabatically deformed to the trivial Hamiltonian $h_0$
without breaking time-reversal symmetry.

In the next subsection, the physical properties of the $Z_2$
non-trivial insulator defined here will be studied. We will see
that the $Z_2$ non-trivial insulator defined here has non-trivial
edge dynamics, and corresponds to the QSH insulator studied in the
literature
\cite{kane2005A,bernevig2006a,bernevig2006d,koenig2007}. Compared
to the former definition of the $Z_2$ topological
classification\cite{kane2005A,kane2005B,moore2007}, our definition
has the advantage of providing a direct relationship between the
topological quantum number and the physical response properties of
the system.

\subsection{Physical properties of the $Z_2$ nontrivial
insulators}
Similar to the $(3+1)$-d case, the topological
properties of a $Z_2$ nontrivial insulator lead to non-trivial
edge state dynamics described by the effective theory
(\ref{Seff2dG2}), or equivalently, the response equation
(\ref{response2dQSH}). The edge of a $Z_2$ nontrivial insulator is
equivalent to a domain wall between a nontrivial insulator and a
trivial insulator (since the vacuum can be considered as a trivial
insulator with a large gap). Thus, in the following we will focus
on the domain wall between a nontrivial system with Hamiltonian
$h_1({\bf k})$ and a trivial system with Hamiltonian $h_0$.

As discussed in the last subsection, an interpolation $h({\bf
k},\theta)$ can be defined between $h_0$ and $h_1$ satisfying
$h({\bf k},0)=h_0,~h({\bf k},\pi)=h_1({\bf k})$ and $T^\dagger
h(-{\bf k},-\theta)T=h^T({\bf k},\theta)$. Since $h_1$ is
nontrivial, $h({\bf k},\theta)$ has to break time-reversal
symmetry for general $\theta$ to adiabatically connect $h_1$ to
$h_0$. Making use of $h({\bf k},\theta)$, two different interfaces
between $h_1$ and $h_0$ can be defined. Consider a spatially
dependent $\theta$ given by
$$
\theta(x,y)=\frac\pi2\left[1-\tanh\left(\frac{y}\xi\right)\right].
$$
Then the spatially dependent Hamiltonian $h({\bf k},\theta(y))$
and $h({\bf k},-\theta(y))$ both describe a spatial domain wall
between $h_1$ (for $y\ll -\xi$) and $h_0$ (for $y\gg \xi$). The
only difference between these two Hamiltonians are the
time-reversal symmetry breaking terms around the interface. Now
consider a more complicated interface, with
$$
h({\bf k},{\bf x})=\left\{\begin{array}{cc}h({\bf
k},\theta(y)),&x<0\\h({\bf k},-\theta(y)),&x>0\end{array}\right.,
$$
as shown in Fig. \ref{fig:QSHfractional} (a). In such a system,
the time-reversal symmetry on the interface is broken in {\em
opposite} ways for $x>0$ and $x<0$, in the sense that $h^T({\bf
k},(x,y))=T^\dagger h(-{\bf k},(-x,y))T$. Now we study the charge
localized around the point $x=0,y=0$. For a loop $C$ enclosing
this point as shown in Fig. \ref{fig:QSHfractional} (a), the
charge in the region $A$ enclosed by $C$ is given by Eq.
(\ref{response2dQSH}) as
\begin{eqnarray}
Q&=&\frac1{2\pi}\int_A
d^2x\left(\partial_x\Omega_y-\partial_y\Omega_x\right)=\frac1{2\pi}\oint_C{\bf
\Omega}\cdot d{\bf l}.\nonumber
\end{eqnarray}
When the size of the loop is large enough compared to the boundary
width $\xi$, such a loop integration is equivalent to an
integration over $\theta$ from $0$ to $2\pi$, which leads to
$Q=\oint \Omega_\theta d\theta/2\pi=P_3[h({\bf k},\theta)]$.
According to the definition of a $Z_2$ non-trivial insulator in
the last section, $P_3[h({\bf k},\theta)]=1/2~{\rm mod}~1$ for any
interpolation $h({\bf k},\theta)$ between $h_0$ and $h_1$.
Consequently, the charge confined on the domain wall is
$Q=(n+1/2)e$ with $n$ an integer depending on the details of the
interface\cite{qi2007}.

\begin{figure}[tbp]
\includegraphics[width=3.5in]{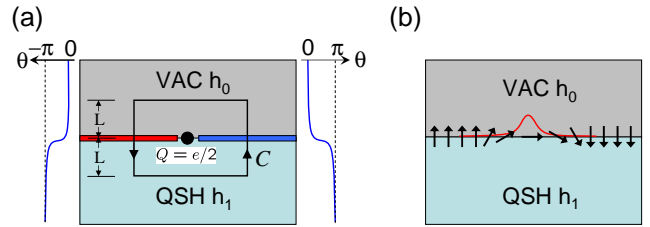}
\caption{(a) Illustration of an interface between the vacuum (VAC)
with Hamiltonian $h_0({\bf k})$ and a QSH insulator (QSH) with
Hamiltonian $h_1({\bf k})$. An interpolation $h({\bf
k},\theta),\theta\in[0,2\pi]$ can be defined between $h_0$ and
$h_1$. On the left (right) half of the interface marked by red
(blue), $\theta$ has a domain wall from $0$ to $-\pi$ ($\pi$). A
fractional charge $Q=e/2$ is trapped on the domain wall between
the red and blue interfaces, which can be calculated by an
 integration of $\Omega$ along the loop $L$ (see text).
(b) Physical realization of the domain wall between two interfaces
in (a) by an anti-phase domain wall of magnetic field or
magnetization. The red curve shows schematically the charge
density distribution.} \label{fig:QSHfractional}
\end{figure}

To summarize, a time-reversal symmetry breaking term can be
applied at the interface of trivial and non-trivial insulators.
For a given interface described by Hamiltonian $h({\bf k},y)$, its
time-reversal partner, $h'({\bf k},y)=Th^T(-{\bf k},y)T^\dagger,$
describes a different connecting condition at the interface. If
the 1D interface is described by $h$ in one region and by $h'$ in
another region, then the domain wall between these two regions
will trap a half-charge as a consequence of the non-trivial
topology. To understand such a domain wall better, we can consider
the case with a magnetic field as the time-reversal symmetry
breaking term on the interface. When the magnetic field has an
anti-phase domain wall as shown in Fig. \ref{fig:QSHfractional}
(b), a half-charge must be trapped on the domain wall.

Moreover, one can also obtain the distribution of 1D charge
density and current density on the interface by integrating the
Eq.  (\ref{response2dQSH}) only along the $y$-direction:
\begin{eqnarray}
\rho_{\rm 1d}(x)&=&\frac1{2\pi}\int_{-L}^L
dy\left(\partial_x\Omega_y-\partial_y\Omega_x\right)\nonumber\\
\j_{\rm 1d}(x)&=&\frac1{2\pi}\int_{-L}^L
dy\left(\partial_y\Omega_t-\partial_t\Omega_y\right)\nonumber
\end{eqnarray}
in which $L$ is a cut-off in the $y$-direction, satisfying $L\gg
\xi$ so that the contribution to $\rho_{\rm 1D}$ and $j_{\rm 1D}$
from the region $|y|>L$ is negligible. According to the
definitions (\ref{A2d}) and (\ref{pullback}) of the effective
gauge vector potential $\Omega_\mu$, we know that
$\Omega_\mu(x,y,t)\rightarrow 0$ for a point deep in the QSH or
VAC region, {\em i.e.}, when $|y|\rightarrow \infty$. Thus the
expression of 1D density and current can be simplified to
\begin{eqnarray}
\rho_{\rm 1D}(x,t)&=&\partial_xP_3(x,t),~j_{\rm
1D}(x,t)=-\partial_tP_3(x,t)\nonumber\\
\label{GWforQSH}
\end{eqnarray}
with $P_3(x,t)=\int_{-L}^Ldy\Omega_y(x,y,t)/2\pi$ the
magneto-electric polarization defined in $(k_x,k_y,y)$ space. Eq.
(\ref{GWforQSH}) is exactly the Goldstone-Wilczek
formula\cite{goldstone1981} describing the charge
fractionalization effect in the $(1+1)$-d Dirac model, and $2\pi
P_3(x,t)$ plays the role of the phase angle of the Dirac mass
term. When the interfaces on the left and right sides of the
domain wall are related by time-reversal symmetry, the change of
$2\pi P_3$ through the domain wall must be $(2n+1)\pi$, which
gives the half charge on the domain wall.

Such a relation between the interface and the $(1+1)$-d Dirac
model can be understood more intuitively in terms of the edge
effective theory\cite{qi2007}. Similar to the relation between the
edge theory of a $(4+1)$-d topological insulator and the reduced
edge theory of a $(3+1)$-d $Z_2$ non-trivial insulator, we can
obtain the edge theory of a $(2+1)$-d $Z_2$ nontrivial insulator
from that of a $(3+1)$-d nontrivial insulator. The interpolation
$h({\bf k},\theta)$ between $h_1$ and $h_0$ can be viewed as the
Hamiltonian of a $(3+1)$-d TRI insulator, in which $\theta$ plays
the role of $k_z$. Consider a specific point of the $(2+1)$-d
boundary of the $(3+1)$-d system, say  $y=0$. According to the
discussion in Sec. \ref{sec:3dZ2}, an odd number of $(2+1)$-d
Dirac fermions are propagating on the boundary of the $(3+1)-d$
system. Due to time-reversal symmetry there must be an odd number
of Dirac cones on the four time-reversal symmetric points, as
shown in Fig. \ref{fig:surface4dT} (b). For a slice at $y=0$, the
wavevector of the surface state is $(k_x,\theta)$. Consequently,
when $\theta$ is considered to be a parameter, the surface energy
spectrum $E(k_x,\theta)$ for a given $\theta$ describes the
dispersion of $(1+1)$-d edge states of $(2+1)$-d insulators.
Specifically, $\theta=0$ corresponds to the vacuum Hamiltonian
$h_0$, which cannot support any non-trivial edge states. Thus, the
Dirac cones at the time-reversal symmetric points can only appear
at $(k_x,\theta)=(0,\pi)$ and $(\pi,\pi)$. To have the minimal
\emph{odd} number of Dirac cones in the $(k_x,\theta)$ BZ, there
must be one gapless Dirac cone at $(0,\pi)$ or $(\pi,\pi)$, but
not both, as shown in Fig. \ref{fig:surface3dT}. In other words,
the edge effective theory of a $(2+1)$-d nontrivial insulator is
given by a gapless Dirac theory
\begin{eqnarray}
H=\int \frac{dk}{2\pi}v\sum_{\sigma=\pm 1}\sigma
k\psi^\dagger_{k\sigma}\psi_{k\sigma}
\end{eqnarray}
where $\sigma=\pm 1$ means  left and right movers, respectively.
This edge theory agrees with the former descriptions of QSH edge
states\cite{kane2005B,wu2006,xu2006} and shows the equivalence of
the $Z_2$ nontrivial insulator defined in this section and the QSH
insulator.

\begin{figure}[tbp]
\includegraphics[width=3.5in]{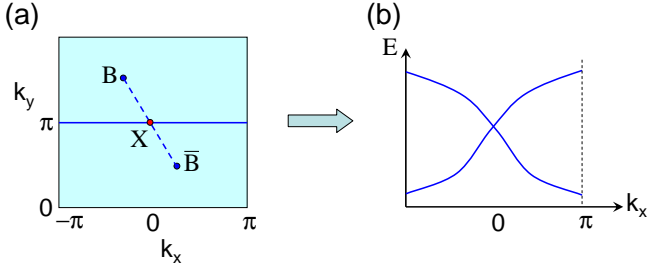}
\caption{Illustration of the dimensional reduction from the
surface of a $(3+1)$-d topological insulator (a) to the edge of a
$(2+1)$-d QSH insulator (b). The red and blue points in (a) are
the positions of gapless $(2+1)$-d Dirac cones in the surface BZ.
The blue line at $k_y=\pi$ defines the edge theory of a QSH
insulator, as shown in Fig. (b).} \label{fig:surface3dT}
\end{figure}

Once the edge theory is obtained, it is easy to understand the
charge fractionalization proposed above. Due to Kramers's
degeneracy, any TRI perturbation cannot open a gap at the edge.
Only when a magnetic field, or other time-reversal symmetry
breaking term is applied, can a mass term
$m_x\sigma_x+m_y\sigma_y$ be generated in the edge theory.
Time-reversal symmetry also guarantees that the mass induced by
opposite magnetic fields is exactly opposite. This implies that an
anti-phase domain wall of the magnetic-field corresponds to a
sign-change of the mass of the Dirac fermion. Thus, the edge state
theory is described by the well-known Jackiw-Rebbi
model\cite{jackiw1976} or equivalently, the Su-Schrieffer-Heeger
model\cite{su1979}. The study of the fractional charge in the edge
theory approach and its experimental consequences have been
presented in Ref. \onlinecite{qi2007}. Thus, we have seen that the
effective theory we obtained from dimensional reduction agrees
with the edge theory analysis, just like in the $(3+1)$-d case.
The effective theory correctly describes the half-charge
associated with a magnetic domain wall, which is a direct physical
manifestation of the $Z_2$ quantum number.

A quantized charge pumping effect always accompanies a fractional
charge effect and can be realized when a \emph{time-dependent}
T-breaking field is applied at the edge. If the system is described
by a time-dependent Hamiltonian $h({\bf k},{\bf x},t)$ which
satisfies $h({\bf k},{\bf x},t=0)=h({\bf k},\theta(y))$ and $h({\bf
k},{\bf x},t=T)=h({\bf k},-\theta(y))$ with $\theta(y)$ the domain
wall configuration discussed above, then the charge pumped through
the interface during the time $t\in[0,T]$ is given by
$$
Q_{\rm pump}=\int_0^T dt j_{\rm
1D}=-\left(P_3(T)-P_3(0)\right)=-\left(n+\frac12\right).
$$
In the example of an applied magnetic field, such a pumping process
implies that a half charge is pumped when a magnetic field rotates
adiabatically from ${\bf B}$ to $-{\bf B}$, as shown in Fig.
\ref{fig:QSHpumping}. The experimental proposal of such a charge
pumping effect is also discussed in Ref. \onlinecite{qi2007}.

\begin{figure}[tbp]
\includegraphics[width=2.5in]{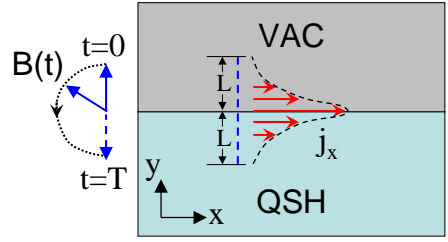}
\caption{Illustration of quantized charge pumping on the boundary of
a QSH insulator induced by a rotating magnetic field. During the
time period $t\in[0,T]$ the magnetic field is rotated from $B$ to
$-B$, and a half charge $Q=\int dt\int_{-L}^{L}dy j_x=e/2$ is pumped
along $x$ direction. $L$ is taken as a cut-off with $j_x$
vanishingly small for $|y|>L$.} \label{fig:QSHpumping}
\end{figure}

Besides providing a quantized response property of the QSH
insulators, the fractional charge and charge pumping effects
proposed here are  a $(1+1)$-d version of electro-magnetic
duality. In $(3+1)$-d, the electro-magnetic duality gives rise to
the Witten effect\cite{witten1979}, where a magnetic monopole
carries a charge $\Theta/2\pi$ and becomes a ``dyon" when a
topological $\Theta$-term is introduced in the
Lagrangian\cite{julia1975,witten1979}. Such an effect can occur in
a $(3+1)$-d topological insulator where the charge of the dyon is
$1/2$, as studied in Sec. \ref{sec:phys3d}. In comparison, the
magnetic domain wall on the boundary of $(2+1)$-d QSH insulator
can be considered as a topological point defect of magnetic field
in $(1+1)$-d, which also carries a half-charge. In this sense, if
we consider the magnetic domain wall as a dynamical degree of
freedom of the system, ({\em e.g.}, when the magnetic domain wall
is generated by a ferromagnetic stripe on top of the 2D QSH
system) it can be considered as the $(1+1)$-d manifestation of
dyons.

Interestingly, such an analogy can also be generalized to
$(2+1)$-d, where the topological defect of magnetic field is a
flux tube. Recently it has been shown that a $\pi$ flux tube
threaded into a QSH insulator carries either charge $\pm e$, spin
$0$ or charge $0$, spin $1/2$, where the spin $0$ ($1/2$) is
generically defined as a Kramers' singlet (doublet) under
time-reversal symmetry.\cite{ran2008,qi2008} In other words, the
$\pi$ flux tube becomes a dyon-like object and realizes
spin-charge separation in $(2+1)$-d. Such a spin-charge separation
phenomenon also provides an alternative definition of the $Z_2$
topological insulators in $(2+1)$-d\cite{qi2008}.

In summary, we have shown how a $Z_2$ classification of $(2+1)$-d
TRI insulators is obtained, and how the physical properties of the
$Z_2$ non-trivial insulator are described by the effective theory
derived from dimensional reduction. Together with the discussion
of $(3+1)$-d topological insulators in Sec.
\ref{sec:dimreduction3d}, we have seen that the nontrivial
topology and its consequences in both $(3+1)$-d and $(2+1)$-d TRI
systems have their origin in the nontrivial second Chern number in
$(4+1)$-d. The dimensional reduction series $(4+1)$-d $\rightarrow
(3+1)$-d $\rightarrow (2+1)$-d is in exact analogy of the lower
dimensional one $(2+1)$-d $\rightarrow (1+1)$-d $\rightarrow
(0+1)$-d. In next section, we will develop the unified framework
of dimensional reduction in generic dimensions, which contains the
two series as simplest examples.

\section{Unified theory of topological insulators}\label{sec:unifiedCS}

\subsection{Phase space Chern-Simons theories}\label{sec:phCS}

Up to now, we have systematically studied several related
topological phenomena, including the $(2+1)$-d QH insulator with
nontrivial first Chern number, the $(4+1)$-d topological insulator
with nontrivial second Chern number, and their dimensional
reductions. Comparing Sec. \ref{sec:firstChern} with Secs.
\ref{sec:secondChern}-\ref{sec:dimreduction2d} one can easily see
the exact analogy between the two series of topological
insulators: the $(2+1)$-d and $(4+1)$-d {\em fundamental}
topological insulators are characterized by an integer---the first
and second Chern number, respectively. Under a discrete symmetry
(particle-hole symmetry for the $(2+1)$-d family and time-reversal
symmetry for the $(4+1)$-d family), a $Z_2$ topological
classification can be defined for the lower dimensional descendent
systems; the physical properties of which can be described by
effective theories obtained from the dimensional reduction
procedure. The main facts about these topological phenomena are
summarized in Table \ref{tab:twofamilies}. In this section, we
will show that the effective theories for all these systems share
a universal form when written in {\em phase space}.
\begin{table*}
 \begin{tabular}[t]{|c|c|c|c|c|c|}
 \hline
 &Dimension&\begin{minipage}[t]{1.5in}Topological quantum number\end{minipage}&Effective theory &
 \begin{minipage}[t]{1in}Symmetry requirement\end{minipage}&Physical
 properties\\
 \hline
 Family 1&2+1&\begin{minipage}[t]{1.5in} $1^{\rm st}$ Chern number $C_1\in\mathbb{Z}$\end{minipage}&
 $\mathcal{L}=\frac{C_1}{2\pi}A_\mu\epsilon^{\mu\nu\rho}\partial_\nu A_\rho$&Not required&\begin{minipage}[t]{1.5in} QH
 effect; axial anomaly on the boundary\end{minipage}\\
 \cline{2-6}
 &1+1&\begin{minipage}[t]{1.5in}$1^{\rm st}$ Chern parity
 $N_1=(-1)^{C_1[h_0,h]}\in Z_2$\end{minipage}&$\mathcal{L}=P_1
 \epsilon^{\mu\nu}\partial_\mu A_\nu$&C&\begin{minipage}[t]{1.5in} Half charge on the boundary\end{minipage}\\
 \cline{2-6}
 &0+1&\begin{minipage}[t]{1.5in}$1^{\rm st}$ Chern parity $N_0=N_1[h_0,h]\in Z_2$\end{minipage}&$\mathcal{L}={\rm{Tr}}[a_0]$&C&Not
 applicable\\
 \hline
 Family 2&4+1&\begin{minipage}[t]{1.5in}$2^{\rm nd}$ Chern number $C_2\in\mathbb{Z}$\end{minipage}&
 $\mathcal{L}=\frac{C_2}{24\pi^2}A_\mu\epsilon^{\mu\nu\rho\sigma\tau}\partial_\nu A_\rho\partial_\sigma A_\tau$&Not required&\begin{minipage}[t]{1.5in}
 4DQH effect; chiral anomaly on the boundary\end{minipage}\\
 \cline{2-6}
 &3+1&\begin{minipage}[t]{1.5in}$2^{\rm nd}$ Chern parity
 $N_3=(-1)^{C_2[h_0,h]}\in Z_2$\end{minipage}&$\mathcal{L}=
 \frac{1}{4\pi}P_3\epsilon^{\mu\nu\sigma\tau}\partial_\mu A_\nu \partial_\sigma A_\tau$&T&\begin{minipage}[t]{1.5in}
 Half QH effect on the boundary, TME effect\end{minipage}\\
 \cline{2-6}
 &2+1&\begin{minipage}[t]{1.5in}$2^{\rm nd}$ Chern parity $N_2=N_3[h_0,h]\in Z_2$\end{minipage}&$\mathcal{L}=
 \frac1{2\pi}A_\mu\epsilon^{\mu\nu\rho}\partial_\nu \Omega_\rho$&T&
 \begin{minipage}[t]{1.5in} QSH effect; half charge at anti-phase
 domain wall on the boundary\end{minipage}
 \\\hline
 \end{tabular}
  \caption{Summary on the properties of the topological insulators. In the
  effective Lagrangians, the indices $\alpha,\beta..=0,1$, $i,j..=0,1,2$,
  $\mu,\nu..=0,1,2,3$ and $a,b..=0,1,..,4$. The vector $a_i$ ($a_\mu$,
  {\em et al.})
  stands for the gauge vector of the external electro-magnetic field, and
  $A_i$ that for the Berry phase gauge field. $C$ and $T$ stands for particle-hole symmetry and time-reversal symmetry, respectively.
  The quantities $P_1, P_3,
  \Omega_\rho$ are defined by Eqs. (\ref{P1}), (\ref{P3}) and
  (\ref{A2d}). See text of Sec. \ref{sec:unifiedCS} for explanations of the effective actions.
  }\label{tab:twofamilies}
\end{table*}

As a simple example, we first consider the effective theory
(\ref{Seff2d}) of the QH effect. Expanding the expression of the
first Chern number explicitly, Eq. (\ref{Seff2d}) can be expressed
as
\begin{eqnarray}
S=\frac1{4\pi}\int \frac{d^2k}{2\pi}\epsilon^{ij}{\rm
Tr}\left[\partial_i a_j\right]\int dt d^2x \epsilon^{\mu\nu\rho}
A_\mu\partial_\nu A_\rho\label{Seff2d2}
\end{eqnarray}
in which $i,j$ are indices $1,2$ in  momentum space and
$\mu,\nu,\rho=0,1,2$ are space-time indices. Here and below,
$A_\mu$ and $a_i$ stand for the external electromagnetic gauge
field in real space and Berry's phase gauge field in momentum
space, respectively. The trace is carried over all occupied energy
levels. If we define the phase-space coordinate as ${\bf
q}=(t,x,y,k_x,k_y)$ and the phase-space gauge potentials ${\bf
A}=(A_0,A_1,A_2,0,0)$,~${\bf a}=(0,0,0,a_1,a_2)$, then the action
above is equivalent to the following second Chern-Simons term:
\begin{eqnarray}
S_{2+1}=\frac1{8\pi^2}\int d^5q\epsilon^{ABCDE}A_A\partial_B
A_C{\rm Tr}\left[\partial_D a_E\right]\label{phcs2d}
\end{eqnarray}\noindent where all capital roman indices
\emph{e.g.} $A,B,C\ldots$ run over the appropriate phase space
coordinates. Since in this system ${\bf A}$ and ${\bf a}$ are always
orthogonal to each other, such a reformulation seems trivial.
However, it turns out to be helpful when considering the dimensional
reduction procedure. As discussed earlier, dimensional reduction of
the $(2+1)$-d system to $(1+1)$-d is defined by replacing, say,
$k_y+A_y$ by a parameter $\theta(x,t)$, which in general is
space-time dependent. Four changes are induced by this substitution:
\begin{enumerate}
\item $A_y$ is replaced by $\theta(x,t)-\theta_0$ with $\theta_0$ playing the role of $k_y$\\
\item
 $\partial/\partial k_y$ is
replaced by $\partial/\partial\theta$\\
\item The Berry's phase gauge field
$a_{k_y}^{\alpha\beta}=-i\left\langle \alpha;{\bf
k}\right|\partial_{k_y}\left|\beta;{\bf k}\right\rangle$ is
replaced by $a_{\theta}^{\alpha\beta}=-i\left\langle
\alpha;k_x,\theta\right|\partial_{\theta}\left|\beta;k_x,\theta\right\rangle$\\
\item The integrations $\int dy$ and $\int dk_y/2\pi$ are removed
from the effective action.
\end{enumerate}
By making these substitutions, the effective action of the
$(1+1)$-d system can be obtained. To help understand the general
dimensional reduction procedure, we show the derivation of the
$(1+1)$-d effective theory explicitly. For simplicity, one can
start from Eq. (\ref{Seff2d2}). Note that the gauge field $A_\mu$
depends only on the $(1+1)$-d coordinates $(t,x)$ after
dimensional reduction.  Consequently, in the Chern-Simons form
$\epsilon^{\mu\nu\tau}A_\mu\partial_\nu A_\tau$ the terms
containing $\partial_y$ are identically zero, so that
$$\epsilon^{\mu\nu\tau}A_\mu\partial_\nu
A_\tau=2\left(A_t\partial_xA_y-A_x\partial_tA_y\right)=
2\left(A_t\partial_x\theta-A_x\partial_t\theta\right)$$ after an
integration by parts. The effective action (\ref{Seff2d2}) after
making all  four substitutions above is expressed as
\begin{eqnarray}
S_{\rm 1+1}&=&\frac1{2\pi}\int dk_x{\rm
Tr}\left[\partial_{k_x}a_\theta-\partial_\theta
a_{k_x}\right]\nonumber\\
& &\cdot \int
dtdx\left(A_t\partial_x\theta-A_x\partial_t\theta\right).\nonumber
\end{eqnarray}

With a smooth space-time dependent $\theta(x,t)$ field, the
eigenstates $\left|\alpha;k_x,\theta\right\rangle$ can be
considered as space-time dependent ``local" eigenstates, whose
space-time dependence originates only from that of $\theta(x,t)$.
In this way, each state
$\left|\alpha;k_x,\theta(x,t)\right\rangle=\left|\alpha;k_x,x,t\right\rangle$
is defined in the full phase space $(t,x,k_x)$, and the Berry's
phase gauge field can gain real-space components defined as
${a}_\mu^{\alpha\beta}=-i\left\langle \alpha
k_x,x,t\right|\partial_\mu\left|\beta,k_x,x,t\right\rangle=a_\theta
\partial_\mu \theta$, in which $\mu=t,x$. Similarly, the space-time derivative of
$a_{k_x}$ is given by $\partial_\mu {a}_{k_x}=\partial_\theta
a_{k_x}\partial_\mu \theta$. By making use of these observations,
the effective action can be simplified to
\begin{eqnarray}
S_{\rm 1+1}&=&\frac1{2\pi}\int dk_x\int
dtdx\left\{A_t{\rm Tr}\left[\partial_{k_x}a_x-\partial_xa_{k_x}\right]\right.\nonumber\\
& &\left.-A_x{\rm
Tr}\left[\partial_{k_x}a_t-\partial_ta_{k_x}\right]\right\}.\label{phcs1dexpand}
\end{eqnarray}
By generalizing the definition of the gauge vector potential
$(A_t,A_x)$ to the phase space vector $(A_t,A_x,0)$, the equation
above can be expressed as the mixed Chern-Simons term in the phase
space:
\begin{eqnarray}
S_{\rm 1+1}=\frac1{2\pi}\int d^3q \epsilon^{ABC}A_A {\rm
Tr}\left[\partial_B a_C\right]\label{phcs1d}
\end{eqnarray}
which describes an inhomogeneous $(1+1)$-d insulator. Note that
momentum derivatives acting on $(A_t,A_x)$ vanish. This effective
action agrees our discussion in Sec. \ref{sec:dimreduction1d} as
one can see by taking $\frac{\delta S_{1+1}}{\delta A_A}$ and
comparing the resulting response equations with Eq.
(\ref{phasespace1d}).

It should be noted that such a phase space formalism is only
applicable when the space-time variation of $\theta$ is smooth and
can be approximated by a constant in the neighborhood of a
space-time point $(t,x,y)$. More quantitatively, the characteristic
frequency $\omega$ and wavevector $k$ of $\theta(x,y,t)$ should
satisfy
\begin{eqnarray}
\hbar\omega,~\hbar vk\ll E_g,\label{adiaCondition}
\end{eqnarray}
where $v$ is a typical velocity scale of the system. For example,
in the lattice Dirac model (\ref{DiracH}), $v$ is the speed of
light (which is normalized to be $1$ in Eq. (\ref{DiracH})). Under
condition (\ref{adiaCondition}) the space-time variation of the
$\theta$ field does not generate excitations across the gap, and
the system can be viewed locally as a band insulator with a
``local Hamiltonian" $h(k_x;x,t)$.

Carrying out such a procedure once more to the action
(\ref{phcs1d}) one can obtain the $(0+1)$-d action. We will show
the derivation explicitly. First one must take
$a_x=0,~\partial_xa_{k_x}=0$ in Eq. (\ref{phcs1dexpand}) since
nothing can depend on the spatial $x$-coordinate after it is
dimensionally reduced. Next we replace $A_x$ by the parameter
$\phi(t)-\phi_0$, which leads to
\begin{eqnarray}
S_{\rm 0+1}&=&-\int dt\left(\phi-\phi_0\right){\rm
Tr}\left[\partial_\phi a_t-\partial_ta_{\phi}\right]\nonumber\\
&=&-\int dt\left(\partial_t\phi
\tilde{a}_\phi+\tilde{a}_t\right).\nonumber
\end{eqnarray}\noindent where the integration $\int dxdk_x/2\pi$ has been removed, and
$a_t,~a_{k_x}$ in $(1+1)$-d are relabelled as
$\tilde{a}_t,~\tilde{a}_\phi$ for later convenience and to obtain
the second equality, an integration by parts is carried out. It
should be noted that $\partial_\phi$ comes from $\partial/\partial
k_y$, which actually means $\partial/\partial\phi_0$ since it is
$\phi_0$ that is replacing $k_y.$ Compared to the $(2+1)$-d
$\rightarrow (1+1)$-d dimensional reduction, the difference here
is that the wavefunctions are, in general, already time-dependent
in $(1+1)$-d. This comes from the dimensional reduction from
$(2+1)$-d. Consequently, the Berry's phase gauge potential in the
$(0+1)$-d system consists of two terms as shown below:
\begin{eqnarray}
{a}_t^{\alpha\beta}&=&-i\left\langle
\alpha;t,\phi(t)\right|\partial_t\left|\beta;t,\phi(t)\right\rangle\nonumber\\
&=& -i\left\langle
\alpha;t,\phi\right|\left[\left(\frac{\partial}{\partial
t}\right)_\phi+\frac{\partial\phi}{\partial
t}\left(\frac{\partial}{\partial\phi}\right)_t\right]\left|\beta;t,\phi\right\rangle\nonumber\\
&=&\tilde{a}_t+\partial_t\phi \tilde{a}_\phi\nonumber
\end{eqnarray}
in which $(\partial/\partial_t)_\phi$ means to take the
$t$-derivative while keeping $\phi$ constant. Both of these terms
are necessary for the correct topological response and similar
terms (including spatially dependent ones) will be present in all
higher dimensions when more than one reduction is carried out.
Combining the two equations above we finally obtain
\begin{eqnarray}
S_{0+1}=\int dt\; {\rm{Tr}}\left[{a}_t \right]
\end{eqnarray}
which has the form of a ``zero-th" Chern-Simons term, and describes
the ``$(0+1)$-d insulator" discussed in Sec. \ref{sec:0dZ2}, {\em
i.e.}, a single-site fermion system. In this case the only gauge
invariant quantity remaining is the Berry's phase the single-site
system obtains during a closed path of adiabatic evolution.

For the second family of topological insulators we discussed, the
effective theory (\ref{Seff4d}) can be expressed in the following
phase space form:
\begin{eqnarray}
S_{4+1}&=&\frac1{192\pi^4}\int
d^9q \epsilon^{ABCDEFGHI}A_A\partial_B A_C\partial_D A_E\nonumber\\
& &\cdot{\rm Tr}\left[D_F a_G D_H a_I\right]\label{phcs4d}
\end{eqnarray}
in which the covariant derivative $D_B=\partial_B+ia_B$ is
introduced for the non-abelian Berry's phase gauge field. The
dimensional reduction to $(3+1)$-d can be performed similarly to
the $(2+1)$-d case. Denoting the $4$-th spatial dimension as $w$,
which is the dimension to be reduced, then any term with
$\partial_w$ vanishes, and so does $a_w$. Consequently, the only
non-vanishing terms in the effective action (\ref{phcs4d}) are
those with $A_w$, which now is replaced by the parameter
$\theta(x,y,z,t)$. On the other hand, one of $F,G,H,I$ in the form
$D_F a_G D_H a_I$ has to be ${k_w}$, which is now replaced by
$\theta$. In summary the theory can be rewritten as
\begin{eqnarray}
S_{3+1}&=&\frac 3{96\pi^3}\int
d^7q \epsilon^{\mu\nu\sigma\tau}\epsilon^{ijk}A_\mu\partial_\nu A_\sigma\partial_\tau \theta\nonumber\\
& &\cdot{\rm Tr}\left[D_{\theta} a_i D_j a_k+{\rm
cycl.}\right]\nonumber
\end{eqnarray}
where $\mu,\nu..=0,1,2,3$ and $i,j,k=1,2,3$ are space-time and
momentum indices of $(3+1)$-d system, and ${\rm cycl.}$ denotes
the other three terms obtained from cyclicly permuting $\theta$
and $i,j,k$. The integration $\int dk_wdw/2\pi$ has been removed,
and a prefactor $3$ appears due to the fact that there are three
$A_A$'s in the effective action (\ref{phcs4d}). In the same way as
in $(2+1)$-d to $(1+1)$-d case, $\partial_\mu \theta D_\theta$ can
be replaced by $D_\mu$, so that the effective theory of the
$(3+1)$-d insulator is finally obtained:
\begin{eqnarray}
S_{3+1}&=&\int \frac{d^7q}{32\pi^3}\epsilon^{AB..G}A_A\partial_B
A_C{\rm Tr}\left[D_D a_E D_F a_G\right].\nonumber\\
\label{phcs3d}
\end{eqnarray}
According to the definition of $P_3$ in Eq. (\ref{P3}) we know that
$$\partial_\ell P_3=\frac1{8\pi^2}\int d^3k\epsilon^{\theta ijk}{\rm
Tr}\left[f_{\ell i}f_{jk}\right],$$ which shows the equivalence of
the action (\ref{phcs3d}) to the action (\ref{Seff3dP3}) we derived
earlier.

Now from the two examples of $(2+1)$-d $\rightarrow (1+1)$-d
$\rightarrow (0+1)$-d and $(4+1)$-d $\rightarrow (3+1)$-d, one can
easily obtain the general rule of dimensional reduction to the
phase-space Chern-Simons theories. For a $(d+1)$ dimensional
system, the phase space dimension is $2d+1$, and the dimensional
reduction of corresponding phase-space Chern-Simons theory is
defined as
\begin{enumerate}
\item Remove a term $\partial_AA_B$ from the action and correspondingly
replace the $(2d+1)$-dimensional anti-symmetric tensor by the one in
$(2d-1)$ dimensions.\\
\item Remove the integration $\int dx_ddk_d/2\pi$ from the action when $x_d,k_d$ are the spatial and momentum indices
to be reduced. \\
\item Multiply the action by a factor $n$ when the power of external gauge field $A_A$ in the original action is $n$.
\end{enumerate}

Following these rules, the effective action for $(2+1)$-d TRI
insulator can be easily obtained by one more step of dimensional
reduction from Eq. (\ref{phcs3d}):
\begin{eqnarray}
S_{2+1}&=&\int \frac{d^5q}{8\pi^2}\epsilon^{ABCDE}A_A{\rm
Tr}\left[D_B a_C D_D a_E\right]\label{phcs2dQSH}
\end{eqnarray}
where the coefficient is determined by $8\pi^2=32\pi^3/(2\cdot
2\pi)$. By considering the space-time and momentum indices
separately, one can easily confirm that Eq. (\ref{phcs2dQSH}) is
equivalent to Eq. (\ref{Seff2dG2}) we obtained earlier.

In summary, we have shown that all the known topological
insulators are described by a Chern-Simons effective theory in
phase space, and the topological theories in different dimensions
can be related by the dimensional reduction procedure. It is
straightforward to generalize this formalism to arbitrary
dimensions. As shown in Appendix \ref{app:C2green}, the relation
between the non-linear response function (\ref{C2green}) and the
corresponding Chern number in momentum space can be generically
proven for any odd space-time dimension. The effective theory of
such a $(2n+1)$-d topological insulator is given
by\cite{golterman1993}
\begin{widetext}
\begin{eqnarray}
S_{2n+1}&=&\frac{C_n}{(n+1)!(2\pi)^n}\int
d^{2n+1}x\epsilon^{\mu_1\mu_2..\mu_{2n+1}}
A_{\mu_1}\partial_{\mu_2}A_{\mu_3}..\partial_{\mu_{2n}}A_{\mu_{2n+1}}
\label{CS2n}
\end{eqnarray}
with the $n$-th Chern number in momentum space defined as
\begin{eqnarray}
C_n&=&\frac1{n!2^n(2\pi)^n}\int {d^{2n}k}\epsilon^{i_1 i_2..
i_{2n}}{\rm Tr}\left[f_{i_1 i_2}f_{i_3 i_4}..f_{i_{2n-1}
i_{2n}}\right].\label{Cn}
\end{eqnarray}
Thus the $(4n+1)$-d phase space formula for this effective action
can be written as
\begin{eqnarray}
S_{2n+1}&=&\frac{1}{n!(n+1)!(2\pi)^{2n}}\int d^{4n+1}q \epsilon^{A_1
A_2...A_{4n+1}}A_{A_1}\partial_{A_2}A_{A_3}..
\partial_{A_{2n}}A_{A_{2n+1}}{\rm Tr}\left[D_{A_{2n+2}}a_{A_{2n+3}}..
D_{A_{4n}}a_{A_{4n+1}}\right]\label{phcs2nd}
\end{eqnarray}

\begin{figure*}[tbp]
\includegraphics[width=6in]{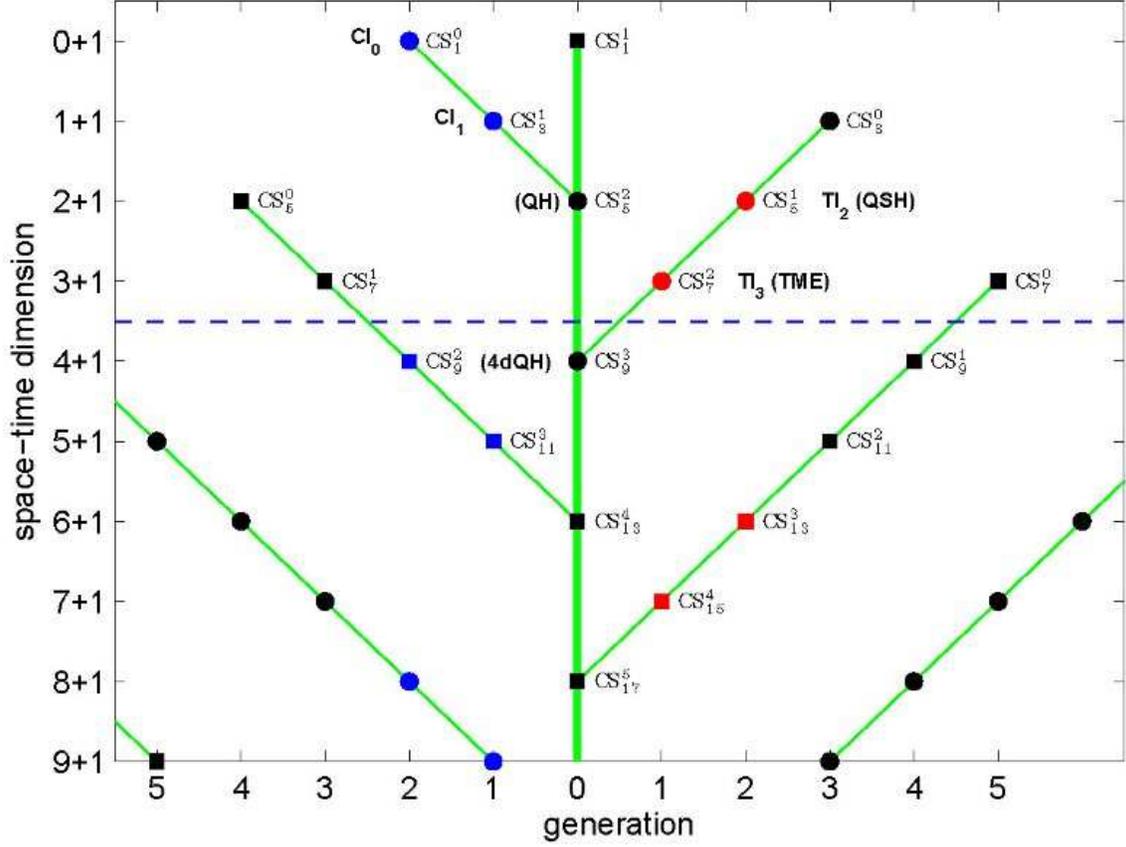}
\caption{The family tree of topological insulators. The black
points on the ``trunk" ({\em i.e.}, $0$-th generation) stand for
the fundamental topological insulators in odd space-time
dimensions characterized by a nontrivial Chern number in momentum
space. The blue and red markers show the descendants of the
$(4n-1)$-d and $(4n+1)$-d insulators, respectively. Physical
effects associated with some of the prominent topological
insulators are indicated in parenthesis. A $Z_2$ classification is
defined for each blue circle (square) under (pseudo) particle-hole
symmetry $C$ ($\tilde{C}$), and for each red circle (square) under
(pseudo) time-reversal symmetry $T$ ($\tilde{T}$). The definitions
of $C,~\tilde{C},~T,~\tilde{T}$ are given in Sec. \ref{sec:Z2nd}.
For all the physically realizable systems with spatial dimensions
$d\leq 3$, the names of topological insulators are labelled, where
${\rm CI_{n}}$ (${\rm TI_{n}}$) stands for a particle-hole
symmetric (TRI) topological insulator in $n+1$ dimension. The
black circles and squares stand for other topological phenomena
obtained from dimensional reduction, which are also described by
the phase space Chern-Simons theories but do not correspond to
$Z_2$ topological insulators. The phase space Chern-Simons theory
${\rm CS}_{2n+1}^t$ (as defined in Eq. (\ref{phcs2n}))
corresponding to each topological phenomenon is also specified on
the figure.
 }\label{fig:familytree}
\end{figure*}

Following the general rules of dimensional reduction procedure
discussed above, one can obtain the effective actions for
lower-dimensional topological insulators as ``descendants" of the
topological theory (\ref{phcs2nd}). From the examples discussed
above it can be seen that the number of Berry's phase gauge vectors
$a_i$ in the effective action remains invariant during dimensional
reduction, while the number of external gauge field insertions
$A_\mu$ decreases by one at each step of dimensional reduction.
After $n+1$ reductions we obtain an effective action of an
$((n-1)+1)$-d system that contains no $A_i$. Just like in the
$(0+1)$-d case, such an effective action does not result in any
response equation of the system, but only describes the Berry's
phase the system obtains during adiabatic evolution. Obviously the
dimensional reduction cannot be carried out again on such an
$((n-1)+1)$-d system. Thus, the $(2n+1)$-d topological insulator
with a nontrivial $n$-th Chern number only has $(n+1)$
``descendants" under dimensional reduction. It is straightforward to
show that the $m$-th descendant ($1\leq m\leq n+1$) of the
$(2n+1)$-d topological insulator has the effective action
\begin{eqnarray}
S_{2n+1-m}^{(m)}&=&\frac{\left(^{2n+1-m}_n\right)}{(2n+1-m)!}\int
\frac{d^{4n+1-2m}q}{(2\pi)^{2n-m}}\epsilon^{A_1 A_2..A_{4n+1-2m}}
A_{A_1}\partial_{A_2}A_{A_3}..
\partial_{A_{2n-2m}}A_{A_{2n+1-2m}}\nonumber\\
&\cdot{\rm Tr}&\left[D_{A_{2n+2-2m}}a_{A_{2n+3-2m}}..
D_{A_{4n-2m}}a_{A_{4n+1-2m}}\right]\nonumber\\
&\equiv &{\rm CS}_{4n-2m+1}^{n-m+1}\label{phcs2n}
\end{eqnarray}
in which ${\rm CS}_s^t$ stands for the mixed Chern-Simons action in
$s$ phase-space dimensions with $t$ powers of the external $A_A$
field. Specifically, ${\rm CS}_s^0$ is a pure non-Abelian
Chern-Simons term of the Berry's phase gauge field $a_A$, which
cannot be reduced to a function of $f_{AB}=D_A a_B-D_B a_A$ alone.
Thus we have seen that a whole family of topological phenomena are
described by phase-space Chern-Simons theories with different $s,t$
values. In a given spatial dimension $d$, all possible topological
phenomena in band insulators are given by the actions ${\rm
CS}_{2d+1}^t$ with all possible values of $t$. It should be noted
that the external gauge field $A_A$ is only defined in real space,
meaning that $A_A$ or $\partial_B A_A$ are both vanishing if $A$ or
$B$ is a momentum index $A,B=d+2,d+3,...,2d+1$. Since in ${\rm
CS}_{2d+1}^t$ there are $t$ $A_A$'s and at least $t-1$ partial
derivative operators acting on $A_A$'s, the Chern-Simons action
${\rm CS}_{2d+1}^t$ vanishes if $2t-1>d+1$. Consequently, there are
in total $[d/2]+2$ available Chern-Simons terms in the phase space
of a $(d+1)$-dimensional system, which are
$${\rm CS}_{2d+1}^t,~t=0,1,...,[d/2]+1.$$ Here $[d/2]$ denotes the
maximal integer that does not exceed $d/2$. For example, in
$(2+1)$-d there are three available phase space Chern-Simons
terms, two of which are ${\rm CS}_{5}^2$ describing a QH
insulator, and ${\rm CS}_{5}^1$ describing a QSH insulator. The
third one is given by
\begin{eqnarray}
{\rm CS}_5^0&=&\frac1{3!(2\pi)^2}\int d^5q\epsilon^{ABCDE}{\rm Tr}\left[a_{A}\partial_B a_C\partial_D a_E\right.\nonumber\\
& &\left.+\frac32a_A a_B a_C\partial_D a_E+\frac35a_A a_B a_C a_D
a_E\right]\nonumber
\end{eqnarray}
which contains no $A_A$ and thus does not describe any
electromagnetic response properties of the system. The information
contained in the effective action ${\rm CS}_5^0$ is the Berry's
phase the system obtains during adiabatic evolution, just like the
effective action of the $(0+1)$-d system ${\rm CS}_1^0=\int dt
a_0$. We have grouped the phase-space Chern-Simons theories  based
on the parent theories and their descendants. The relationships
are summarized in a ``family-tree" in Fig. \ref{fig:familytree}.
Similar to the generalization of the $(2+1)$-d QH insulator to any
odd space-time dimension, the $Z_2$ topological insulators we have
studied can also be generalized to higher dimensions, which will
be explained in the next subsection.

Before moving on to that, we would like to point out an
interesting mathematical fact about the phase-space Chern-Simons
theories. For an $(n+1)$ dimensional system with $N$ occupied
bands, a $U(N)$ gauge vector potential can be defined in phase
space as
$$\mathcal{A}^\lambda_A=\lambda A_A+a_A,$$ with $A_A$ being the
external gauge potential and $a_A$ being the Berry phase gauge
potential. The non-Abelian Chern-Simons term for
$\mathcal{A}^\lambda_A$ can be expressed as
\begin{eqnarray}
{\rm CS}_{2n+1}(\lambda)\equiv{\rm
CS}_{2n+1}\left[\mathcal{A}^\lambda_A\right]=\frac1{(n+1)!(2\pi)^n}\int
d^{2n+1}q\epsilon^{A_1A_2...A_{2n+1}}{\rm
Tr}\left[\mathcal{A}_{A_1}^\lambda\partial_{A_2}\mathcal{A}_{A_3}^\lambda...\partial_{A_{2n}}\mathcal{A}^\lambda_{2n+1}+\text{N.T.}\right].\label{CSgenerator}
\end{eqnarray}
\end{widetext}
Here N.T. stands for the non-Abelian terms containing commutators
of $\mathcal{A}^\lambda_A$, which can be determined by the
relation of the $(2d+1)$-d \emph{Chern-Simons} form to the
$((2d+1)+1)$-d \emph{Chern} form. For more details, {\it c.f.}
Section 11.5 of Ref. \onlinecite{nakahara1990}. By expanding the
parameterized action ${\rm CS}_{2n+1}(\lambda)$ over $\lambda$,
the following equality can be obtained:
\begin{eqnarray}
{\rm CS}_{2n+1}(\lambda)&=&\sum_{t=0}^{[n/2]+1}\lambda^t{\rm
CS}_{2n+1}^t\nonumber\\
\Rightarrow {\rm CS}_{2n+1}^t&=&\frac1{t!}{\left.\frac{\partial^t
}{\partial \lambda^t}{\rm CS}_{2n+1}(\lambda)\right|}_{\lambda=0}.
\end{eqnarray}
This implies that all  possible phase space Chern-Simons terms can
be obtained from a single ``generating functional" ${\rm
CS}_{2n+1}(\lambda)$. We present (\ref{CSgenerator}) as the
unified theory of all topological insulators.

\subsection{$Z_2$ topological insulator in generic
dimensions}\label{sec:Z2nd}

For the descendants of the $(2+1)$-d and $(4+1)$-d insulators, we
have defined a $Z_2$ classification under the constraint of a
discrete symmetry. For the descendants of the $(2+1)$-d QH
insulator, the $Z_2$ classification is defined for particle-hole
symmetric insulators satisfying Eq. (\ref{Ccondition}):
\begin{eqnarray}
C^\dagger h(-{\bf k})C=-h^T({\bf k}),~C^\dagger
C=C^*C=\mathbb{I}.\label{Ccondition2}
\end{eqnarray}
The key point of this classification is to show that an
interpolation between two particle-hole symmetric Hamiltonians
$h_1(k)$ and $h_2(k)$ forms a closed path when combined with its
particle-hole transformed partner. The Chern number enclosed in
such a closed path always has a certain parity which does not
depend on the choice of the path. In the same way, a $Z_2$
classification of particle-hole symmetric insulators is also
defined in $(0+1)$-d. For the family of $(4+1)$-d insulators it is
the same story except that the particle-hole symmetry is replaced
by time-reversal symmetry
\begin{eqnarray}
T^\dagger h(-{\bf k})T=h^T({\bf k}),~T^\dagger
T=-T^*T=\mathbb{I}.\label{Tcondition2}
\end{eqnarray}
Following this one can easily generalize such $Z_2$
classifications to higher dimensions. To do that, one first needs
to understand what is the difference between $(2+1)$ and $(4+1)$
dimensions that requires the choice of different discrete
symmetries. The easiest way to see such a difference is to study
the transformation of the corresponding Chern-Simons theories
under particle-hole symmetry ($C$) and time-reversal symmetry
($T$). Under particle-hole symmetry, the charge density and charge
current both change sign. The vector potential does as well, as
required by the invariance of the minimal coupling $A_\mu j^\mu$.
In the same way one can obtain the time-reversal property of
$A_\mu$, as summarized below:
\begin{eqnarray}
C:A_\mu\rightarrow
-A_\mu,~T:A_\mu\rightarrow\left\{\begin{array}{cc}A_0\\-A_i.\end{array}\right.
\end{eqnarray}
 In both cases of $C$ and $T$, the momentum operator
$-i\partial_\mu$ has the same transformation property as $A_\mu$.
Based on these facts the transformation properties of the
Chern-Simons Lagrangian (\ref{CS2n}) are
\begin{eqnarray}
C: S^{\rm CS}_{2n+1}&\rightarrow& (-1)^{n+1}S^{\rm
CS}_{2n+1}\nonumber\\
T: S^{\rm CS}_{2n+1}&\rightarrow& (-1)^nS^{\rm
CS}_{2n+1}.\label{CTofCS}
\end{eqnarray}
Thus, we see that $S^{\rm CS}_{4n+1}$ is $T$-even but $C$-odd,
while $S^{\rm CS}_{4n-1}$ is $T$-odd but $C$-even. In other words,
a $(4n+1)$-d topological insulator has to break particle-hole
symmetry but can be time-reversal invariant, just like the case of
$(4+1)$-d; a $[(4n-2)+1]$-d topological insulator has to break
time-reversal symmetry but can be particle-hole symmetric, just
like the case of $(2+1)$-d. Consequently, for the descendants of
$(4n+1)$-d ($(4n-1)-d$) topological insulators, it is only
possible to define $Z_2$ topological classifications by the
dimensional reduction procedure under the constraint of $T$ ($C$)
symmetry.

Naively, it seems that the procedure we introduced to define the
$Z_2$ classification by dimensional reduction could be applied
recursively to all the descendants of a $(2n+1)$-d topological
insulator. However, this turns out to be incorrect. As an example,
we can study the $(1+1)$-d TRI insulator as a descendant of the
$(2+1)$-d QSH insulator(not the $(2+1)$-d QH insulator). In the
dimensional reduction from $(3+1)$-d to $(2+1)$-d discussed in
Sec. \ref{sec:2dZ2}, we define an interpolation $h({\bf
k},\theta)$ between two $(2+1)$-d TRI Hamiltonians $h_1({\bf k})$
and $h_2({\bf k})$. When the interpolation $h({\bf k},\theta)$ is
required to satisfy time-reversal symmetry (Eq.
(\ref{InterpQSH})), it corresponds to the Hamiltonian of a
$(3+1)$-d topological insulator, for which a $Z_2$ index
$N_3[h({\bf k},\theta)]$ can be defined. In Sec. \ref{sec:2dZ2} we
have shown that $N_3[h({\bf k},\theta)]$ does not depend on the
choice of the interpolation $h({\bf k},\theta)$, which thus
provides a criteria on whether $h_1({\bf k})$ and $h_2({\bf k})$
are topologically equivalent. If we carry out the same procedure
on $(1+1)$-d TRI insulators, it seems that for two Hamiltonians
$h_1(k)$ and $h_2(k)$ a $Z_2$ topological classification can be
defined in the same way. To see if this is true, one can again
take the lattice Dirac model as an example. The single-particle
Hamiltonian of $(2+1)$-d $4\times 4$ lattice Dirac model with
time-reversal symmetry is written as
\begin{eqnarray}
h_{\rm 2D}({\bf k})&=&\Gamma^1\sin k_x+\Gamma^2\sin
k_y\nonumber\\
& &+\Gamma^0\left[m+c\left(\cos k_x+\cos k_y\right)\right]
\end{eqnarray}
which is in the topological nontrivial phase for $0<m<2|c|$ or
$-2|c|<m<0$. By dimensional reduction, $h_{\rm 2D}({\bf k})$ can
be considered as the interpolation between two $(1+1)$-d TRI
Hamiltonians $h_1(k)=h_{\rm 2D}(k,0)$ and $h_2(k)=h_{\rm
2D}(k,\pi)$. Thus if the $Z_2$ classification procedure applied to
$(1+1)$-d systems, $h_1(k)$ and $h_2(k)$ should be topologically
distinct when $h_{\rm 2D}({\bf k})$ is in the nontrivial phase. In
other words, it should be impossible to define another
interpolation $h_0(k,\theta)$ between $h_1(k)$ and $h_2(k)$, which
satisfies $T^\dagger
h_0(k,\theta)T=h_0^*(-k,-\theta)=h_0^*(-k,\theta),~\forall
\theta$. However, such an interpolation can in fact be constructed
as follows:
\begin{eqnarray}
h_0(k,\theta)&=&\Gamma^1\sin k+\Gamma^{02}\sin^2
\theta\nonumber\\
& &+\Gamma^0\left(m+c\cos k_x+c\cos\theta\right)
\end{eqnarray}
in which $\Gamma^{02}=i\Gamma^0\Gamma^2$ is even under
time-reversal. The existence of two topologically distinct
interpolations $h_0(k,\theta)$ and $h_{\rm 2D}(k,\theta)$ shows
that it is not possible to define a $Z_2$ classification of
$(1+1)$-d TRI insulators in the same way as in $(2+1)$-d and
$(3+1)$-d. The main reason for the failure is that the proof in
Sec. (\ref{sec:2dZ2}) requires  the parameterized Hamiltonian
manifold to be \emph{simply connected.} In other words, an
interpolation $h({\bf k},\theta,\varphi)$ can always be defined
for two interpolations $h({\bf k},\theta)$ and $h'({\bf
k},\theta)$. Similar arguments do not work in the classification
of $(1+1)$-d Hamiltonians because it may not be possible to
adiabatically deform one path to the other. In the example of the
lattice Dirac model, the paths $h_{\rm 2D}$ and $h_0$ cannot be
adiabatically connected, because the combined path
$$g(k,\theta)=\left\{\begin{array}{cc}h_{\rm
2D}(k,\theta),&\theta\in[0,\pi]\\h_0(k,2\pi-\theta),&\theta\in[\pi,2\pi]\end{array}\right.$$
is a $(2+1)$-d Hamiltonian that breaks time-reversal symmetry and
has a nontrivial first Chern number $C_1=-1$. Consequently, the
path $g$ cannot be contracted to a point, and the definition of a
path-independent $Z_2$ invariant fails.

From this example we have seen that the definition of a $Z_2$
topological classification for the descendants of $(2n+1)$-d
topological insulators fails when the dimension is reduced to
$((2n-3)+1)$-d, since the lower Chern number $C_{n-1}$ is defined
for each closed path of $((2n-3)+1)$-d Hamiltonians, thus
obstructing the adiabatic connection between two different paths.
In other words, the $Z_2$ topological insulators as descendants of
$(2n+1)$-d topological insulators can only be defined in
$((2n-1)+1)$ and $((2n-2)+1)$-d . Since a bulk topological
insulator always corresponds to a topologically protected gapless
edge theory, the validity of a $Z_2$ classification can also be
justified by studying the stability of edge theories. As discussed
in Sec.\ref{sec:Dirac5d}, the boundary theory of a $(4+1)$-d
topological insulator with second Chern number $C_2$ contains of
$|C_2|$ flavors of chiral (Weyl) fermions. In the simplest example
of the lattice Dirac model (\ref{Dirac5d}) with $-4c<m<-2c$, the
boundary single-particle Hamiltonian is
\begin{eqnarray}
H_{\partial(4+1)}=v{\bf \vec{\sigma}\cdot\vec{p}}\nonumber
\end{eqnarray}
which is topologically stable since no mass term is available for
the edge system. Under dimensional reduction the boundary theory
of a $(3+1)$-d $Z_2$ nontrivial insulator is simply given by
taking $p_z=0$ in the above equation:
$$H_{\partial(3+1)}=v\left(\sigma_xp_x+\sigma_yp_y\right)$$
which is stable \emph{in the presence of time-reversal symmetry}
since no T-invariant mass terms are available. The same analysis
shows the stability of the edge theory of a $(2+1)$-d topological
insulator, given by $H_{\partial(2+1)}=v\sigma_zp_z$. When
dimensional reduction is carried out once more, we obtain the
$(0+1)$-d edge of the $(1+1)$-d insulator described by
$H_{\partial(1+1)}=0.$ This just describes a Kramers's pair of
localized states on the boundary. Since such a pair of mid-gap
states can be easily removed by a constant energy shift without
breaking time-reversal symmetry, the $(1+1)$-d TRI insulator does
not have a topologically nontrivial class. Different edge state
stabilities for effective theories in different dimensions are
illustrated in Fig. \ref{fig:familytree}.

Such an edge theory analysis can be easily generalized to higher
dimensions. The boundary states of a $(2n+1)$-d topological
insulator with nontrivial Chern number are described by a
$((2n-1)+1)$-d chiral fermion theory with the Hamiltonian
\begin{eqnarray}
H_{2n-1}({\bf p})=v\sum_{i=1}^{2n-1}p_i\Gamma^i\label{chiralnd}
\end{eqnarray}
in which $\Gamma^i$ are $2^{n-1}\times 2^{n-1}$ matrices forming a
representation of the $so(2n-1)$ Clifford algebra. The boundary
theory of the $m$-th descendant of the $(2n+1)$-d system is given by
simply taking $p_i=0$ for $i=2n-m,2n-m+1,...,2n-1$. The symmetry
properties and stability of the theory can be studied by studying
the properties of the $\Gamma^a$ matrices. Here we will display the
conclusions of the edge state analysis, with the details presented
in Appendix \ref{app:edgestab}:
\begin{widetext}
\begin{enumerate}
\item
The chiral Hamiltonians (\ref{chiralnd}) in different dimensions
satisfy different discrete symmetry properties, as listed below:
\begin{eqnarray}
\left\{\begin{array}{ccc}C^\dagger H_{2n-1}({\bf
p})C=-H_{2n-1}(-{\bf
p}),&C^*C=\mathbb{I},&n=4m-3,~m\in\mathbb{N}\\T^\dagger
H_{2n-1}({\bf p})T=H_{2n-1}(-{\bf
p}),&T^*T=-\mathbb{I},&n=4m-2,~m\in\mathbb{N}\\
\tilde{C}^\dagger H_{2n-1}({\bf p})\tilde{C}=-H_{2n-1}(-{\bf
p}),&\tilde{C}^*\tilde{C}=-\mathbb{I},&n=4m-1,~m\in\mathbb{N}\\
\tilde{T}^\dagger H_{2n-1}({\bf p})\tilde{T}=H_{2n-1}(-{\bf
p}),&\tilde{T}^*\tilde{T}=\mathbb{I},&n=4m,~m\in\mathbb{N}\end{array}\right.
\end{eqnarray}
\item Only the first two descendants of $H_{2n-1}$, {\em i.e.},
$H({\bf p})=\sum_{i=1}^{2n-2}p_i\Gamma^i$ and $H({\bf
p})=\sum_{i=1}^{2n-3}p_i\Gamma^i$, are topologically stable under
the constraint of the discrete symmetry in the given dimension
($C,\tilde{C},T$ or $\tilde{T}$). Consequently, the $Z_2$
topologically nontrivial insulators  descending from the
$(2n+1)$-d topological insulator only exist in $((2n-1)+1)$-d and
$((2n-2)+1)$-d.
\end{enumerate}
\end{widetext}
The edge state analysis confirms our insight from the bulk
picture, that is,  the $Z_2$ topological classification is only
well defined for the first two generations of descendants of the
$(2n+1)$-d topological insulator. Moreover, it also provides more
information about the discrete symmetries in different dimensions.
In the $(6+1)$-d topological insulator, and its descendants, the
correct discrete symmetry is $\tilde{C}$ which is similar to
particle-hole symmetry $C$ but satisfies
$\tilde{C}^*\tilde{C}=-\mathbb{I}$. This is necessary since a
usual particle-hole symmetry cannot be defined for the $(5+1)$-d
chiral fermion $\sum_{i=1}^5p_i\Gamma^i$. Such a symmetry
$\tilde{C}$ can be called a ``pseudo particle-hole symmetry".
Similarly, in the $(8+1)$-d topological insulator, and its
descendants, the discrete symmetry is a ``pseudo time-reversal
symmetry" satisfying $T^*T=\mathbb{I}$. In Fig.
\ref{fig:familytree}, the dimensions with ``true" $C$ or $T$
symmetry are labelled with filled circles, and those with
$\tilde{C}$ or $\tilde{T}$ symmetry are labelled with squares.

This paper is partly inspired by work in high-energy physics. The
study of topological insulators with non-trivial boundary states
is analogous to the generation of massless fermions on higher
dimensional domain walls \cite{kaplan1992,creutz2001}. In a high
energy context these surface states would be subsequently used to
mimic chiral fermions in lattice gauge theories. In our case they
become the gapless boundary liquids that generate novel transport
properties  and characterize the topological stability of the
state. Our picture of a $(4+1)$-d topological insulator
characterized by the Chern-Simons term has a very special meaning
in high-energy physics when reduced to $(3+1)$-d. The Chern-Simons
term becomes a
 $\theta$-term which is related to the so-called
vacuum angle. If $\theta$ is a constant then this term does not
contribute to the equations of motion.  In our case the
$\theta$-term has a solitonic structure with a domain wall at the
surface of the topological insulator. Inside the insulator
$\theta$ jumps by $\pi$ which still preserves CP if the original
vacuum does. Thus, the only effect on the system is a non-zero
boundary term at the $\theta$ domain wall. From a high-energy
perspective we have introduced an axionic domain wall and a
topological insulator exists in one domain while a trivial
insulator exists in the other.

The dimensional reduction procedure we introduced here is not new
to physics, and was first used in the 1920's in an attempt to
unify gravity and electro-magnetism in $(4+1)$-d by Kaluza and
Klein\cite{kaluza1921,klein1926}. Basically, the dimensional
reduction amounts to compactifying the ``extra dimension" with
periodic boundary conditions, {\it e.g.} on a circle, and
shrinking its radius to zero. The compactification creates a tower
of modes labelled by a discrete index, but as the circle is shrunk
only one low-energy mode from each field remains (which can be
seen by taking the Fourier transform of a field on a circle with
periodic boundary conditions). These ``zero-modes" become the
propagating fields in the lower dimensional space. Carrying out
this procedure via our method or by compactification yields the
same results. The adiabatic parameter we introduced is connected
with the flux threading the higher dimensional circle. One could
imagine compactifying using higher dimensional manifolds, such as
a sphere or something more exotic, and threading various fluxes
through non-trivial cycles of the compact space. The zero-mode
structures of these manifolds are more complicated and we will not
deal with them here, but perhaps other interesting theories can
arise.

Additionally, we have unearthed a ladder of topological insulators
with gapless fermionic boundary states whose stability depends on
the presence of a discrete symmetry. The discrete symmetries
repeat modulo $8$, which should be no surprise since their form is
derived from the representation theory of real Clifford algebras
which exhibit Bott periodicity with period $8$\cite{lawson1989}.
Due to this periodicity we can analyze the types of gapless
fermions allowed to exist at the boundaries of topological
insulators. Any boundary can support Dirac fermions and any even
dimensional space-time boundary can have Weyl (chiral) fermions as
well. For Majorana fermions we are restricted to boundaries with
spacetime dimensions $\{(1+1),(2+1),(3+1),(7+1),(8+1)\}\;
{\rm{mod}} 8.$ Finally, there is a special representation, the
Majorana-Weyl fermion, which is a real fermion with a definite
handedness which can only exist in $(1+1){~\rm mod~}8$ dimensions
\emph{i.e.} $(1+1),(9+1),(17+1),(25+1)\ldots$\cite{lawson1989}. It
is this type of fermion which appears at the edge of $(p+ip)$
superconductors\cite{read2000,ivanov2001,stone2004}. However,
because this representation is missing in $(5+1)$-d a $(6+1)$-d
``(p+ip)- superconductor" which obeys the ``pseudo-particle hole"
symmetry will not have single Majorana-Weyl boundary states. This
can also be seen by the fact that $\tilde{C}^2=-1$ which means
there would have to be at the minimum two Majorana-Weyl fermions
at the boundary due to a Kramers'-like theorem. Beginning with the
parent $(2n+1)-d$ topological insulator we see that the boundary
theories of itself and its stable descendants are massless
fermions in $((2n-1)+1),((2n-2)+1)$ and $((2n-3)+1)$ dimensions
respectively. Theories of massless fermions often result in field
theory anomalies and there is a deep connection between this
boundary theory ladder and the corresponding anomaly
ladder\cite{callan1985,nakahara1990}.

In summary, we have provided a unified framework to describe a
whole family of topological insulators in generic dimensions. All
the topological effects are described by phase-space Chern-Simons
theories, which either describe the topological insulators with
nontrivial Chern number in odd space-time dimensions, or describe
their lower dimensional descendants through dimensional reduction.
$Z_2$ topologically nontrivial insulators exist in $((2n-1)+1)$
and $((2n-2)+1)$ space-time dimensions and are protected by a
given discrete symmetry that is preserved by the parent $(2n+1)$-d
topological insulator. We found that in $(2n+1)$-d there are two
types of topological insulators, one of which is characterized by
the Chern number $C_n$ and the other by the $Z_2$ invariant, as a
descendant of $((2n+2)+1)$-d topological insulator. In comparison,
in $((2n-1)+1)$-d there is only one type of topological insulator,
which is characterized by a $Z_2$ invariant, as a descendant of
$(2n+1)$-d topological insulator. There are many tantalizing
connections of our work with well-developed sectors of high-energy
physics, and diving deeper into these subjects is sure to benefit
both condensed-matter and high-energy physics.

\section{Conclusion and discussions}

In conclusion we have constructed the topological field theory of
TRI insulators. We showed that the fundamental TRI insulator
naturally exists in $4+1$-d, and the effective topological field
theory is the Chern-Simons theory in $4+1$-d. We introduced the
concept of dimensional reduction for microscopic fermion models,
where some spatial dimensions are compactified and the associated
momentum variables are replaced by adiabatic fields. This method
enables us to obtain the topological field theory for the 3D and
2D TRI insulators from the dimensional reduction of the $4+1$-d
Chern-Simons field theory. In particular, we obtain the ``axion"
field theory for the 3D insulator, and many experimental
consequences follow directly from this topological field theory.
The most striking prediction is the TME effect, where an electric
field induces a magnetic field along the same direction, with a
universal constant of proportionality quantized in odd multiples
of the fine structure constant. The role of the ``axion", or the
adiabatic field, is played by a magneto-electric polarization,
whose change is quantized when an adiabatic process is completed.
The topological field theory for the 2D TRI insulator involves two
adiabatic fields, and this theory directly predicts the fractional
charge of a magnetic domain wall at the edge of the QSH insulator.
These topological effects illustrate the predictive power of the
topological field theory constructed in this work, and many more
experimental consequences can be obtained by the proper
generalization of the concepts introduced here.

Our work also presents the general classification of topological
insulators in various dimensions. The fundamental topological
insulators are described by the topological Chern-Simons field
theory, and the effective topological field theory of their
descendants can be obtained by the procedure of dimensional
reduction. The descendent topological insulators are generally
classified by discrete symmetries like the charge conjugation and
the time reversal symmetries. This way, the Chern number
classification of fundamental topological insulators and the $Z_2$
classifications of their descendants is unified. Finally, we
present a framework in terms of the Chern-Simons field theory in
phase space, which gives a unified theory of all topological
insulators and contains all experimentally observable topological
effects.

We would like to thank Dr. B. A. Bernevig for many insightful
discussions on this subject and collaborations at the early stage of
this project. We would like to thank H. D. Chen, S. Kachru, C. X.
Liu, J. Maciejko, M. Mulligan, S. Raghu, S. Shenker and J. Zaanen
for helpful discussions.  This work is supported by the NSF under
grant numbers DMR-0342832 and the US Department of Energy, Office of
Basic Energy Sciences under contract DE-AC03-76SF00515.

\appendix

\section{Conventions}\label{app:conventions}
\begin{table*}
 \begin{tabular}[t]{l|r}
 \hline
 symbol&explanation
 \\\hline
 $(n+1)$-d&space-time dimension $(n+1)$\\
 $n$D&spatial dimension $n$\\
 $\mu,\nu,\rho,\sigma,...$&space-time or frequency-momentum indices $0,1,..,n$\\
 $i,j,k,l,...$&spatial or momentum indices $1,2,..,n$\\
 $A,B,C,D,...$&phase space indices $0,1,...n,n+1,..2n+1$\\
 $a,b,c,d,...$&indices for anticommuting $\Gamma^a$ matrices and
 corresponding coefficients $d_a$\\
 $\alpha,\beta,\gamma,\delta...$&energy band indices\\
 $A_\mu,~F_{\mu\nu}$&gauge vector potential and gauge curvature of
 external electro-magnetic field \\
 $a_i,~f_{ij}$&(generally non-Abelian) Berry phase gauge
 vector potential and gauge curvature in momentum space\\
 $A_A,~a_A$&electro-magnetic or Berry phase gauge vector potential
 defined in phase space, respectively\\
 \hline
 \end{tabular}
  \caption{Table of conventions used in the paper.}\label{tab:conventions}
\end{table*}

Due to the special importance of dimensionality in this work we
tried to be very consistent with our dimension and index
conventions. In addition, gauge fields of all types appear and we
have selected a convention for the electro-magnetic vector
potential and the adiabatic (Berry's phase) connection. The
conventions we used in this paper are summarized in Table
\ref{tab:conventions} and also explained below.

For spacetime conventions we have chosen the form $(n+1)$-d where
$n$ can be $0.$ For spatial dimensions only we use nD with a
capital D.

For indices, space-time and frequency-momentum indices share the
same convention. Greek indices from the middle of the alphabet
such as $\mu,\nu,\rho,\sigma,\tau$ run from $0$ to the spatial
dimension in the current context. Examples being $0,1$, $0,1,2$
\emph{etc.} Additionally, when indexing momentum space objects $0$
is frequency and $1,2,3\ldots$ are the momentum components in the
$1,2,3,\ldots$ directions. Latin indices from the middle of the
alphabet such as $i,j,k,\ell$ are purely spatial indices and run
from $1$ onward to the spatial dimension of the current context.
For momentum space objects they index the spatial momenta
\emph{e.g} $k_x,k_y,k_z,k_w\ldots.$ We always use the Einstein
summation convention unless stated otherwise. Since we are
considering flat space we make no distinction between raised and
lowered indices.

In some special contexts we will need two more sets of indices.
Several of the Hamiltonian models we use can be written in terms
of the  $2\times 2$ or $4\times 4$ Dirac matrices. In these cases
where you see vectors $d^a$ indexed by lowercase Latin letters
from the beginning of the alphabet they run from $1,2,3$ and
$0,1,2,3,4$ respectively. Finally, for cases where the indices
don't just run over coordinate or momentum space separately, but
instead cover ``phase-space" coordinates, we use capital Latin
letters from the beginning of the alphabet such as $A,B,C,D,E.$
These run over the phase space variables in the current context.
Some examples being $q^{A}=(t,x,y,k_x,k_y),$ $q^A=(t,x,k_x),$ or A
running over $(k_x,k_y,\theta,\phi).$

For orbital, band, or state labels we use Greek letters from the
beginning of the alphabet such as $\alpha,\beta,\gamma.$

For the electro-magnetic $U(1)$ gauge field we use $A^{\mu}$ where
$A$ is capitalized. For the Berry's phase gauge field we use
$a^{i}$ with a lower case $a.$ Note the different index labels.
The electromagnetic gauge field has a $0$-component while the
Berry's phase has no ``frequency"-component. For the curvatures we
use $F_{\mu\nu}$ and $f_{ij}$ respectively.

\section{Derivation of Eq.
(\ref{2ndtknn})}\label{app:C2green}

In this appendix we will prove the conclusion (\ref{2ndtknn}). As
is briefly sketched in Sec. \ref{sec:secondtknn}, the
demonstration consists of three steps: (1) Topological invariance
of Eq. (\ref{C2green}); (2) Any Hamiltonian $h({\bf k})$ is
adiabatically connected to an $h_0({\bf k})$ with the form of Eq.
(\ref{projectorH}); (3) For such an $h_0({\bf k})$ the non-linear
correlation function (\ref{C2green}) is equal to the second Chern
number. In the following we will demonstrate these three steps
separately.

\subsection{Topological invariance of Eq. (\ref{C2green})}

To prove the topological invariance of Eq. (\ref{C2green}), we
just need to prove any infinitesimal deformation of the Green's
function $G({\bf k},\omega)$ leads to a vanishing variation of
$C_2$. Under a variation of $G({\bf k},\omega)$ we have
\begin{widetext}
\begin{eqnarray}
\delta\left(G\partial_\mu G^{-1}\right)&=&\delta G\partial_\mu
G^{-1}+G\partial_\mu\left(\delta G^{-1}\right)=\delta
G\partial_\mu G^{-1}-G\partial_\mu\left(G^{-1}\delta G
G^{-1}\right)
\nonumber\\
&=&-G\left(\partial_\mu G^{-1}\right)\delta G
G^{-1}-\partial_\mu\left(\delta G\right)G^{-1}.\nonumber
\end{eqnarray}
Thus the variation of $C_2$ is

\begin{eqnarray}
\delta C_2&=&-\frac {\pi^2}{15}\epsilon^{\mu\nu\rho\sigma\tau}\int
\frac{d^4kd\omega}{\left(2\pi\right)^5}{\rm
tr}\left[\delta\left(G\partial_\mu
G^{-1}\right)\left(G\partial_\nu G^{-1}\right)
\left(G\partial_\rho G^{-1}\right)\left(G\partial_\sigma G^{-1}\right)\left(G\partial_\tau G^{-1}\right)\right]\nonumber\\
&=&\frac {\pi^2}{15}\epsilon^{\mu\nu\rho\sigma\tau}\int
\frac{d^4kd\omega}{\left(2\pi\right)^5}{\rm
tr}\left[\left(G\partial_\mu G^{-1}\delta G
G^{-1}\right)\left({G\partial_\nu G^{-1}}\right)
\left(G\partial_\rho G^{-1}\right)\left(G\partial_\sigma G^{-1}\right)\left(G\partial_\tau G^{-1}\right)\right]\nonumber\\
& &+\frac {\pi^2}{15}\epsilon^{\mu\nu\rho\sigma\tau}\int
\frac{d^4kd\omega}{\left(2\pi\right)^5}{\rm
tr}\left[\left(\partial_\mu \delta G
G^{-1}\right)\left(G\partial_\nu G^{-1}\right)
\left(G\partial_\rho G^{-1}\right)\left(G\partial_\sigma G^{-1}\right)\left(G\partial_\tau G^{-1}\right)\right]\nonumber\\
&=&\frac {\pi^2}{15}\epsilon^{\mu\nu\rho\sigma\tau}\int
\frac{d^4kd\omega}{\left(2\pi\right)^5}{\rm tr}\left[\partial_\mu
\left(G^{-1}\delta G\right)\left(\partial_\nu G^{-1}G\right)
\left(\partial_\rho G^{-1}G\right)\left(\partial_\sigma G^{-1}G\right)\left(\partial_\tau G^{-1}G\right)\right]\nonumber\\
&=&\frac {\pi^2}{15}\epsilon^{\mu\nu\rho\sigma\tau}\int
\frac{d^4kd\omega}{\left(2\pi\right)^5}\partial_\mu {\rm
tr}\left[\left(G^{-1}\delta G\right)\left(\partial_\nu
G^{-1}G\right)
\left(\partial_\rho G^{-1}G\right)\left(\partial_\sigma G^{-1}G\right)\left(\partial_\tau G^{-1}G\right)\right]\nonumber\\
& &-\frac {\pi^2}{15}\epsilon^{\mu\nu\rho\sigma\tau}\int
\frac{d^4kd\omega}{\left(2\pi\right)^5}\left\{{\rm
tr}\left[\left(G^{-1}\delta G\right)\partial_\mu\left(\partial_\nu
G^{-1}G\right)
\left(\partial_\rho G^{-1}G\right)\left(\partial_\sigma G^{-1}G\right)\left(\partial_\tau G^{-1}G\right)\right]\right.\nonumber\\
& &+{\rm tr}\left[\left(G^{-1}\delta G\right)\left(\partial_\nu
G^{-1}G\right)
\partial_\mu\left(\partial_\rho G^{-1}G\right)\left(\partial_\sigma G^{-1}G\right)\left(\partial_\tau G^{-1}G\right)\right]\nonumber\\
& &+{\rm tr}\left[\left(G^{-1}\delta G\right)\left(\partial_\nu
G^{-1}G\right)
\left(\partial_\rho G^{-1}G\right)\partial_\mu\left(\partial_\sigma G^{-1}G\right)\left(\partial_\tau G^{-1}G\right)\right]\nonumber\\
& &\left.+{\rm tr}\left[\left(G^{-1}\delta
G\right)\left(\partial_\nu G^{-1}G\right)
\left(\partial_\rho G^{-1}G\right)\left(\partial_\sigma G^{-1}G\right)\partial_\mu\left(\partial_\tau G^{-1}G\right)\right]\right\}\nonumber\\
&=&\frac {\pi^2}{15}\epsilon^{\mu\nu\rho\sigma\tau}\int
\frac{d^4kd\omega}{\left(2\pi\right)^5}\partial_\mu{\rm
tr}\left[\left(G^{-1}\delta G\right)\left(\partial_\nu
G^{-1}G\right)
\left(\partial_\rho G^{-1}G\right)\left(\partial_\sigma G^{-1}G\right)\left(\partial_\tau G^{-1}G\right)\right]\nonumber\\
&\equiv &0.
\end{eqnarray}
\end{widetext}

Thus Eq. (\ref{C2green}) is topologically invariant. The
topological invariance of the second Chern number defined in Eq.
(\ref{2ndtknn}) is a well-known mathematical fact. Such a
topological invariance is quite helpful for showing the
equivalence between the second Chern number and Eq.
(\ref{C2green}).

\subsection{Adiabatic deformation of arbitrary $h({\bf k})$ to
$h_0({\bf k})$}

An adiabatic deformation $h(k,t),~t\in[0,1]$ can be written down,
which connects an arbitrary gapped Hamiltonian $h({\bf k})$ to a
``maximally degenerate Hamiltonian" $h_0({\bf k})$ in the form of
Eq. (\ref{projectorH}). Any single particle Hamiltonian $h({\bf
k})$ can be diagonalized as
\begin{eqnarray}
h({\bf k})=U({\bf k})D({\bf k})U^\dagger({\bf k})\nonumber
\end{eqnarray}
with $U({\bf k})$ unitary and $h_0({\bf k})={\rm
diag}\left[\epsilon_1({\bf k}),\epsilon_2({\bf
k}),...,\epsilon_N({\bf k})\right]$ the diagonal matrix of energy
eigenvalues. Without loss of generality, the chemical potential
can be defined to be zero, and the eigenvalues can be arranged in
ascending order. For an insulator with $M$ bands filled, one has
\begin{eqnarray}
\epsilon_1({\bf k})&\leq& \epsilon_2({\bf k})\leq ...\leq
\epsilon_M({\bf k})<0\nonumber\\
&<&\epsilon_{M+1}({\bf k})\leq...\leq \epsilon_N({\bf k}).
\end{eqnarray}

For $t\in[0,1]$, define
\begin{eqnarray}
E_\alpha({ {\bf
k}},t)=\left\{\begin{array}{cc}\epsilon_\alpha({\bf
k})(1-t)+\epsilon_Gt,&1\leq \alpha\leq M\\
\epsilon_\alpha({\bf k})(1-t)+\epsilon_Et,&M<\alpha\leq
N\end{array}\right.
\end{eqnarray} and
$D_0({{\bf k}},t)={\rm diag}\left[E_1({\bf k},t),E_2({\bf
k},t),...,E_N({\bf k},t)\right]$, then we have
\begin{eqnarray}
D_0({\bf k},0)=D({\bf k}),~D_0({\bf k},1)=\left(\begin{array}{cc}
\epsilon_G\mathbb{I}_{M\times
M}&\\&\epsilon_E\mathbb{I}_{N-M\times
N-M}\end{array}\right).\nonumber
\end{eqnarray}
As long as $\epsilon_G<0<\epsilon_H$, $D_0({\bf k},t)$ remains
gapped for $t\in[0,1]$. Thus by defining
\begin{eqnarray}
h({\bf k},t)=U({\bf k})D_0({\bf k},0)U^\dagger({\bf k})
\end{eqnarray}
we obtain an adiabatic interpolation between $h({\bf k},0)=h({\bf
k})$ and $h({\bf k},1)=U({\bf k})D_0({\bf k},1)U^\dagger({\bf k})$.
Since the matrix $U({\bf k})$ can be written in the eigenstates of
$h({\bf k})$ as $U({\bf k})=\left(\left|1,{\bf
k}\right\rangle,\left|2,{\bf k}\right\rangle,...,\left|N,{\bf
k}\right\rangle\right)$, we have
\begin{eqnarray}
h({\bf k},1)&=&\epsilon_G\sum_{\alpha=1}^M\left|\alpha,{\bf
k}\right\rangle\left\langle \alpha,{\bf
k}\right|+\epsilon_E\sum_{\beta=M+1}^N\left|\beta,{\bf
k}\right\rangle\left\langle \beta,{\bf k}\right|\nonumber\\
&=&\epsilon_GP_G({\bf k})+\epsilon_EP_E({\bf
k})\label{projectorHapp}
\end{eqnarray}

In summary, we have proven that each gapped Hamiltonian $h({\bf
k})$ can be adiabatically connected to a Hamiltonian with the form
of Eq. (\ref{projectorH}).

\subsection{Calculation of correlation function (\ref{C2green}) for
$h_0({\bf k})$}

 For Hamiltonian
of the form (\ref{projectorHapp}) the Green's function is written
in the simple form:
\begin{eqnarray}
G({\bf k},\omega)&=&\left[\omega+i\delta-\epsilon_GP_G({\bf
k})-\epsilon_EP_E({\bf k})\right]^{-1}\nonumber\\
&=&\frac {P_G({\bf k})}{\omega+i\delta-\epsilon_G}+\frac{P_E({\bf
k})}{\omega+i\delta-\epsilon_E}.
\end{eqnarray}
On the other hand, we have
\begin{eqnarray}
\frac{\partial G^{-1}({\bf k},\omega)}{\partial
\omega}&=&1\nonumber\\
\frac{\partial G^{-1}({\bf k},\omega)}{\partial
k_i}&=&-\epsilon_G\frac{\partial P_G({\bf k})}{\partial
k_i}-\epsilon_E\frac{\partial P_E({\bf k})}{\partial
k_i}\nonumber\\
&=&\left(\epsilon_E-\epsilon_G\right)\frac{\partial P_G({\bf
k})}{\partial k_i}
\end{eqnarray}\noindent where $i=1,2,3,4.$
Thus Eq. (\ref{C2green}) can be written
\begin{widetext}
\begin{eqnarray}
C_2=-\frac{\pi^2}{3}\epsilon^{ijk\ell}\int
\frac{d^4kd\omega}{\left(2\pi\right)^5}\sum_{n,m,s,t=1,2}\frac{{\rm
Tr}\left[P_n\frac{\partial P_G}{\partial k_i}P_m\frac{\partial
P_G}{\partial k_j}P_s\frac{\partial P_G}{\partial
k_k}P_t\frac{\partial P_G}{\partial
k_\ell}\right]\left(\epsilon_E-\epsilon_G\right)^4}{\left(\omega+i\delta-\epsilon_n\right)^2\left(\omega+i\delta-\epsilon_m\right)
\left(\omega+i\delta-\epsilon_s\right)\left(\omega+i\delta-\epsilon_t\right)}\label{C2forh0}
\end{eqnarray}
in which $\epsilon_{1,2}=\epsilon_{G,E}$ and $P_{1,2}({\bf
k})=P_{G,E}({\bf k})$, respectively. From the identity
$P_E+P_G\equiv \mathbb{I}$ and $P_E^2=P_E,~P_G^2=P_G$, we obtain
\begin{eqnarray}
P_E\frac{\partial P_G}{\partial k_i}&=&-\frac{\partial
P_E}{\partial k_i}P_G=\frac{\partial P_G}{\partial
k_i}P_G,~P_G\frac{\partial P_G}{\partial k_i}=-P_G\frac{\partial
P_E}{\partial k_i}=\frac{\partial P_G}{\partial k_i}P_E
.\label{Pidentity}
\end{eqnarray}
Consequently, $P_G\partial_i P_GP_G=P_E\partial_i P_GP_E=0$, so
that the trace in Eq. (\ref{C2forh0}) can be nonzero only when
$n\neq m,~m\neq s,~s\neq t,~t\neq n$. In other words, only two
terms are left out of the $16$ terms summed over in Eq.
(\ref{C2forh0}):
\begin{eqnarray}
C_2&=&-\frac{\pi^2}{3}\epsilon^{ijk\ell}\int
\frac{d^4kd\omega}{\left(2\pi\right)^5}\left\{\frac{{\rm
Tr}\left[P_G\frac{\partial P_G}{\partial k_i}P_E\frac{\partial
P_G}{\partial k_j}P_G\frac{\partial P_G}{\partial
k_k}P_E\frac{\partial P_G}{\partial
k_\ell}\right]}{\left(\omega+i\delta-\epsilon_G\right)^3
\left(\omega+i\delta-\epsilon_E\right)^2}+\frac{{\rm
Tr}\left[P_E\frac{\partial P_G}{\partial k_i}P_G\frac{\partial
P_G}{\partial k_j}P_E\frac{\partial P_G}{\partial
k_k}P_G\frac{\partial P_G}{\partial
k_\ell}\right]}{\left(\omega+i\delta-\epsilon_G\right)^2
\left(\omega+i\delta-\epsilon_E\right)^3}\right\}\left(\epsilon_E-\epsilon_G\right)^4
.\nonumber
\end{eqnarray}
Carrying out the integral over $\omega$ and using identities
(\ref{Pidentity}) again, we obtain
\begin{eqnarray}
C_2&=&\frac1{8\pi^2}\int d^4k\epsilon^{ijk\ell}{\rm
Tr}\left[P_E\frac{\partial P_G}{\partial k_i}\frac{\partial
P_G}{\partial k_j}P_E\frac{\partial P_G}{\partial
k_k}\frac{\partial P_G}{\partial k_\ell}\right].\label{C2inP}
\end{eqnarray}
\end{widetext}

Now we will show this is just the second Chern number. The Berry
phase gauge field is defined by
\begin{eqnarray}
a_i^{\alpha\beta}({\bf k})&=&-i\left\langle \alpha,{\bf
k}\right|\frac{\partial}{\partial k_i }\left|\beta,{\bf
k}\right\rangle\nonumber
\end{eqnarray}
in which $\alpha,\beta=1,2,..,M$ are the occupied bands. The
$U(M)$ gauge curvature is given by
\begin{eqnarray}
f^{\alpha\beta}_{ij}&=&\partial_i a^{\alpha\beta}_j-\partial_j
a^{\alpha\beta}_i+i\left[a_i,a_j\right]^{\alpha\beta}\nonumber\\
&=&-i\left(\frac{\partial \left\langle \alpha,{\bf
k}\right|}{\partial k_i}\frac{\partial \left|\beta,{\bf
k}\right\rangle}{\partial
k_j}-\left(i\leftrightarrow j\right)\right)\nonumber\\
& &+i\left(\frac{\partial \left\langle \alpha,{\bf
k}\right|}{\partial k_i}\sum_{\gamma=1}^{M}\left|\gamma,{\bf
k}\right\rangle\left\langle \gamma,{\bf k}\right|\frac{\partial
\left|\beta,{\bf k}\right\rangle}{\partial
k_j}-\left(i\leftrightarrow j\right)\right)\nonumber\\
&=&-i\left(\frac{\partial \left\langle \alpha,{\bf
k}\right|}{\partial k_i}P_E({\bf k})\frac{\partial
\left|\beta,{\bf k}\right\rangle}{\partial
k_j}-\left(i\leftrightarrow j\right)\right)\nonumber
\end{eqnarray}
in which $\left(i\leftrightarrow j\right)$ means the term with
$i,j$ exchanged. In operator form, we have
\begin{eqnarray}
\sum_{\alpha,\beta=1}^M\left|\alpha,{\bf k}\right\rangle
f_{ij}^{\alpha\beta}\left\langle \beta,{\bf
k}\right|=-i\frac{\partial P_G({\bf k})}{\partial
k_i}P_E\frac{\partial P_G({\bf k})}{\partial
k_j}-\left(i\leftrightarrow j\right).\nonumber
\end{eqnarray}
Thus Eq. (\ref{C2inP}) can be written in terms of the Berry phase
curvature as\cite{demler1999,murakami2004}
\begin{eqnarray}
C_2=\frac1{32\pi^2}\int d^4k\epsilon^{ijk\ell}{\rm
Tr}\left[f_{ij}f_{k\ell}\right].
\end{eqnarray}

In conclusion, the equivalence of the non-linear correlation
function (\ref{C2green}) and the second Chern number
(\ref{2ndtknn}) has been proven for the specific models of the
form Eq. (\ref{projectorH}). Due to the topological invariance of
both Eq. (\ref{C2green}) and the second Chern number, such a
relation holds for any band insulator in $(4+1)$-d.

The procedure in this section can be easily generalized to any odd
space-time dimensions. In $(2n+1)$-d space-time, a $n$-th Chern
number $C_n$ is defined in the BZ of a band-insulator, which
appears as a topological response coefficient to an external gauge
field. Equivalently, the coefficient of the $n$-th Chern-Simons
term in the effective action of an external gauge field obtained
from integrating out the fermions\cite{golterman1993}.

\section{The winding number in the non-linear response of Dirac-type
models}\label{app:dirac4}

Starting from the symmetric form, in terms of general Green's
functions, of Eq. (\ref{C2green}) we want to calculate $C_2$ for
the generalized lattice Dirac model $H(k)=d_a(k)\Gamma^a$. The
Green's function for this model can be simplified
to\cite{golterman1993}
\begin{equation}
G(k,\omega)=\frac{\omega+d^a(k)\Gamma^a}{(\omega^2-d^a(k)d_a(k))}.
\end{equation} Inserting this into Eq. (\ref{C2green}) yields the
following calculation
\begin{widetext}
\begin{eqnarray}
C_2&=&-\frac{\pi^2}{3}\epsilon_{ijkl}\int \frac{d^4 k
d\omega}{(2\pi)^5}\frac{1}{(\omega^2-|d|^2)^5}{\rm{Tr}}\left[(\omega+d_m\Gamma^m)(\omega+d_n\Gamma^n)(\partial_i
d^o\Gamma^o)(\omega+d_p\Gamma^p)(\partial_j d^q\Gamma^q)\right.
\nonumber \\
&\times& \left. (\omega +d_r\Gamma^r)(\partial_k
d^s\Gamma^s)(\omega+d_t\Gamma^t)(\partial_l
d^u\Gamma^u)\right]\nonumber\\
&=&-\frac{\pi^2}{3}\epsilon_{ijkl}\int \frac{d^4 k
d\omega}{(2\pi)^5}\frac{\partial_i d^o \partial_j d^q\partial_k
d^s\partial_l
d^u}{(\omega^2-|d|^2)^5}{\rm{Tr}}\left[(\omega+d_m\Gamma^m)(\omega+d_n\Gamma^n)\Gamma^o(\omega+d_p\Gamma^p)\Gamma^q(\omega
+d_r\Gamma^r)\Gamma^s(\omega+d_t\Gamma^t)\Gamma^u\right]\nonumber
\end{eqnarray}\end{widetext}
\noindent where $m,n,o,p,q,r,s,t,u=0,1,2,3,4.$ Among all the terms
in the bracket $[~]$, the only ones with non-zero traces are those
of $5,7,$ and $9$ $\Gamma$ matrices\cite{murakami2004} and the
complete trace simplifies nicely to $-4\epsilon_{toqsu} d^t
(\omega^2-|d|^2)^2$, which simplifies $C_2$ as
\begin{eqnarray}
C_2&=&\frac{\pi^2}{3}\epsilon_{ijkl}\int \frac{d^4 k
d\omega}{(2\pi)^5}4\epsilon_{toqsu}\frac{d^t\partial_i d^o
\partial_j d^q\partial_k d^s\partial_l d^u}{(\omega^2-|d|^2)^3}\nonumber\\
&=&\frac{\pi^2}{4}\epsilon_{ijkl}\int \frac{d^4 k
}{(2\pi)^4}\epsilon_{toqsu}\frac{d^t\partial_i d^o
\partial_j d^q\partial_k d^s\partial_l d^u}{|d|^5}\nonumber\\
&=&\frac{3}{8\pi^2}\int d^4k\epsilon_{toqsu}{\hat{d}^t\partial_x
\hat{d}^o
\partial_y \hat{d}^q\partial_z \hat{d}^s\partial_w \hat{d}^u}.
\end{eqnarray}\noindent Thus we have proved that the winding number given in the expression
(\ref{windingS4}) is equal to the second Chern number defined in Eq.
(\ref{C2green}) for generic $(4+1)$-d Dirac models.

\section{Stability of edge theories in
 generic dimensions}\label{app:edgestab}

In this appendix, we will determine the existence or absence of
$Z_2$ topological insulators in generic dimensions by studying the
stability of boundary theories. First of all, the boundary theory
of a topological insulator in $(2n+1)$-d with nontrivial $n$-th
Chern number $C_n$ is $|C_n|$ copies of chiral fermions:
\begin{eqnarray}
H={\rm
sgn}\left(C_n\right)\sum_{s=1}^{|C_n|}\sum_{a=1}^{2n-1}\psi_s^\dagger
vp_a\Gamma^a\psi_s
\end{eqnarray}
The $2n-1$ $\Gamma^a$ matrices are $2^{n-1}$ dimensional and form
an ${\rm so}(2n-1)$ Clifford algebra. We will study the lattice
Dirac model, since other systems with the same Chern number can be
obtained by an adiabatic deformation from the lattice Dirac model
and the topological stability of edge states does not depend on
the adiabatic deformation, as will be discussed below. For
simplicity, in the following we will focus on the case $C_n=1$, in
which case the single particle Hamiltonian can be written as
$h({\bf p})=vp_a\Gamma^a$. We have the following theorem about the
symmetry of the Hamiltonian:

\begin{itemize}
\item \textsc{Theorem I.} In a $2^{n-1}$ dimensional
representation of an ${\rm so(2n-1)}$ Clifford algebra generated
by $\Gamma^a,~a=1,2,..2n-1$, a unitary matrix $M_{(n)}$ can be
defined such that
\begin{eqnarray}
M_{(n)}^\dagger \Gamma^aM_{(n)}&=&(-1)^{n-1}\Gamma^{a*},~\forall
a=1,2,..,2n-1\nonumber\\ \label{MsymmetryGamma}\\
M_{(n)}^*M_{(n)}&=&(-1)^{[n/2]}\mathbb{I}\label{Msquare}
\end{eqnarray}
Consequently the chiral Hamiltonian $h({\bf p})=vp_a\Gamma^a$
satisfies
\begin{eqnarray}
M_{(n)}^\dagger h({\bf p})M_{(n)}=(-1)^nh^*(-{\bf
p}).\label{Msymmetry}
\end{eqnarray}
\end{itemize}
In Eq. (\ref{Msquare}) $[n/2]$ means the maximal integer that does
not exceed $n/2$.

Theorem I can be proved by induction. First, the $\Gamma^a$
matrices for $n=2$ are the Pauli matrices
$\Gamma^a=\sigma_a,a=1,2,3$, and $M_{(2)}=i\sigma_2$ satisfies the
theorem. Suppose the theorem is true for case $n$ with the matrix
$M_{(n)}$ and the $2n-1$ Gamma matrices $\Gamma^a_{(2n-1)}$, then
the $\Gamma^a_{(2n+1)}$ matrices for the ${\rm so}(2n+1)$ Clifford
algebra can be generated by
\begin{eqnarray}
\Gamma^a_{(2n+1)}=\left\{\begin{array}{cc}\Gamma^a_{(2n-1)}\otimes
\tau_y,&a=1,2,..2n-1\\\mathbb{I}\otimes
\tau_x,&a=2n\\\mathbb{I}\otimes \tau_z,&a=2n+1\end{array}\right. .
\end{eqnarray}
It is straightforward to check the anticommutation relations.
Defining
\begin{eqnarray}
M_{(n+1)}=\left\{\begin{array}{cc}M_{(n)}\otimes
i\tau_y,&n\text{~odd}\\M_{(n)}\otimes
\mathbb{I},&n\text{~even}\end{array}\right.
\end{eqnarray}
we find $M_{(n+1)}$ satisfies the Theorem I.

From Eqs. (\ref{Msquare}) and (\ref{Msymmetry}) one finds
different properties of $M_{(n)}$ in different dimensions. (i) For
$n=4k-3,~k\in\mathbb{Z}$, $M_{(n)}$ reverses the sign of energy,
and satisfies $M_{(n)}^*M_{(n)}=\mathbb{I}$, which can be
identified as particle-hole symmetry; (ii) for
$n=4k-2,~k\in\mathbb{Z}$, $M_{(n)}$ preserves the energy and
satisfies $M_{(n)}^*M_{(n)}=-\mathbb{I}$, which can be identified
as the time-reversal symmetry; (iii) for $n=4k-1,~k\in\mathbb{Z}$,
$M_{(n)}$ reverses the sign of energy but satisfies
$M_{(n)}^*M_{(n)}=-\mathbb{I}$, which we call ``psuedo"
particle-hole symmetry and denote by $\tilde{C}$; (iv) for
$n=4k,~k\in\mathbb{Z}$, $M_{(n)}$ preserves the energy but
satisfies $M_{(n)}^*M_{(n)}=\mathbb{I}$, which behaves like the
time-reversal symmetry of an integer-spin particle, and we call
``pseudo" time-reversal symmetry $\tilde{T}$.

In the following, we will denote the symmetries $C$, $\tilde{C}$,
$T$ and $\tilde{T},$ defined by Eq. (\ref{Msymmetry}), by
$M$-symmetry. Now we study what other terms can be added in the
Hamiltonian without breaking the $M$ symmetry. Given $\Gamma^a$
for the ${\rm so(2n-1)}$ case, all the $2^{n-1}\times 2^{n-1}$
Hermitian matrices can be expanded in the basis
\begin{eqnarray}
\left\{\mathbb{I},\Gamma^{a_1a_2...a_m}=i^{m(m-1)/2}\Gamma^{a_1}\Gamma^{a_2}...\Gamma^{a_m},~m=1,2,..n-1\right\}.\nonumber
\end{eqnarray}
As expected, the total number of matrices forming the basis is
$1+\sum_{m=1}^{n-1}\left(_{2n-1}^m\right)=2^{2n-2}$. The $M_{(n)}$
transformation property of $\Gamma^{a_1a_2..a_m}$ can be determined
by that of $\Gamma^a$ as
\begin{eqnarray}
M_{(n)}^\dagger\Gamma_{(n)}^{a_1a_2..a_m}M_{(n)}=(-1)^{m(2n+m-3)/2}
{\Gamma_{(n)}^{a_1a_2..a_m*}}\label{MGammaab}
\end{eqnarray}

If we have a constant term $m_{a_1a_2..a_m}\Gamma^{a_1a_2..a_m}$ in
the Hamiltonian without breaking the $M$ symmetry, the following
condition must be satisfied:
\begin{eqnarray}
M^\dagger\Gamma^{a_1a_2..a_m}
M=(-1)^n{\Gamma^{a_1a_2..a_m*}}.\label{MGammaab2}
\end{eqnarray}
Eqs. (\ref{MGammaab}) and (\ref{MGammaab2}) thus require
\begin{eqnarray}
(-1)^{m(2n+m-3)/2}=(-1)^{n}\label{constraintM}
\end{eqnarray}
so that $(m-1)(2n-2+m)/2$ must be odd. This condition can be
satisfied by several possibilities: (1) when $m$ is odd, $(m-1)/2$
must be odd, which means $m=4k-1,~k\in\mathbb{N}$. (2) When $m$ is
even, $n-1+m/2$ must be odd, which means $m=4k$ if $n$ is even, or
$m=4k-2$ if $n$ is odd. In summary, the terms available in the
Hamiltonian are given by
\begin{eqnarray}
m&=&\left\{\begin{array}{cc}4k-1\text{~or~}4k,&n\text{~even}\\
4k-1\text{~or~}4k-2,&n\text{~odd}\end{array}\right.,\label{mconstraint}
\end{eqnarray}
in which $k\in\mathbb{N}$ and $m$ is also bounded by $1\leq m<n$.
For example, the first non-trivial case is $n=2$ in which
$\Gamma^a,a=1,2,3$ are Pauli matrices. Since $n$ is even, $m$ is
required to be $4k-1$ or $4k$, in which the lowest value is $m=3$.
Consequently there is no mass term available. When $n=3$
($2n-1=5$) the $\Gamma^a,a=1,2,..,5$ matrices are the usual Dirac
matrices, and the only value of $m$ satisfying Eq.
(\ref{mconstraint}) is $m=2$. In other words, the terms
$\Gamma^{ab}=i\Gamma^a\Gamma^b$ do not break the corresponding
symmetry---the pseudo particle-hole symmetry $\tilde{C}$.

Though there are all these constant terms available, the perturbed
Hamiltonian
$h(k)=k_a\Gamma^a+\sum_mm_{a_1a_2..a_m}\Gamma^{a_1a_2..a_m}$ remains
gapless because each $\Gamma^{a_1a_2..a_m}$ commutes with some
$\Gamma^a$. If $m$ is even, it commutes with $\Gamma^b,~b\neq
a_s~\forall s=1,..,m$. If $m$ is odd, it commutes with
$\Gamma^{a_s},~\forall s=1,..,m$. For the Hamiltonian with only one
constant term $m_{a_1a_2...a_m}\Gamma^{a_1a_2...a_m}$, one can take
$p_i=0$ for all $i$ except for $i=a$, where $a$ is chosen such that
$\Gamma^a$ commutes with $\Gamma^{a_1a_2..a_m}$. Due to the
commutativity, the Hamiltonian can be diagonalized along the $a$-th
axis, with the eigenvalues $p_a\pm m$. Consequently, we know the
Hamiltonian is gapless. In other words, we have shown that no
perturbation with the $2^{n-1}$ band theory can open a gap for the
$2n$ dimensional chiral fermion, which agrees with the topological
stability of the $(2n+1)$-d bulk system characterized by the $n$-th
Chern number.

Starting from the $(2n+1)$-d topological insulator, dimensional
reduction procedures can be carried out to obtain a $((2n-1)+1)$-d
topological insulator. Correspondingly, the edge theory of the
$((2n-2)+1)$-d topological insulator is given by the dimensional
reduction of the $((2n-1)+1)$-d chiral fermion. If a nontrivial
topological insulator can be defined, it has the boundary theory
$h({\bf p})=\sum_{a=1}^{2n-2}p_a\Gamma^a$. Compared to the
$((2n-1)+1)$-d chiral fermion, one $\Gamma^a$ matrix is absent in
the theory. In the same way, the boundary theory of lower
dimensional descendants can be obtained by removing more momenta and
$\Gamma^a$ matrices from the chiral fermion Hamiltonian. Obviously,
if too few $\Gamma^a$ matrices are left, a mass term will be
available, which anti-commutes with all the rest of the $\Gamma^a$
matrices and thus can make gap the whole edge spectrum. The upper
critical dimension where the edge states become unstable is
determined by the maximal number of $\Gamma^a$ matrices that
anti-commute with some $\Gamma^{a_1a_2..a_m}$. For $m$ even, the
maximal number of $\Gamma^a$'s that anticommute with
$\Gamma^{a_1a_2..a_m}$ is $m$, while for $m$ odd, the maximal number
is $2n-1-m$. On the other hand, the available values of $m$ are
defined by the constraint Eq. (\ref{mconstraint}). By studying the
cases $n=4k-3,4k-2,4k-1,4k,~k\in\mathbb{N}$ separately, we obtain
that the space-time dimension in which the chiral theory becomes
unstable is given by $d=2n-3$. For example, in the case $n=3$
discussed earlier, the mass terms $\Gamma^{ab}$ are permitted by the
pseudo particle-hole symmetry. When the dimension is reduced from
$(5+1)$-d to $(2+1)$-d, only two $\Gamma^a$ matrices are used in the
$k$-linear terms, so that some $\Gamma^{ab}$ can be found which
anti-commutes with the gapless Hamiltonian and thus can make the
surface theory gapped. Consequently, topologically stable boundary
theories as descendants of $((2n-1)+1)$-d chiral fermions can only
exist in $((2n-2)+1)$ and $((2n-3)+1)$ dimensions. Correspondingly,
the $Z_2$ topological insulators as descendants of the $(2n+1)$-d
topological insulator (with $n$-th Chern number) can only be defined
in $((2n-1)+1)$ and $((2n-2)+1)$ dimensions.


\begin{thebibliography}{79}
\bibitem[{foo({\natexlab{a}})}]{footnoteGamma}
\bibinfo{note}{We use the capital $\Gamma^a$ here since they are defined
  differently from the usual Dirac matrix $\gamma^a,~a=0,1,2,3,5$. The relation
  between $\Gamma^a$ and $\gamma^a$ is $\Gamma^0=\gamma^0,~\Gamma^{1,2,3,4}=
  -i\gamma^0\gamma^{1,2,3,5}$.}
\bibitem[{foo({\natexlab{b}})}]{footnoteGamma2}
\bibinfo{note}{More explicitly, if we choose the representation
  $\Gamma^{0}=\sigma^3\otimes \mathbb{I}$, $\Gamma^{1,2,3}=\sigma^1\otimes
  \tau^{1,2,3}$ and $\Gamma^4=\sigma^2\otimes \mathbb{I}$, then the
  time-reversal transformation matrix $T$ is defined as $T=\mathbb{I}\otimes
  i\tau^2$.}
\end{thebibliography}
\end{document}